\documentstyle[11pt,fullpage,doublespace,epsf]{article}
 \DeclareMathSizes{11}{19}{13}{9}   

\makeatletter
\@addtoreset{equation}{section}
\makeatother

\begin{document}

\title{Riemann Hypothesis and Short Distance Fermionic Green's Functions}
\author{Michael McGuigan\\Brookhaven National Laboratory\\Upton NY 11973\\mcguigan@bnl.gov}
\date{}
\maketitle

\begin{abstract}We show that the Green's function of a two dimensional fermion with a modified dispersion relation and short distance parameter $a$
is given by the Lerch zeta function. The Green's function is defined on a cylinder of radius R and we show that the condition $R = a$
 yields the Riemann zeta function as a quantum transition amplitude for the fermion. We formulate the Riemann hypothesis physically as a nonzero condition on the transition amplitude between two special states associated with the point of origin and a point half way around the cylinder each of which are fixed points of a $Z_2 $
transformation. By studying partial sums we show that that the transition amplitude formulation is analogous to neutrino mixing in a low dimensional context. We also derive the thermal partition function of the fermionic theory and the thermal divergence at temperature $1/a.$ In an alternative harmonic oscillator formalism we discuss the relation to the fermionic description of two dimensional string theory and matrix models. Finally we derive various representations of the Green's function using energy momentum integrals, point particle path integrals, and string propagators. \end{abstract}

\section{Introduction}

The Riemann Hypothesis (that the Riemann zeta function has nontrivial zeros only when the real part of its argument is one half) embodies one of the deepest mysteries in mathematics. Mathematics and physics have similar origins and many seemingly abstract results in mathematics turn out to have physical realizations. Because of the depth of the Riemann hypothesis one may hope not only that  a physical realization of the hypothesis may lead to its solution, but also that its formulation in physical terms may uncover physical ideas perhaps related to short distance physics and string theory. There are several examples of a physical realization of mathematical developments leading to physical advances. The relation of the Atiyah-Singer index theorem to anomalies, instantons, and chiral fermions is one example, the relation of Chern-Simons gauge theory and knot theory to three manifold theory is another, also one has the Frenkel-Kac vertex operator construction of infinite Lie algebras and the relation to the heterotic string. Thus one can expect that the physical realization of mathematical problems like the Riemann hypothesis has benefit for both mathematics and physics.

The search for a physical realization of the Riemann Hypothesis has a long history going back to Hilbert and Polya who looked for a quantum hermitian operator with eigenvalues at the Riemann zeros. More recent approaches include the relation of the Riemann zeros to a Jost function in scattering theory \cite{Khuri:2001yd}, the relation of the zeta function to quantum chaos \cite{Berry}, and the relation of the Riemann hypothesis to noncommutative geometry \cite{Connes:2004es}. More information about recent approaches to the Riemann hypothesis are described in \cite{Elizalde:2001ee}.

In this paper this we give a new physical realization of the Riemann hypothesis in terms of a  $1 + 1$
 dimensional fermionic theory whose short distance dynamics are modified in a definite way. The short distance dynamics introduce a new parameter of dimension length or inverse energy. Within this theory we derive the propagator or Green's function which gives the transition amplitude to observe the fermion at any point in space-time $(x,t)$
given that its initial position and time. When the spatial coordinate $x$
 livetakes its values on a circle $S^1 $
 of circumference $L = 2\pi R$
 we show that this Green's function is proportional to the Lerch zeta function with arguments $(\displaystyle\frac{x}{{2\pi R}},\displaystyle\frac{{it}}{a},\displaystyle\frac{R}{a})$
. Only when the ratio $\displaystyle\frac{L}{{2\pi a}} = \displaystyle\frac{R}{a} = 1$
, that is when the radius of the universe equals the new short distance scale , do we obtain the Riemann zeta function, so in this sense the Riemann zeta function probes short distance behavior of the Green's function. By considering initial states of the fermionic particle that are statistically distributed, (essentially Boltzmann distributed with respect to energy and a mixing parameter $\sigma $
) we show that the transition amplitude is modified by the Lerch zeta function with parameters $(\displaystyle\frac{x}{{2\pi R}},\displaystyle\frac{{it}}{a} + \displaystyle\frac{\sigma }{a},\displaystyle\frac{R}{a})$
. The Riemann hypothesis is formulated physically as saying that the transition probability to a position $\displaystyle\frac{L}{2}$
 half way across the $1 + 1$
 dimensional universe is strictly nonzero for all times and values of the mixing parameter in the range $\displaystyle\frac{1}{2} < \displaystyle\frac{\sigma }{a} < 1$
.

 This paper is organized as follows. In section 2 we give a review of certain Dirichlet series and zeta functions that will occur in this paper. In section 3 we discuss the transition amplitudes and Green's function of  the 1+1 dimensional fermionic theory in the first quantized and second quantized field theory point of view and explain the relation to the zeta function. In section 4 we introduce the notion of a Boltzmann distributed initial state and introduce the mixing parameter $\sigma $
. We formulate the Riemann hypothesis as saying that the transition rate is never zero for the transition from initial state parameterized by $\sigma $
and $x = 0$
 to a final position state at $x = \displaystyle\frac{L}{2} = \pi R$
. We discuss a simplified analysis in terms of partial sums and show how the behavior of the transition probabilities is reflected in the phenomena of neutrino mixing. We also discuss the $\sigma $
 parameter from several points of view including the integral over theta functions, target space duality, and space-time reflections. In section 5 we discuss the statistical mechanics of the nonstandard 1+1 dimensional fermionic theory and its relation to multiplicative number theory and thermal divergence. In section 6 we discuss an alternate formulation in terms a nonstandard oscillator and discuss the relation to the fermionic form of two dimensional string theory and matrix models. In section 7 we discuss the relationship between ground ring operators, algebraic curves, modular forms and L-series. In section 8 we discuss other representations of the fermionic Green's function including point particle path integrals, energy momentum integrals, and string modifications to the Green's function. We also investigate the relation of the Wick rotated Green's function relation to discrete field theory. In section 9 we discuss the main conclusions of the paper. In Appendix A we discuss the permutation of the position, time, momentum and energy and how this effects the physical picture of the zeta function. In appendix B we discuss the one particle partition function in more detail. In appendix C we discuss a similar treatment for the scalar particle.

\section{Review of Dirichlet functions}

In this section we briefly review various Dirichlet functions used in this paper. For more details the reader can consult texts such as \cite{Titchmarsh}, \cite{Edwards}, \cite{Ivic}, \cite{Laurincikas} and \cite{bateman}.
Dirichlet functions are defined through series expansions of the form $f(t,\sigma ) = \sum\limits_{n = 1}^\infty  {a_n \displaystyle\frac{1}{{n^{it + \sigma } }}} $. Some prominent examples of Dirichlet functions are the Riemann zeta function, Dirichlet eta function, lambda function and beta function defined by:
	\begin{equation}
\begin{array}{l}
 \zeta (it + \sigma ) = \sum\limits_{n = 1}^\infty  {n^{ - (it + \sigma )} }      :\sigma  > 1 \\ 
 \eta (it + \sigma ) = \sum\limits_{n = 1}^\infty  {n^{ - (it + \sigma )} ( - 1)^{n + 1} }  =  - (2^{1 - (it + \sigma )}  - 1)\zeta (it + \sigma ) \\ 
 \lambda (it + \sigma ) = \sum\limits_{n = 0}^\infty  {(2n + 1)^{ - (it + \sigma )} }  = (1 - 2^{ - (it + \sigma )} )\zeta (it + \sigma ) \\ 
 \beta (it + \sigma ) = \sum\limits_{n = 0}^\infty  {(2n + 1)^{ - (it + \sigma )} ( - 1)^n }  \\ 
  \\ 
 \end{array}
\end{equation}
In terms of these functions the Riemann hypothesis is equivalent to $\eta  (it + \sigma ) \ne 0$
for $\displaystyle\frac{1}{2} < \sigma  < 1$
. Dirichlet functions obey a functional relation that relates its value at $s = it + \sigma $
to its value at $1 - s =  - it + (1 - \sigma )$
. The extension of the Riemann hypothesis from $\displaystyle\frac{1}{2} < \sigma  < 1$
 to the whole critical strip $0 < \sigma  < 1$
 is straight forward using the functional relation for eta:
	\begin{equation}
\eta (s) = 2\pi ^{s - 1} \sin (\displaystyle\frac{{\pi s}}{2})\Gamma (1 - s)\eta (1 - s)
\end{equation}
or in terms of the functional equation for the Riemann zeta function:
\begin{equation}
\begin{array}{l}
 \zeta (s) = 2^s \pi ^{s - 1} \sin (\displaystyle\frac{{\pi s}}{2})\Gamma (1 - s)\zeta (1 - s) \\ 
 \zeta (1 - s) = 2(2\pi )^{ - s} \cos (\displaystyle\frac{{\pi s}}{2})\Gamma (s)\zeta (s) \\ 
 \pi ^{ - s/2} \Gamma (s/2)\zeta (s) = \pi ^{ - (1 - s)/2} \Gamma ((1 - s)/2)\zeta (1 - s) \\ 
 \end{array}
\end{equation}

In addition to the series representation for Dirichlet functions one has integral representations. For example the eta and Riemann zeta function are given by:
	\begin{equation}
\eta (s) =  - \displaystyle\frac{1}{{\Gamma (s)}}\int\limits_{ - \infty }^\infty  {dE\displaystyle\frac{1}{{\exp (e^{ - E} ) + 1}}} e^{sE}  =  - \displaystyle\frac{1}{{\Gamma (s)}}\int\limits_0^\infty  {dw\displaystyle\frac{1}{{\exp w + 1}}} w^{s - 1}     :0 < \sigma  < 1
\end{equation}
and 
\begin{equation}
\zeta (s) = \displaystyle\frac{1}{{\Gamma (s)}}\int\limits_{ - \infty }^\infty  {dE\displaystyle\frac{1}{{\exp (e^{ - E} ) - 1}}} e^{sE}  = \displaystyle\frac{1}{{\Gamma (s)}}\int\limits_0^\infty  {dw\displaystyle\frac{1}{{\exp w - 1}}} w^{s - 1}      :\sigma  > 1
\end{equation}
The Dirichlet functions can also be represented as an integral over theta functions. For example the Dirichlet eta function can be written as:
\begin{equation}
\eta (s) = \displaystyle\frac{1}{{2^s  - 1}}\pi ^{\displaystyle\frac{s}{2}} \displaystyle\frac{1}{{\Gamma (\displaystyle\frac{s}{2})}}\int\limits_0^\infty  {\displaystyle\frac{{d\tau }}{\tau }} \tau ^{s/2} (\theta _4 (0|i\tau ) + \theta _2 (0|i\tau ) - \theta _3 (0|i\tau ))
\end{equation}
and the Riemann zeta function can be expressed as:
\begin{equation}
\zeta (s) = \pi ^{\displaystyle\frac{s}{2}} \displaystyle\frac{1}{{\Gamma (\displaystyle\frac{s}{2})}}\int\limits_0^\infty  {\displaystyle\frac{{d\tau }}{\tau }} \tau ^{s/2} (\theta _3 (0|i\tau ) - 1)       :\sigma  > 2
\end{equation}

Several other functions can be defined which place these Dirichlet functions in a more generalized context. The Riemann zeta function can be seen as a special case of the Hurwitz zeta function defined by:
	\begin{equation}
\zeta _H (it + \sigma ,\alpha ) = \sum\limits_{n = 0}^\infty  {(n + \alpha )^{ - (it + \sigma )} } 
\end{equation}
which is in turn a special case of the Lerch zeta function defined by:
	\begin{equation}
\phi (x,it + \sigma ,\alpha ) = \sum\limits_{n = 0}^\infty  {(n + \alpha )^{ - (it + \sigma )} e^{2\pi inx} } 
\end{equation}
The Lerch zeta function also has the integral representation for $\sigma  > 0$
 :
	\begin{equation}
\begin{array}{l}
 \phi (x,it + \sigma ,\alpha ) = \displaystyle\frac{1}{{\Gamma (it + \sigma )}}\int\limits_{ - \infty }^\infty  {dEe^{ - itE - \sigma E} \displaystyle\frac{1}{{\exp ((\alpha  - 1)e^{ - E}  + 2\pi ix) - \exp (\alpha e^{ - E} )}}}  \\ 
                        = \displaystyle\frac{1}{{\Gamma (s)}}\int\limits_0^\infty  {dww^{s - 1} \displaystyle\frac{{e^{ - \alpha w} }}{{1 - \exp ( - w + 2\pi ix)}}}  \\ 
 \end{array}
\end{equation}
and in terms of theta functions is expressed as:
\begin{equation}
\begin{array}{l}
 \phi (x,s,\alpha ) + e^{ - 2\pi ix} \phi ( - x,s,1 - \alpha ) = \displaystyle\frac{{\pi ^{s/2} }}{{\Gamma (s/2)}}\int\limits_0^\infty  {d\tau   \tau ^{s/2 - 1/2} \sum\limits_{n\varepsilon Z} {e^{ - \pi (n + \alpha )^2 \tau } e^{2\pi inx} } }  \\ 
  \\ 
 \end{array}
\end{equation}
The functional relation of the Lerch zeta function is:
	\begin{equation}
\begin{array}{l}
 \phi (x,1 - s,\alpha ) = \sum\limits_n {\displaystyle\frac{{e^{2\pi inx} }}{{(n + \alpha )^{1 - s} }}}  =  \\ 
            (2\pi )^{ - s} \Gamma (s)e^{2\pi i(\displaystyle\frac{1}{4}s - \alpha x)} \sum\limits_n {\displaystyle\frac{{e^{ - 2\pi in\alpha } }}{{(n + x)^s }}}  + (2\pi )^{ - s} \Gamma (s)e^{2\pi i(\displaystyle\frac{1}{4}s - \alpha x)} \sum\limits_n {\displaystyle\frac{{e^{2\pi in\alpha } }}{{(n + 1 - x)^s }}}  \\ 
               = (2\pi )^{ - s} \Gamma (s)e^{2\pi i(\displaystyle\frac{1}{4}s - \alpha x)} \phi ( - \alpha ,s,x) + (2\pi )^{ - s} \Gamma (s)e^{2\pi i(\displaystyle\frac{1}{4}s - \alpha x)} \phi (\alpha ,s,1 - x) \\ 
 \end{array}
\end{equation}
Another useful function is given by Laurincikas as:
	\begin{equation}
\varphi (x,s) = \sum\limits_{n = 1}^\infty  {e^{2\pi ixn} n^{ - s} } 
\end{equation}
for $\sigma  > 1$
with integral representations:
\begin{equation}
\begin{array}{l}
 \varphi (x,s) = \displaystyle\frac{{\pi ^{ - s} }}{{\Gamma (s/2)}}\int\limits_0^\infty  {d\tau (\tau ^{s/2} \sum\limits_{n > 0} {e^{ - \pi \tau (n)^2 } e^{2\pi inx} } } ) \\ 
 \varphi (x,s) = \displaystyle\frac{1}{{\Gamma (s)}}\int\limits_{ - \infty }^\infty  {dEe^{ - sE} \displaystyle\frac{1}{{\exp (2\pi ix) - \exp (e^{ - E} )}}}\\
  = \displaystyle\frac{1}{{\Gamma (s)}}\int\limits_0^\infty  {dw(w^{s - 1} \displaystyle\frac{1}{{\exp (2\pi ix) - \exp (w)}}} ) \\ 
 \end{array}
\end{equation}
This function is given by the Lerch function for $\alpha  = 1$
as $\varphi (x,s) = \phi (x,s,1)$.

In terms of the Lerch Zeta function the Dirichlet functions are given by:
\begin{equation}
\begin{array}{l}
 \phi (x = 0,it + \sigma ,\alpha  = 1) = \zeta (it + \sigma ) \\ 
 \phi (x = \displaystyle\frac{1}{2},it + \sigma ,\alpha  = 1) =  - \eta (it + \sigma ) \\ 
 \phi (x = 0,it + \sigma ,\alpha  = \displaystyle\frac{1}{2}) = 2^{ - (it + \sigma )} \lambda (it + \sigma ) \\ 
 \phi (x = \displaystyle\frac{1}{2},it + \sigma ,\alpha  = \displaystyle\frac{1}{2}) = 2^{ - (it + \sigma )} \beta (it + \sigma ) \\ 
 \end{array}
\end{equation}
and in terms of the Lerch function the Riemann hypothesis is 
\begin{equation}
\phi (it + \sigma ,x = \displaystyle\frac{1}{2},\alpha  = 1) =  - \eta  (it + \sigma ) \ne 0
\end{equation}
  for $\displaystyle\frac{1}{2} < \sigma  < 1$. $\zeta (s)$
and $\beta (s)$ are examples of Dirichlet $L$
 functions and are denoted by $L_{ + 1} $
and $L_{ + 4} $ respectively. Other Dirichlet $L$
 functions can be defined from linear combinations of the Lerch function at fractional values of $\alpha$. Clearly the Lerch zeta is a key mathematical object. In the next three sections we describe how to represent this zeta function and its parameters as a transition amplitude of a 2D fermionic quantum field theory.

\section{Propagator, transition amplitudes and Green's function associated with the Lerch zeta function} 

In the one particle description the propagator, transition amplitude or Green's function of a quantum system is represented by:
\begin{equation}
A(1 \to 1) = G(x,t;x',t') = \left\langle {x,t|\left. {x',t'} \right\rangle  = } \right.\left\langle {x'} \right.|U(t - t')|\left. x \right\rangle  = \left\langle {x'} \right.|e^{ - i\hat E(t - t')} |\left. x \right\rangle 
\end{equation}
Here $\hat E$
is the one particle energy operator. In a second quantized description one represents this Green's function as the two point function:

	\begin{equation}
G(x,t;x',t') = \left\langle {0|T\hat \psi (x,t)} \right. \bar \psi (x',t')|\left. 0 \right\rangle 
\end{equation}
where $\hat \psi (x,t)$
 is a second quantized field. The wave function is promoted to an operator and $|\left. 0 \right\rangle $
is the vacuum state. In this paper we will always choose $t' < t$
so that the expression will be implicitly time ordered. The one particle and second quantized descriptions are related because the one particle state is given by $|\left. {x,t} \right\rangle  = \psi ^\dag  (x,t)|\left| 0 \right\rangle $. For a free particle, the transition amplitude is completely determined by the dispersion relation between energy and momentum as well as the spatial boundary conditions and whether it is a boson or fermion. 

We consider four cases of dispersion relations and derive the associated fermionic Green's functions. These are: 

(1)$E = p$
, a right moving particle moving at the speed of light. 

(2) $E = p$
, a right moving particle with $x$
taking values on a spatial circle $S^1 $
of circumference $L = 2\pi R$
.
 
(3) $E = \displaystyle\frac{1}{a}\log (ap + 1)$
, a right moving particle with logarithmic dispersion and 

(4) $E = \displaystyle\frac{1}{a}\log (ap + 1)$
, a right moving particle with spatial coordinate $x$
taking values on a spatial circle $S^1 $
of circumference $L$. 

\noindent The first two cases are standard and are included for pedagogical purposes

\subsection*{Case (1): $E = p$}

The dispersion relation $E = p$
 with $p \ge 0$ represents the usual case of a right moving massless particle on space-time $R \times R$. 
Classically we denote $E$
as the one particle energy, $p$
the momentum, $v$
the velocity, $x$
 position and $t$
time. The basic classical relations are given by the formulas:
\begin{equation}
\begin{array}{l}
 {\rm{Energy: }}\;\;\;\;       E = cp \\ 
 {\rm{Momentum:}}\;\;\;\;p = \displaystyle\frac{E}{c} \\ 
 {\rm{Velocity:}}\;\;\;\;       v = \displaystyle\frac{{\partial E}}{{\partial p}} = c \\ 
 {\rm{Position: }}\;\;\;\;     x = ct \\ 
 {\rm{Time: }}\;\;\;\;           t = \displaystyle\frac{1}{c}x \\ 
 \end{array}
\end{equation}
The dispersion relation is given by the first two lines in (3.3) and depicted in figure 1.

\begin{figure}[htbp]

   \centerline{\hbox{
   \epsfxsize=2.0in
   \epsffile{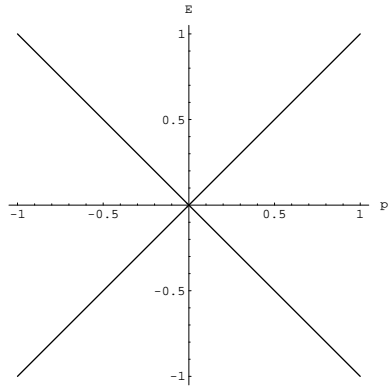}
     }
  }
  \caption{Dispersion relation for Case (1) $E=p$ or $E=-p$ } 
          
  \label{fig1}

\end{figure}

For the most part we shall choose units where $c = 1$
and $\not h = 1$
however sometimes we will include them for illustration. Quantum mechanically the dispersion relation becomes the operator equation $\hat E = \hat p$
. The momentum operator then has momentum eigenstates denoted by $
\hat p\left. {|p} \right\rangle  = p\left. {|p} \right\rangle$. In the position basis we have spatial wave functions: $\psi _p (x) = \left\langle {x|\left. p \right\rangle } \right. = e^{ipx}$. The time dependent wave function is given by: $\psi _p (x,t) = \left\langle {x|e^{ - iHt} } \right.|\left. p \right\rangle  = e^{ - iE_p t} e^{ipx}  = e^{ - ipt} e^{ipx} $
and the transition amplitude is expressed as:
\begin{equation}A_1  = \left\langle {x',t'|\left. {x,t} \right\rangle  = } \right.\left\langle {x'} \right.|U(t' - t)|\left. x \right\rangle  = \left\langle {x'} \right.|e^{ - iH(t' - t)} |\left. x \right\rangle  = \sum\limits_p {e^{ - iE_p (t - t')} \psi (x)\psi ^* (x')} 
\end{equation}
Using the specific form of the wave function in the position basis we have:
\begin{equation}
\begin{array}{l}
 A_1  = \displaystyle\frac{1}{{2\pi }}g_1 (x - x',s) = G_1 (x,t;x',t') = \sum\limits_k {e^{ - iE_k (t' - t)} \psi (x)\psi^{*}(x')}  \\ 
     \;\;\;\; = \displaystyle\frac{1}{{2\pi }}\int\limits_0^\infty  {dpe^{ - ip(t - t') + ip(x - x')} }  = \displaystyle\frac{1}{{2\pi }}\displaystyle\frac{i}{{(x - x') - (t - t')}} \\ 
 \end{array}
\end{equation}
So that:
\begin{equation}
g_1 (x,s) = 2\pi G_1 (x,t;0,0) = \displaystyle\frac{i}{{x + is/a}}
\end{equation}
where $s = it/a$.

To see that this is the correct form of the Green's function note that Wick rotating to Euclidean space $\bar G(x,\bar t,0,0) = \partial _{x - i\bar t} K(x,\bar t,0,0) = 2\partial _{x - i\bar t} \displaystyle\frac{1}{{4\pi }}\log (x^2  + \bar t^2 ) = \displaystyle\frac{1}{{2\pi }}\displaystyle\frac{1}{{x + i\bar t}}$  a standard result in 2d field theory. Here  $K(x,t;0,0)$
 is the scalar two point Green's function and $G(x,t;0,0)$
is the two point chiral fermionic Green's function. Wick rotating back to real space we obtain (3.6). Note the usual notation for $G$
and $K$
is $S$
and $D$
, we don't use this notation as it sometimes conflicts with notation for the action and Dirac operator.

To obtain the Green's function in the second quantized field theory formalism one can write the Dirac equation for a chiral fermion in 1+1 dimensions as:
\begin{equation}
\begin{array}{l}
 (i\gamma ^0 \partial _0  + i\gamma ^1 \partial _1 )\displaystyle\frac{{(1 + \gamma _3 )}}{2}\left( \begin{array}{l}
 \psi  \\ 
 0 \\ 
 \end{array} \right) = 0 \\ 
  \\ 
 \end{array}
\end{equation}
We use the representation of Gamma matrices given by:
\begin{equation}\gamma ^0  = \left( {\begin{array}{*{20}c}
   0 & { - i}  \\
   i & 0  \\
\end{array}} \right),\;         \gamma ^1  = \left( {\begin{array}{*{20}c}
   0 & i  \\
   i & 0  \\
\end{array}} \right),\;           \gamma _3  = \gamma ^0 \gamma ^1  = \left( {\begin{array}{*{20}c}
   1 & 0  \\
   0 & { - 1}  \\
\end{array}} \right)\end{equation}
The Dirac equation reduces to $
(\partial _t  + \partial _x )\psi  = 0
$ with solutions $
\psi (x,t) = e^{ - ipt} e^{ipx} 
$. These wave functions are normalized by:	\begin{equation}
(\psi ,\chi ) = \int {dx\bar \psi \gamma _0 \chi } 
\end{equation}
In the second quantization form of the propagator the Dirac field becomes an operator with mode expansion:\begin{equation}
\hat \psi (x,t) = \int\limits_{ - \infty }^\infty  {dp } \mathord{\buildrel{\lower3pt\hbox{$\scriptscriptstyle\frown$}} 
\over b} _p e^{ - itp + ixp} 
\end{equation}
where the operators $b_p $
obey the anticommutation relations $\begin{array}{l}
  \\ 
 \{ b_p ,b_{p'} \}  = \delta (p + p') \\ 
  \\ 
 \end{array}$. The second quantized form of the Green's function is:
\begin{equation}
\begin{array}{l}
 G_1 (x,t,0,0) = \left\langle {0|T\psi (x,t)\psi (0,0)|\left. 0 \right\rangle } \right. = \sum\limits_{p,p' = 0}^\infty  {\left\langle {0|b_p } \right.} e^{ - itp + ixp} b_{ - p'} |\left. 0 \right\rangle  \\ 
       \;\;\;\;              = \sum\limits_{p = 0}^\infty  {e^{ - itp + ixp} }  = \int\limits_0^\infty  {\displaystyle\frac{{dp}}{{2\pi }}e^{ - itp + ixp}  = \displaystyle\frac{1}{{2\pi }}} \displaystyle\frac{i}{{(x - t)}} \\ 
 \end{array}
\end{equation}
in agreement with (3.5).

The Weyl Dirac equation follows from the action:\begin{equation}
I = \int {dtdx(\bar \psi } (\gamma ^0 \partial _t  + \gamma ^1 \partial _x )\psi 
\end{equation}
 and the Hamiltonian derived from (3.12) is given by $H = \int {dx\bar \psi \gamma ^1 \partial _x \psi }$. In Fourier space the Hamiltonian is written $H = \int\limits_0^\infty  {dppb_{ - p} b_p }$.
.

\subsection*{Case (2) $E = \displaystyle\frac{{2\pi n}}{L}$}

This is identical to case (1) except the spatial direction takes its values on a circle and  space-time is a 1+1 dimensional cylinder $R \times S^1 $
. As is well known on such a space-time the momentum is quantized as the wave function obeys the periodic boundary condition $\psi (x + L) = \psi (x)$
. Then the basic equations are:
	\begin{equation}
\begin{array}{l}
 {\rm{Energy:}}\;\;\;\;         E = cp = \displaystyle\frac{{c2\pi n}}{L} \\ 
 {\rm{Velocity:}}\;\;\;\;         v = \displaystyle\frac{{\partial E}}{{\partial p}} = c \\ 
 {\rm{Position:  }}\;\;\;\;      x = ct \\ 
 {\rm{Time: }}\;\;\;\;             t = \displaystyle\frac{x}{c} \\ 
  \\ 
 \end{array}
\end{equation}
Classically all particles are right moving and travel with the velocity of light. If a set of particles leave position 0 and time 0 they all arrive at position $x$ at the same time regardless of energy. In the quantum case the above dispersion relation becomes the operator equation $\hat E = \hat p$. The momentum eigenstates are denoted by $\hat p\left. {|p} \right\rangle  = p\left. {|p} \right\rangle$. In the position basis we have: $\left\langle {x|\left. p \right\rangle } \right. = \displaystyle\frac{1}{{L^{1/2} }}e^{ipx}$. Now $x \varepsilon   S^1$ so that momentum is quantized $p = \displaystyle\frac{{2\pi n}}{L}$
with time independent staes
\begin{equation}
\psi _n (x) = \left\langle {x|\left. {p = \displaystyle\frac{{2\pi n}}{L}} \right\rangle } \right. = \displaystyle\frac{1}{{L^{1/2} }}e^{i\displaystyle\frac{{2\pi n}}{L}x} 
\end{equation}
which satisfies the periodic boundary condition \begin{equation}
\psi _n (x + L) = \psi (x)
\end{equation} 
The transition amplitude or Green's function is then given by:
\begin{equation}
\begin{array}{l}
 A_2  = \displaystyle\frac{1}{{2\pi }}g_2 (x - x',s) = G_2 (x,t;x',t') = \sum\limits_n {e^{ - iE_n (t' - t)} \psi _n (x)\psi _n^* (x')}  \\ 
       = \sum\limits_{n = 0}^\infty  {e^{ - i\displaystyle\frac{{2\pi n}}{L}(t - t') + i\displaystyle\frac{{2\pi n}}{L}(x - x')}  = } \displaystyle\frac{1}{L}\displaystyle\frac{1}{{1 - e^{i\displaystyle\frac{{2\pi }}{L}((x - x') - (t - t'))} }}\\ = \displaystyle\frac{1}{{2\pi R}}\displaystyle\frac{1}{{1 - e^{i\displaystyle\frac{1}{R}((x - x') - (t - t'))} }} \\ 
 \end{array}
\end{equation}
so that the Green's function is written:
\begin{equation}
g_2 (x,s) = 2\pi G_2 (x,t;0,0) =   \displaystyle\frac{1}{R}\displaystyle\frac{1}{{1 - e^{i\displaystyle\frac{1}{R}(x + is/a)} }}
\end{equation}
with $s = it/a$
. Note as $L \to \infty $
the Green's function reduces to that of Case (1) $G_1 (x,t;0,0) = \displaystyle\frac{1}{{2\pi }}\displaystyle\frac{i}{{x - t}}$.

In the second quantized formalism the chiral Dirac equation is given by:
\begin{equation}
\begin{array}{l}
 (i\gamma ^0 \partial _0  + i\gamma ^1 \partial _1 )\displaystyle\frac{{(1 + \gamma _3 )}}{2}\left( \begin{array}{l}
 \psi  \\ 
 0 \\ 
 \end{array} \right) = 0 \\ 
  \\ 
 \end{array}
\end{equation}
with solutions given by:\begin{equation}
\psi (x,t) = \displaystyle\frac{1}{{L^{1/2} }}e^{ - it\displaystyle\frac{{2\pi n}}{L} + ix\displaystyle\frac{{2\pi n}}{L}} 
\end{equation} 
These solutions obey the periodic boundary condition $\psi (x) = \psi (x + L)$
. The mode expansion is:
 \begin{equation}
\hat \psi (x,t) = \sum\limits_{n = 0}^\infty  {b_n e^{ - it\displaystyle\frac{{2\pi n}}{L} + ix\displaystyle\frac{{2\pi n}}{L}} } 
\end{equation}
where the operators obey $\{ b_n ,b_{ - n'} \}  = \delta _{nn'} $
. The Hamiltonian is\begin{equation}
H = \displaystyle\frac{{2\pi }}{L}\sum\limits_{n = 1}^\infty  {nb_{ - n} b_n } 
\end{equation}The Green's function in the second quantized form is given by:
\begin{equation}
\begin{array}{l}
 G_2 (x,t,0,0) = \left\langle {0|T\psi (x,t)\psi (0,0)|\left. 0 \right\rangle } \right. = \displaystyle\frac{1}{L}\sum\limits_{m,n = 0}^\infty  {\left\langle {0|b_n } \right.} e^{ - it\displaystyle\frac{{2\pi n}}{L} + ix\displaystyle\frac{{2\pi n}}{L}} b_{ - m} |\left. 0 \right\rangle  \\ 
                      = \displaystyle\frac{1}{L}\sum\limits_{n = 0}^\infty  {e^{ - it\displaystyle\frac{{2\pi n}}{L} + ix\displaystyle\frac{{2\pi n}}{L}} }  = \displaystyle\frac{1}{L}\displaystyle\frac{1}{{1 - e^{ - it\displaystyle\frac{{2\pi }}{L} + ix\displaystyle\frac{{2\pi }}{L}} }} \\ 
 \end{array}
\end{equation}
in agreement with (3.16).

\subsection*{Case (3): $E = \displaystyle\frac{1}{a}\log (ap + 1)$} 

In case (3) the energy varies logarithmically with the momentum. This is a nonstandard dispersion relation. However it reduces to the standard case (1) $E = p$
in the limit $a \to 0$
. The relation of (E vs p) is plotted in figure 2. 

\begin{figure}[htbp]

   \centerline{\hbox{
   \epsfxsize=2.0in
   \epsffile{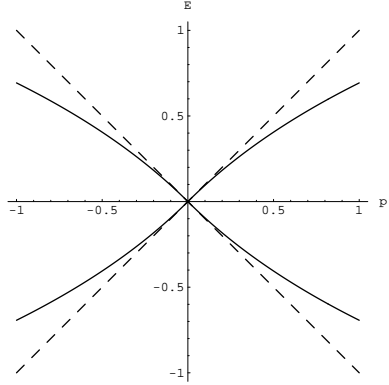}
     }
  }
  \caption{E vs p for Case (3) logarithmic dispersion. } 
          
  \label{fig2}

\end{figure}

We shall mainly be concerned with the upper right quadrant associated with a right moving particle with positive energy. We can think of $a$
as a new length scale that modifies physics at short distances.
 
The classical theory is described by the equations:
\begin{equation}
\begin{array}{l}
 {\rm{Energy:}}\;\;\;\;         E = \displaystyle\frac{1}{a}\log (ap + 1) \\ 
 {\rm{Momentum: }}\;\;\;\;  p = \displaystyle\frac{1}{a}(e^{aE}  - 1) \\ 
 {\rm{Velocity:  }}\;\;\;\;       v = \displaystyle\frac{{\partial E}}{{\partial p}} = \displaystyle\frac{c}{{ap + 1}} = ce^{ - aE}  \\ 
 {\rm{Position: }}\;\;\;\;       x = \displaystyle\frac{c}{{ap + 1}}t = e^{ - aE} ct \\ 
 {\rm{Time: }}\;\;\;\;              t = \displaystyle\frac{{ap + 1}}{c}x = e^{aE} \displaystyle\frac{x}{c} \\ 
 \end{array}
\end{equation}
These equations yield the nonintuitive relation that the more energy or momentum one puts into a particle the slower it goes. That velocity is inversely related to momentum and vanishes exponentially with energy. So if a set of particles are released at time 0 and space 0 it will take exponentially more time to arrive at point $x$ depending on the energy. Nevertheless as long as one stays at energy and momentum far less than $\displaystyle\frac{1}{a}$
the usual picture of case (1) emerges. In figure 3 and 4 we plot the velocity as a function of momentum and energy.

\begin{figure}[htbp]

   \centerline{\hbox{
   \epsfxsize=3.0in
   \epsffile{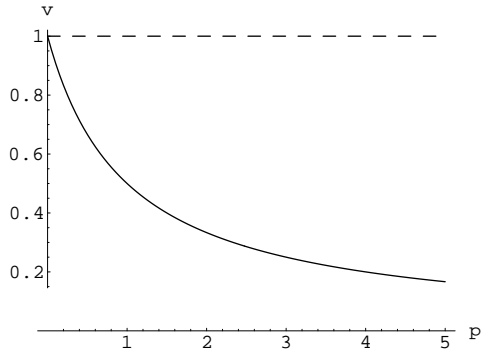}
     }
  }
  \caption{Velocity as a function of momentum for Case (3) $v = \displaystyle\frac{c}{ap+1}$ } 
          
  \label{fig3}

\end{figure}

\begin{figure}[htbp]

   \centerline{\hbox{
   \epsfxsize=3.0in
   \epsffile{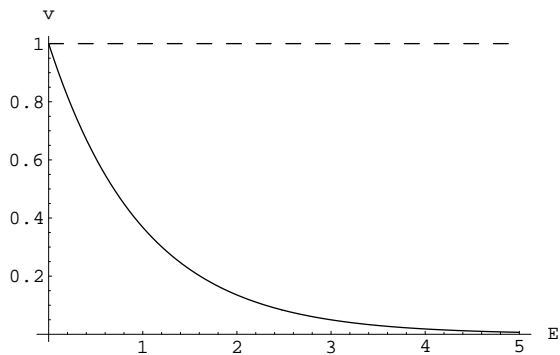}
     }
  }
  \caption{Velocity as a function of energy for Case (3) $v = ce^{-aE}$ } 
          
  \label{fig4}

\end{figure}

One way to motivate the nonstandard dispersion relation is the following. Consider first the Lagrangian $L$  that is sometimes used as a starting point for the description of a fermion on a Euclidean semilattice $aZ \times R$.
	\begin{equation}
L = \psi (x,\bar t)(\displaystyle\frac{1}{a}(\psi (x,\bar t + a) - \psi (x,\bar t - a)) + i\partial _x \psi (x,\bar t))
\end{equation}
We can write this using the exponential of the derivative operator as:
\begin{equation}
L = \psi (x,\bar t)(\displaystyle\frac{1}{a}(e^{a\partial _{\bar t} } \psi (x,\bar t) - e^{ - a\partial _{\bar t} } e\psi (x,\bar t)) + i\partial _x \psi (x,\bar t))
\end{equation}
Now Wick rotating back to real time using $\bar t =  - it$
we have formally:
\begin{equation}
L = \psi (x,t)(\displaystyle\frac{1}{a}(e^{ia\partial _t } \psi (x,t) - e^{ - ia\partial _t } e\psi (x,t)) + i\partial _x \psi (x,t))
\end{equation}
After Fourier transforming the fields we obtain the dispersion relation: \begin{equation}
\displaystyle\frac{1}{a}\sinh (aE) - p = 0
\end{equation}
or inverting the relation $E = \displaystyle\frac{1}{a}\sinh ^{ - 1} (ap) = \displaystyle\frac{1}{a}\log (ap + \sqrt {1 + (ap)^2 } )$.
If instead of (3.24) if one started with
\begin{equation}
L = \psi (x,\bar t)(\displaystyle\frac{1}{a}(\psi (x,\bar t + a) - \psi (x,\bar t)) + i\partial _x \psi (x,\bar t))
\end{equation}
we can again write this using the exponential of the derivative operator as:
\begin{equation}
L = \psi (x,\bar t)(\displaystyle\frac{1}{a}(e^{a\partial _{\bar t} } \psi (x,\bar t) - \psi (x,\bar t)) + i\partial _x \psi (x,\bar t))
\end{equation}
Now again Wick rotating back to real time using $\bar t =  - it$
we have formally:
\begin{equation}
L = \psi (x,t)(\displaystyle\frac{1}{a}(e^{ia\partial _t } \psi (x,t) - \psi (x,t)) + i\partial _x \psi (x,t))
\end{equation}
Fourier transforming this Lagrangian leads to the dispersion relation \begin{equation}\displaystyle\frac{1}{a}(e^{aE}  - 1) - p = 0\end{equation}
or inverting the relationship $E = \displaystyle\frac{1}{a}\log (ap + 1)$
 as in (3.23). We shall return to a euclidean quantum field description in section 8. For now we will continue with a first quantized description based on (3.1).

In the first quantized description the spatial wave functions are the same as in Case (1) $\left\langle {x|\left. p \right\rangle } \right. = e^{ipx}.$ The time dependence is modified however and from $i\partial _t \psi  = E_p \psi $
we have:\begin{equation}
\psi _p (x,t) = e^{ - iE_p t} e^{ipx}  = e^{ - it\displaystyle\frac{1}{a}\log (ap + 1)} e^{ipx}  = (ap + 1)^{i\displaystyle\frac{t}{a}} e^{ipx} 
\end{equation}
Then using the general formula for the transition amplitude or propagator:
\begin{equation}\begin{array}{l}
 A_3  = G_3 (x,t;x',t') = \sum\limits_k {e^{ - iE_k (t' - t)} \psi (x)\psi *(x')}\\
  = \displaystyle\frac{1}{{2\pi }}\int\limits_0^\infty  {dpe^{ - i\displaystyle\frac{1}{a}\log (ap + 1)(t - t') + ip(x - x')} }  \\ 
  = \displaystyle\frac{1}{{2\pi }}\int\limits_0^\infty  {dp(ap + 1)^{ - i\displaystyle\frac{1}{a}(t - t')} e^{ip(x - x')}   }\\
= \displaystyle\frac{1}{{2\pi }}\displaystyle\frac{1}{a}e^{ - i(x - x')/a} \int\limits_1^\infty  {duu^{ - i\displaystyle\frac{1}{a}(t - t')} e^{i(u)(x - x')/a} }  \\ 
  = \displaystyle\frac{1}{{2\pi }}\displaystyle\frac{1}{a}e^{ - i(x - x')/a} \left( {(x - x')/a} \right)^{i\displaystyle\frac{1}{a}(t - t') - 1} \int\limits_{ - i((x - x')/a}^\infty  {dyy^{ - i\displaystyle\frac{1}{a}(t - t') + 1 - 1} e^{ - y} }  \\ 
  = \displaystyle\frac{1}{{2\pi }}\displaystyle\frac{1}{a}e^{ - i(x - x')/a} \left( {(x - x')/a} \right)^{i\displaystyle\frac{1}{a}(t - t') - 1} \Gamma ( - i((x - x')/a, - i\displaystyle\frac{1}{a}(t - t') + 1) \\ 
  \\ 
 \end{array}
\end{equation}
where \begin{equation}
\Gamma (b,x) = \int\limits_x^\infty  {dyy^{b - 1} e^{ - y} } 
\end{equation}
 and we have used the change of variables $u = ap + 1$
and $y =  - i(x - x')u$. Defining $g_3 (x,s) = 2\pi G_3 (x,t;0,0)$
for $s = \displaystyle\frac{{it}}{a}$
we express write the Green's function as:
\begin{equation}g_3 (x,s) = 2\pi G_3 (x,t;0,0) = \displaystyle\frac{1}{a}e^{ - ix/a} (\displaystyle\frac{{ - ix}}{a})^{ - 1 + s} \Gamma (\displaystyle\frac{{ - ix}}{a},1 - s)\end{equation}

\subsection*{Case  ($3'$): $E = \displaystyle\frac{1}{a}\log (ap + 1)$ using $(E,p)$ representation}

This is a variant on the above. In this case the Green's function is given by the $(E,p)$
representation:
	\begin{equation}
\displaystyle\frac{1}{{2\pi }}g_{3'} (x,s) = G_{3'} (x,t - ia;0,0) = \displaystyle\frac{1}{{(2\pi )^2 }}\int {dEdp} \displaystyle\frac{1}{{\displaystyle\frac{1}{a}(e^{aE}  - 1) - p}}e^{ - iE(t - ia)} e^{ipx} 
\end{equation}
Performing the integral over $p$
by contour integration we have:
\begin{equation}
\begin{array}{l}
 g_{3'} (x,s) = 2\pi G_{3'} (x,t + ia;0,0) = \int {dE} e^{ - iE(t + ia)} e^{i(e^{aE}  - 1)x/a}  \\ 
               = \displaystyle\frac{1}{a}\int {dWW^{ - s} e^{i(W - 1)x/a} }  \\ 
 \end{array}
\end{equation}
where we have defined $W = e^{aE} $. Now setting $u =  - iWx/a$
this becomes:
\begin{equation}
\begin{array}{l}
 g_{3'} (x,s) = 2\pi G_{3'} (x,t - ia;0,0) = \displaystyle\frac{1}{a}( - ix/a)^{s - 1} e^{ - ix/a} \int {duu^{ - s} e^{ - u} }  \\ 
               = \displaystyle\frac{1}{a}( - ix/a)^{s - 1} e^{ - ix/a} \Gamma (1 - s) \\ 
 \end{array}
\end{equation}
using $s = it$
we write this as:
\begin{equation}
g_{3'} (x,s) = \displaystyle\frac{1}{a}( - ix/a)^{it - 1} e^{ - ix/a} \Gamma (1 - it)
\end{equation}We discuss the $(E,p)$
 representation and other representations of the Green's function in more detail in section 8.

\subsection*{Second quantized Case (3)}

In the second quantized description of the Green's function one solves the Dirac equation associated to the dispersion relation (3.31) :
\begin{equation}
(\gamma ^0 \displaystyle\frac{1}{a}(e^{ai\partial _0 }  - 1) + i\gamma ^1 \partial _1 )\displaystyle\frac{{(1 + \gamma _3 )}}{2}\left( \begin{array}{l}
 1 \\ 
 0 \\ 
 \end{array} \right)\psi  = 0
\end{equation}
which reduces to the equation:
\begin{equation}
(\displaystyle\frac{1}{{ai}}(e^{ai\partial _0 }  - 1) + \partial _x )\psi  = 0
\end{equation}
Alternatively one can consider an equation of the form:
\begin{equation}
(i\partial _t  + \log (ai\partial _x  + 1)\psi  = 0
\end{equation}
In either case we have solutions of the form:
\begin{equation}
\psi (x,t) = e^{ - i(\log (ap + 1))t/a} e^{ipx} 
\end{equation}
To second quantize one forms a superposition of these solutions and promotes the coefficients of the expansion to operators as:
\begin{equation}
\hat \psi (x,t) = \displaystyle\frac{1}{{\sqrt {2\pi } }}\int\limits_{ - \infty }^\infty  {dp \hat b_p e^{ - it\displaystyle\frac{1}{a}\ln (ap + 1) + ixp}  = \displaystyle\frac{1}{{\sqrt {2\pi } }}\int\limits_{ - \infty }^\infty  {dp \hat b_p (ap + 1)^{ - it/a} e^{ixp} } } 
\end{equation}
As in (3.10) $\hat b_{p > 0} $
are annihilation operators, $b_{p < 0} $
are creation operators so that operating on the vacuum state $\hat b_p |\left. 0 \right\rangle  = \theta ( - p)|\left. 0 \right\rangle $
 and $\left\langle {0|} \right.\hat b_p  = \left\langle {0|} \right.\theta (p).$ The Green's function is then:
\begin{equation}\begin{array}{l}
 2\pi G_3 (x,t,0,0) = 2\pi \left\langle {0|T\psi (x,t)\psi (0,0)|\left. 0 \right\rangle } \right. \\ 
           = \int\limits_{ - \infty }^\infty  {dp dp'\left\langle {0|} \right.\hat b_p \hat b_{p'} |\left. 0 \right\rangle (ap + 1)^{ - it/a} e^{ixp}  = } \int\limits_0^\infty  {dp (ap + 1)^{ - it/a} e^{ixp} }  \\ 
  \\ 
 \end{array}\end{equation}
in agreement with (3.33).

The non standard Dirac equation (3.40) follows from the action:
	\begin{equation}
I = \int {dtdx\bar \psi (\gamma ^0 \partial _t }  + \gamma ^1 ( - i\displaystyle\frac{1}{a}\log (1 + ai\partial _x ))\psi 
\end{equation}
The differential operator \begin{equation}
 D_x=- i\displaystyle\frac{1}{a}\log (1 + ai\partial _x )
\end{equation}
is defined by the series expansion:
	\begin{equation}
 - i\displaystyle\frac{1}{a}\log (1 + ai\partial _x ) = \sum\limits_{n = 1}^\infty  {i^{n - 1} a^{n - 1} (-1)^{n - 1} \partial _x^n } 
\end{equation}
An alternative action that yields the same solutions to the equation of motion is:
\begin{equation}
I = \int {dtdx\bar \psi (\gamma ^0 \displaystyle\frac{{ - i}}{a}(e^{ai\partial _t }  - 1) }  + \gamma ^1 \partial _x )\psi 
\end{equation}
where the differential operator \begin{equation}
D_t=\displaystyle\frac{{ - i}}{a}(e^{ai\partial _t }  - 1)
\end{equation}
is defined through the series expansion:
\begin{equation}
\displaystyle\frac{{ - i}}{a}(e^{ai\partial _t }  - 1) = \sum\limits_{n = 1}^\infty  {a^{n - 1} i^{n - 1} \displaystyle\frac{1}{{n!}}} \partial _t^n 
\end{equation}

\subsection*{Case (4): $E = \displaystyle\frac{1}{a}\log (a\displaystyle\frac{{2\pi n}}{L} + 1)$}

This is the case that is relevant to the Riemann hypothesis.
As in case (2) periodicity for the $x$
field means that it takes its values on $S^1 $
 of radius $L$ and this
implies $p_n  = \displaystyle\frac{{2\pi n}}{L}$
. The dispersion relation is depicted in figure 5. 

\begin{figure}[htbp]

\centerline{\hbox{
   \epsfxsize=2.0in
   \epsffile{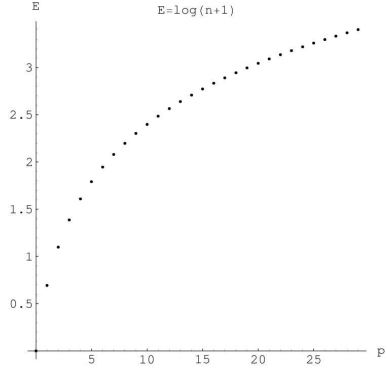}
     }
  }
  \caption{E vs p for Case (4) Logarithmic dispersion with discrete momentum. We have assumed $R=a$.} 
          
  \label{fig5}

\end{figure}

Again we are mainly concerned with the upper right quadrant. The spatial wave functions are $\psi _n (x) = \left\langle {x|\left. {p_n } \right\rangle } \right. = \displaystyle\frac{1}{{L^{1/2} }}e^{i\displaystyle\frac{{2\pi n}}{L}x} $
 and time dependence of the wave functions is given by:\\
$\psi _n (x,t) = \left\langle {x,t|\left. {p_n } \right\rangle } \right. = \displaystyle\frac{1}{{(L)^{1/2} }}e^{ - iE_n t} e^{i\displaystyle\frac{{2\pi n}}{L}x}  = \displaystyle\frac{1}{{(L)^{1/2} }}(n\displaystyle\frac{{2\pi }}{L}a + 1)^{ - it/a} e^{i\displaystyle\frac{{2\pi n}}{L}x} $
. For simplicity we set $x' = t' = 0$
. Then the propagator is written as: 
\begin{equation}
\begin{array}{l}
 A_4  = G_4 (x,t;0,0) = \left\langle {x|e^{ - itH} |\left. {x = 0} \right\rangle } \right. = \sum\limits_{n = 0}^\infty  {e^{ - iE_n t} \psi _n (x)\psi _n^* (0)}  \\ 
  = \displaystyle\frac{1}{{2\pi }}\displaystyle\frac{{2\pi }}{L}\sum\limits_{n = 0}^\infty  {(a\displaystyle\frac{{2\pi n}}{L} + 1)^{ - it/a} e^{i2\pi nx/L} }\\  = \displaystyle\frac{1}{{2\pi }}\displaystyle\frac{{2\pi }}{L}\left( {\displaystyle\frac{L}{{2\pi a}}} \right)^{it/a} \sum\limits_{n = 0}^\infty  {\left( {n + \displaystyle\frac{L}{{2\pi a}}} \right)^{ - it/a} e^{i2\pi nx/L} }  \\ 
  = \displaystyle\frac{1}{{2\pi }}\displaystyle\frac{{2\pi }}{L}\left( {\displaystyle\frac{L}{{2\pi a}}} \right)^{it/a} \phi (\displaystyle\frac{x}{L},\displaystyle\frac{{it}}{a},\displaystyle\frac{L}{{2\pi a}}) \\ 
  = \displaystyle\frac{1}{{2\pi }}\displaystyle\frac{1}{R}(\displaystyle\frac{R}{a})^{it/a} \phi (\displaystyle\frac{x}{{2\pi R}},\displaystyle\frac{{it}}{a},\displaystyle\frac{R}{a}) \\ 
 \end{array}
\end{equation}
In terms of $s = \displaystyle\frac{{it}}{a}$
the Green's function becomes:
\begin{equation}g_4 (x,s) = 2\pi G(x,t;0,0) = \displaystyle\frac{1}{R}(\displaystyle\frac{R}{a})^s \phi (\displaystyle\frac{x}{{2\pi R}},s,\displaystyle\frac{R}{a})\end{equation}

So that we have obtained a physical representation of the Lerch zeta function $\phi (x,it + \sigma ,\alpha )$
where we can interpret the parameter $t$
 as time, $x$
as space and $\alpha $
proportional to radius of the universe $\displaystyle\frac{L}{{2\pi }} = R$
. We discuss the interpretation of the parameter $\sigma $
 in section 4.

In the second quantized description of the Green's function one solves the Dirac equation associated to the dispersion relation (3.31) :
\begin{equation}
(\gamma ^0 \displaystyle\frac{1}{a}(e^{ai\partial _0 }  - 1) + i\gamma ^1 \partial _1 )\displaystyle\frac{{(1 + \gamma _3 )}}{2}\left( \begin{array}{l}
 1 \\ 
 0 \\ 
 \end{array} \right)\psi  = 0
\end{equation}
which reduces to the equation:
\begin{equation}
(\displaystyle\frac{1}{{ai}}(e^{ai\partial _0 }  - 1) + \partial _x )\psi  = 0
\end{equation}
with periodic boundary conditions $\psi (x + L,t) = \psi (x,t)$
 and with solutions of the form:
\begin{equation}
\psi (x,t) = \displaystyle\frac{1}{{(L)^{1/2} }}e^{ - i\ln (a\displaystyle\frac{{2\pi n}}{L} + 1)t} e^{i\displaystyle\frac{{2\pi n}}{L}x} 
\end{equation}
To second quantize one forms a superposition of these solutions and promotes the coefficients of the expansion to operators as:
\begin{equation}
\begin{array}{l}
 \hat \psi (x,t) = \displaystyle\frac{1}{{(L)^{1/2} }}\sum\limits_{n =  - \infty }^\infty  {\hat b_n e^{ - it\displaystyle\frac{1}{a}\log (a\displaystyle\frac{{2\pi n}}{L} + 1) + ix\displaystyle\frac{{2\pi n}}{L}} }  \\ 
             = \displaystyle\frac{1}{{(L)^{1/2} }}\sum\limits_{n =
  - \infty }^\infty  {\hat b_n (a\displaystyle\frac{{2\pi n}}{L} + 1)^{ - it\displaystyle\frac{1}{a}}e^{ + ix\displaystyle\frac{{2\pi n}}{L}} }  \\ 
 \end{array}
\end{equation}
$\hat b_{n > 0} $
are annihilation operators, $b_{n < 0} $
are creation operators so that operating on the vacuum state $\hat b_n |\left. 0 \right\rangle  = \theta ( - n)|\left. 0 \right\rangle $
 and $\left\langle {0|} \right.\hat b_n  = \left\langle {0|} \right.\theta (n)$. The Green's function is then:
\begin{equation}\begin{array}{l}
 G_4 (x,t,0,0) = \left\langle {0|T\psi (x,t)\psi (0,0)|\left. 0 \right\rangle } \right. \\ 
                     = \displaystyle\frac{1}{{2\pi }}\displaystyle\frac{{2\pi }}{L}\sum\limits_{n,n' =  - \infty }^\infty  {\left\langle {0|} \right.\hat b_n \hat b_{n'} |\left. 0 \right\rangle (a\displaystyle\frac{{2\pi n}}{L} + 1)^{ - it/a} e^{ix\displaystyle\frac{{2\pi n}}{L}} }  \\ 
                     = \displaystyle\frac{1}{{2\pi }}\displaystyle\frac{{2\pi }}{L}\sum\limits_{n = 0}^\infty  {(a\displaystyle\frac{{2\pi n}}{L} + 1)^{ - it/a} e^{ix\displaystyle\frac{{2\pi n}}{L}} }  \\ 
  \\ 
 \end{array}\end{equation}
In agreement with (3.52).

\subsection*{Interpretation of $t$ for Dirichlet series}

Using the formula (3.52) and (3.53) various Dirchlet series can be given a physical interpretation in terms of transition amplitudes. For example the Dirichlet eta function is given by:
	\begin{equation}
\begin{array}{l}
 G(x = \displaystyle\frac{L}{2},t;0,0)_{R = a}  = \left\langle {x = \displaystyle\frac{L}{2}} \right.,t|x = 0,t = \left. 0 \right\rangle _{R = a}  \\ 
        = \displaystyle\frac{1}{{2\pi a}}\sum\limits_{n = 0}^\infty  {(n + 1)^{ - it/a} e^{i\pi n} }  =  - \displaystyle\frac{1}{{2\pi a}}\eta(it/a) \\ 
 \end{array}
\end{equation}
when $\alpha  = \displaystyle\frac{L}{{2\pi a}} = 1$
 and Riemann zeta function given by $\zeta(it/a) = (2^{1 - it/a}  - 1)\eta _D (it/a)$. Here $t$ is given the interpretation of time and the one dimensional volume obeys the condition $L = 2\pi a$ or $R = a$.

The rest of the Dirichlet functions discussed in section 2 are given in terms of the Green's function through:
\begin{equation}
\begin{array}{l}
 G_4 (x = 0,t;0,0)_{R = a}  = \displaystyle\frac{1}{{2\pi a}}\phi (\displaystyle\frac{x}{L} = 0,\displaystyle\frac{{it}}{a},\alpha  = 1) = \displaystyle\frac{1}{{2\pi a}}\zeta (it/a) \\ 
 G_4 (x = \pi a,t;0,0)_{R = a}  = \displaystyle\frac{1}{{2\pi a}}\phi (\displaystyle\frac{x}{L} = \displaystyle\frac{1}{2},\displaystyle\frac{{it}}{a},\alpha  = 1) =  - \displaystyle\frac{1}{{2\pi a}}\eta (it/a) \\ 
 G_4 (x = 0,t;0,0)_{R = \displaystyle\frac{a}{2}}  = \displaystyle\frac{1}{{2\pi a}}\phi (\displaystyle\frac{x}{L} = 0,it,\alpha  = \displaystyle\frac{1}{2}) = \displaystyle\frac{1}{{2\pi a}}2^{ - it/a} \lambda (it/a) \\ 
 G_4 (x = \pi a,t;0,0)_{R = \displaystyle\frac{a}{2}}  = \displaystyle\frac{1}{{2\pi a}}\phi (\displaystyle\frac{x}{L} = \displaystyle\frac{1}{2},it,\alpha  = \displaystyle\frac{1}{2}) = \displaystyle\frac{1}{{2\pi a}}2^{ - it} \beta (it/a) \\ 
 \end{array}
\end{equation}
Introducing $\sigma $
through $t \to t + i\sigma $
we relate the Dirichlet functions to the Green's function at analytically continued time.

\subsection*{Generalization with constant vector potential}

In the presence of a gauge potential $A_1 $
 one can use the substitution $p \to p + qA_1 $
 to see the effect on the Green's function. For a spatial $S^1 $
a constant gauge potential cannot be gauged away because of the Aharonov Bohm effect on a nonsimply connected space. The dispersion relation (3.31) is then modified to $E_n  = \displaystyle\frac{1}{a}\log (ap_n  + 1)$
with $p_n  = \displaystyle\frac{{2\pi n}}{L} + qA_1 $
 so that $p \in \displaystyle\frac{{2\pi }}{L}Z + qA_1.$ The transition amplitude or Green's function is modified to :
\begin{equation}
\begin{array}{l}
 \left\langle {x,t} \right.\left| {0,0} \right\rangle  = \displaystyle\frac{1}{{2\pi }}\displaystyle\frac{{2\pi }}{L}\sum\limits_{n = 0}^\infty  {e^{ - iE_n t + ip_n x} }  \\ 
                 = \displaystyle\frac{1}{{2\pi }}\displaystyle\frac{{2\pi }}{L}\left( {\displaystyle\frac{L}{{2\pi a}}} \right)^{it/a} e^{ixqA_1 } \sum\limits_{n = 0}^\infty  {\left( {n + \displaystyle\frac{L}{{2\pi a}} + \displaystyle\frac{{LqA_1 }}{{2\pi }}} \right)^{ - it/a} e^{i2\pi nx/L} }  \\ 
                 = \displaystyle\frac{1}{{2\pi }}\displaystyle\frac{{2\pi }}{L}\left( {\displaystyle\frac{L}{{2\pi a}}} \right)^{it/a} e^{ixqA_1 } \phi (\displaystyle\frac{x}{L},\displaystyle\frac{{it}}{a},\displaystyle\frac{L}{{2\pi a}} + \displaystyle\frac{{LqA_1 }}{{2\pi }}) \\ 
 \end{array}
\end{equation}
In this case the boundary condition on the wave function becomes $\psi (x + L) = e^{iqA_1 L} \psi (x).$ Note that in two dimensions the charge $q$
has dimensions of $[q] = (length)^{ - 1} $
while the vector potential $A_1 $
 is dimensionless. In the presence of constant $A_1 $
 the value of the parameter $\alpha $
in the Lerch zeta function is $\alpha  = \displaystyle\frac{L}{{2\pi a}} + \displaystyle\frac{{LqA_1 }}{{2\pi }} = L(\displaystyle\frac{1}{{2\pi a}} + \displaystyle\frac{{qA_1 }}{{2\pi }})$. 

An especially useful form of the dispersion relation in the presence of a constant gauge potential is written as:
	\begin{equation}
\displaystyle\frac{{e^{aE} }}{a} = \displaystyle\frac{1}{a} + \displaystyle\frac{{n + A}}{R} = \displaystyle\frac{{n + A'}}{R}
\end{equation}
Where we have defined $A = RqA_1 $
 and $A' = \displaystyle\frac{R}{a} + A.$ Two important cases are (i) $A' = 0, A =  - \displaystyle\frac{R}{a}$
and (ii) $A' = \displaystyle\frac{1}{2}, A = \displaystyle\frac{1}{2} - \displaystyle\frac{R}{a}$
 which we shall use in section 4.
The dispersion relation can be used in deriving the $(E,p)$
form of the Green's function given by:
	\begin{equation}
G_{4'} (x,t;0,0)|_{R,A'}  = \displaystyle\frac{1}{{2\pi R}}\sum\limits_{n =  - \infty }^\infty  {\int {\displaystyle\frac{{dE}}{{2\pi }}} \displaystyle\frac{{\displaystyle\frac{{(n + A')}}{R} + \displaystyle\frac{{e^{aE} }}{a}}}{{(\displaystyle\frac{{(n + A')}}{R})^2  - \displaystyle\frac{{e^{2aE} }}{{a^2 }}}}e^{i\displaystyle\frac{n}{R}x} e^{ - iEt} } 
\end{equation}
Introducing the proper time parameter $C$
we have:
\begin{equation}
\begin{array}{l}
 G_{4'} (x,t;0,0)|_{R,A'}  \\ 
        = \displaystyle\frac{1}{{2\pi R}}\sum\limits_{n =  - \infty }^\infty  {\int {dC\displaystyle\frac{{dE}}{{2\pi }}} (\displaystyle\frac{{(n + A')}}{R} + \displaystyle\frac{{e^{aE} }}{a})e^{ - C((\displaystyle\frac{{(n + A')}}{R})^2  - \displaystyle\frac{{e^{2aE} }}{{a^2 }})} e^{i\displaystyle\frac{n}{R}x} e^{ - iEt} }  \\ 
 \end{array}
\end{equation}
Now defining $W = e^{aE} $
we write this as:
\begin{equation}
\begin{array}{l}
 G_{4'} (x,t;0,0)|_{R,A'}  \\ 
        = \displaystyle\frac{1}{{2\pi R}}\sum\limits_{n =  - \infty }^\infty  {\int {dC\displaystyle\frac{{dW}}{{2\pi aW}}} (\displaystyle\frac{{(n + A')}}{R} + \displaystyle\frac{W}{a})e^{ - C((\displaystyle\frac{{(n + A')}}{R})^2  - \displaystyle\frac{{W^2 }}{{a^2 }})} e^{i\displaystyle\frac{n}{R}x} W^{ - it/a} }  \\ 
 \end{array}
\end{equation}
Further defining $u =  - Ca^{ - 2} W^2 $
the Green's function becomes:
\begin{equation}
\begin{array}{l}
 G_{4'} (x,t;0,0)|_{R,A'}  \\ 
     = \displaystyle\frac{1}{{2\pi R}}\sum\limits_{n =  - \infty }^\infty  {\int {dC\displaystyle\frac{{du}}{{4\pi au}}} \displaystyle\frac{{(n + A')}}{R} e^{ - C((\displaystyle\frac{{(n + A')}}{R})^2 } e^{ - u} e^{i\displaystyle\frac{n}{R}x} ( - u)^{ - it/2a} (Ca^{ - 2} )^{it/2a} }  \\
      + \displaystyle\frac{1}{{2\pi R}}\sum\limits_{n =  - \infty }^\infty  {\int {dC\displaystyle\frac{{du}}{{4\pi au}}}  \displaystyle\frac{{( - u)^{1/2} }}{{C^{1/2} }}e^{ - C((\displaystyle\frac{{(n + A')}}{R})^2 } e^{ - u} e^{i\displaystyle\frac{n}{R}x} ( - u)^{ - it/2a} (Ca^{ - 2} )^{it/2a} } 
 \end{array}
\end{equation}
Performing the integral over $u$
and $C$
and expressing the sum over $n$
in terms of the Lerch zeta function we have:
\begin{equation}
\begin{array}{l}
 2\pi i G_{4'} (x,t;0,0)|_{R,A'}  \\ 
       = (\displaystyle\frac{R}{a})^{\displaystyle\frac{{it}}{a} + 1} \displaystyle\frac{1}{R}(\phi (\displaystyle\frac{x}{{2\pi R}},s + 1,A') - e^{ - 2\pi i\displaystyle\frac{x}{{2\pi R}}} \phi ( - \displaystyle\frac{x}{{2\pi R}},s + 1,1 - A'))\cot(\pi s)
 \end{array}
\end{equation}
	
\section{Interpretation of the parameter $\sigma $
}

We have found an interpretation of the parameters $t,x,\alpha $
of the Lerch Zeta function in terms of a fermionic Green's function. What about the parameter $\sigma $
? Certainly this is a crucial parameter for the Dirichlet functions as the Riemann hypothesis is given by $\phi (x = \displaystyle\frac{1}{2},it + \sigma ,\alpha  = 1) \ne 0$
 for $\displaystyle\frac{1}{2} < \sigma  < 1$
 and the functional equation relates $\sigma $
to $1 - \sigma $
. 

One way to interpret $\sigma $
is in terms of scattering amplitudes. After all the two point fermionic Green's function is a special example of a $(1 \to 1)$
 scattering amplitude. Scattering amplitudes because of their analytic nature can be continued in the complex plane and even to other signatures of space-time as have been remarked in \cite{Witten:2003nn}. 

Another way to interpret $\sigma $
is in terms of a one particle partition function. This is related to the Green's function by:
\begin{equation}\begin{array}{l}
 z_{{\rm{one}}} (\beta  = \tau ) = \int {dxG(x,\tau ;x,0)}  \\ 
  = \sum\limits_{n = 0}^\infty  {e^{ - \beta E_n }  = (\displaystyle\frac{R}{a})^{\beta /a} \sum\limits_{n = 0}^\infty  {(n + \displaystyle\frac{R}{a}} )^{ - \beta /a} }  \\ 
 \end{array}\end{equation}
where $\tau $
is a Wick rotated time. 

One can also study the partition function in a Harmonic oscillator formalism nonstandard logarithmic oscillator with energy $E = \displaystyle\frac{1}{a}\log (a\omega (n + \displaystyle\frac{1}{2}) + 1)$
. In this case the partition function is:
\begin{equation}\begin{array}{l}
 z_{{\rm{osc}}} (\beta  = \tau ) = \int {dxG_{{\rm{osc}}} (x,\tau ;x,0)}  \\ 
  = \sum\limits_{n = 0}^\infty  {e^{ - \beta E_n } }  = (a\omega )^{ - \beta /a} \sum\limits_{n = 0}^\infty  {(n + \displaystyle\frac{1}{{a\omega }} + \displaystyle\frac{1}{2})^{ - \beta /a} }  \\ 
 \end{array}\end{equation}
Again because of the analytic properties of the partition function one can study the function in the complex plane and relate its zeros to phase transitions of the theory in the sense of Yang and Lee. 

However one does lose some intuition when introduces $\sigma $
through analytic continuation. After all when some asks what time has elapsed one does not respond $10{\rm{hours}} + i5{\rm{minutes}}$, or if someone asks what the temperature  is one usually doesn't come back with  $(.5 + i14.3)^ \circ  K$. We shall return to the statistical mechanics associated with the Dirichlet series in section 5, and we shall discuss the relation of the Harmonic oscillator formalism to the fermionic description of the 2D string matrix model in section 6. In this section we will investigate some other physical interpretations of the parameter $\sigma $
 to obtain some insight into the effect of the parameter without using direct analytic continuation.

\subsection{$\sigma $
as a mixing parameter}

To begin with consider the first quantized description of the transition amplitude. The superposition of momentum space is used to define the states:
\begin{equation}|\left. x \right\rangle  = \sum\limits_{n = 0}^\infty  {e^{i\displaystyle\frac{{2\pi n}}{L}x} |\left. {p_n } \right\rangle } \end{equation}
Consider a similar set of states parameterized by a parameter $\sigma $
through:
\begin{equation}|\left. \sigma  \right\rangle  = \sum\limits_{n = 0}^\infty  {e^{ - \sigma E_n } |\left. {p_n } \right\rangle }  = e^{ - \sigma \hat E} \left. {|x = 0} \right\rangle \end{equation}
and $|\left. \sigma  \right\rangle $
 has been normalized so that $\left\langle {\sigma |\left. \sigma  \right\rangle } \right. = z(2\sigma )$
 where $z(2\sigma ) = \sum\limits_{n = 0}^\infty  {e^{ - 2\sigma E_n } } $. The probability of measuring a momentum value $p_n $
from an initially prepared state $|\left. \sigma  \right\rangle $
 is given by $P_n  = \displaystyle\frac{{|\left\langle {p_n } \right|\left. \sigma  \right\rangle |^2 }}{{\left\langle \sigma  \right.|\left. \sigma  \right\rangle }} = \displaystyle\frac{{e^{ - 2\sigma E_n } }}{{z(2\sigma )}}$
. These probabilities satisfy $0 < P_n  < 1$
and 
	\begin{equation}
\sum\limits_{n = 0}^\infty  {P_n }  = \displaystyle\frac{1}{{z(2\sigma )}}\sum\limits_{n = 0}^\infty  {e^{ - 2\sigma E_n } }  = 1
\end{equation}
The first quantized description of the transition amplitude from: 
	\begin{equation}
A_H  = \left\langle {x,t|} \right.\left. \sigma  \right\rangle  = \left\langle {x|} \right.e^{ - itH} e^{ - \sigma H} \left. {|x' = 0} \right\rangle  = \sum\limits_{n = 0}^\infty  {e^{ - itE_n  - \sigma E_n } e^{ixp_n } } 
\end{equation}
In the second quantized description we use the correspondence:
\begin{equation}\begin{array}{l}
 |x,\left. t \right\rangle  \to \hat \psi (x,t)|\left. 0 \right\rangle  \\ 
 e^{ - \sigma H} |x,\left. t \right\rangle  = |x,\left. {t - i\sigma } \right\rangle  \to \hat \psi (x,t - i\sigma )|\left. 0 \right\rangle  \\ 
 \end{array}\end{equation}
Then the Green's function becomes:
\begin{equation}
\begin{array}{l}
 G_\sigma  (x,t;x',t') = G(x,t;x',t' + i\sigma ) = \left\langle {0|\psi (x,t)\psi (x',t' + i\sigma )|\left. 0 \right\rangle } \right. \\ 
  = \left\langle {0|\psi (x,t)e^{\sigma H} \psi (x',t')e^{ - \sigma H} |\left. 0 \right\rangle } \right. = \left\langle {0|e^{ - \sigma H} \psi (x,t)e^{\sigma H} \psi (x',t')|\left. 0 \right\rangle } \right. \\ 
  = \left\langle {0|\psi (x,t - i\sigma )\psi (x',t')|\left. 0 \right\rangle } \right. = G(x,t - i\sigma ;x',t') \\ 
 \end{array}
\end{equation}
where $H = \sum\limits_{n = 1}^\infty  {E_n b_{ - n} b_n } $
is the second quantized Hamiltonian. So in the second quantized form the introduction of the mixing parameter $\sigma $
is simply realized by evaluating the Green's function at complex time $t - i\sigma $
. Note the mixing parameter dependent Green's function is not the same as the finite temperature Green's function which is given by: $G_\beta  (x,t;x',t') = \sum\limits_{n =  - \infty }^\infty  {( - 1)^n G(x,t - ni\beta ;x',t')} $
although they both involve the complex time. Further discussion of statistical ensembles and finite temperature Green's function will be given in section 5.

Now we specialize to the Cases considered in section 3. First consider case (2) and (4)

\subsubsection*{Case (2) $E=\displaystyle\frac{{2\pi n}}{L}$} 
For the usual case $H_{(2)} $
we have $E_n  = p_n $
and $p_n  = \displaystyle\frac{{2\pi n}}{L}$
. We define mixing parameter states by:
	\begin{equation}
\begin{array}{l}
 |x,\sigma /\left. 2 \right\rangle  = \sum\limits_{n = 0}^\infty  {e^{ip_n x} e^{ - \sigma E_n /2} |\left. {p_n } \right\rangle }  \\ 
 |x',\sigma /\left. 2 \right\rangle  = \sum\limits_{n = 0}^\infty  {e^{ip_n x} e^{ - \sigma E_n /2} |\left. {p_n } \right\rangle }  \\ 
 \end{array}
\end{equation}
These states are normalized through: 
	\begin{equation}
\left\langle x \right.,\sigma /2|x,\sigma /\left. 2 \right\rangle  = \left\langle x \right.',\sigma /2|x',\sigma /\left. 2 \right\rangle  = z(\sigma ) = \sum\limits_{n = 0}^\infty  {e^{ - \sigma E_n } }  = \displaystyle\frac{1}{{1 - e^{ - \displaystyle\frac{{2\pi \sigma }}{L}} }}
\end{equation}
Because of the simple time evolution of the momentum states, the time dependence of $|x,\sigma /\left. 2 \right\rangle $
is :
	\begin{equation}
|x,\sigma /2,\left. t \right\rangle  = \sum\limits_{n = 0}^\infty  {e^{ip_n x - iE_n t - \sigma E_n /2} |\left. {p_n } \right\rangle } 
\end{equation}
So that the transition amplitude becomes:
\begin{equation}
\left\langle x \right.',\sigma /2|x,\sigma /2,\left. t \right\rangle  = \sum\limits_{n = 0}^\infty  {e^{ip_n (x - x') - iE_n t - \sigma E_n } }  = \displaystyle\frac{1}{{1 - e^{i\displaystyle\frac{{2\pi }}{L}x - i\displaystyle\frac{{2\pi }}{L}t - \displaystyle\frac{{2\pi }}{L}\sigma } }}
\end{equation}
The transition probability is then:
\begin{equation}
\begin{array}{l}
 P(x,t \to x')_\sigma   = \displaystyle\frac{{|\left\langle x \right.',\sigma /2|x,\sigma /2,\left. t \right\rangle |^2 }}{{\left\langle x \right.,\sigma /2|x,\sigma /\left. 2 \right\rangle \left\langle x \right.',\sigma /2|x',\sigma /\left. 2 \right\rangle }} \\ 
  = \displaystyle\frac{{|\left\langle x \right.',\sigma /2|x,\sigma /2,\left. t \right\rangle |^2 }}{{z(\sigma )z(\sigma )}} = \displaystyle\frac{{(1 - e^{ - \displaystyle\frac{{2\pi \sigma }}{L}} )^2 }}{{|1 - e^{i\displaystyle\frac{{2\pi }}{L}x - i\displaystyle\frac{{2\pi }}{L}t - \displaystyle\frac{{2\pi }}{L}\sigma } |^2 }} \\ 
 \end{array}
\end{equation}
This can be further simplified to:
\begin{equation}
P(x,t \to x')_\sigma   = \displaystyle\frac{{(1 - e^{ - \displaystyle\frac{{2\pi }}{L}\sigma } )^2 }}{{(1 - 2\cos (\displaystyle\frac{{2\pi }}{L}(x - x' - t))e^{ - \displaystyle\frac{{2\pi }}{L}\sigma }  + e^{ - \displaystyle\frac{{4\pi }}{L}\sigma } )}}
\end{equation}
The minimum vale of the probability is then given by:
	\begin{equation}
P_{\min }  = \displaystyle\frac{{(1 - e^{ - \displaystyle\frac{{2\pi }}{L}\sigma } )^2 }}{{(1 + e^{ - \displaystyle\frac{{2\pi }}{L}\sigma } )^2 }} = (\coth (\displaystyle\frac{\pi }{L}\sigma ))^2  > 0
\end{equation}
This is strictly positive except in the trivial case of $\sigma  = 0$
when there is no mixing.

\subsubsection*{Case (4): $E=\displaystyle\frac{1}{a}\log (ap_n  + 1)$} For the unusual case we have $E_n  = \displaystyle\frac{1}{a}\log (ap_n  + 1)$
and $p_n  = \displaystyle\frac{{2\pi n}}{L}$. 
The mixing parameter states are defined as:
	\begin{equation}
\begin{array}{l}
 |x,\sigma /\left. 2 \right\rangle  = \sum\limits_{n = 0}^\infty  {e^{ip_n x} e^{ - \sigma E_n /2} |\left. {p_n } \right\rangle }  = \sum\limits_{n = 0}^\infty  {e^{i\displaystyle\frac{{2\pi n}}{L}x} (a\displaystyle\frac{{2\pi n}}{L} + 1)^{ - \sigma /2a} |\left. {p_n } \right\rangle }  \\ 
 |x',\sigma /\left. 2 \right\rangle  = \sum\limits_{n = 0}^\infty  {e^{ip_n x} e^{ - \sigma E_n /2} |\left. {p_n } \right\rangle  = } \sum\limits_{n = 0}^\infty  {e^{i\displaystyle\frac{{2\pi n}}{L}x'} (a\displaystyle\frac{{2\pi n}}{L} + 1)^{ - \sigma /2a} |\left. {p_n } \right\rangle }  \\ 
 \end{array}
\end{equation}
These states are normalized by: 
	\begin{equation}
\begin{array}{l}
 \left\langle x \right.,\sigma /2|x,\sigma /\left. 2 \right\rangle  = \left\langle x \right.',\sigma /2|x',\sigma /\left. 2 \right\rangle  = z(\sigma ) \\ 
  = \sum\limits_{n = 0}^\infty  {e^{ - \sigma E_n } }  = \sum\limits_{n = 0}^\infty  {(a\displaystyle\frac{{2\pi n}}{L} + 1)^{ - \sigma /a} }  = (\displaystyle\frac{L}{{2\pi a}})^{\sigma /a} \zeta _H (\displaystyle\frac{\sigma }{a},\displaystyle\frac{L}{{2\pi a}}) \\ 
 \end{array}
\end{equation}
Because of the simple time evolution of the momentum states, the time dependence of  $|x,\sigma /\left. 2 \right\rangle $
is:
	\begin{equation}
|x,\sigma /2,\left. t \right\rangle  = \sum\limits_{n = 0}^\infty  {e^{ip_n x - iE_n t - \sigma E_n /2} |\left. {p_n } \right\rangle } 
\end{equation}
The transition amplitude is:
\begin{equation}
\begin{array}{l}
 \left\langle x \right.',\sigma /2|x,\sigma /2,\left. t \right\rangle  = \sum\limits_{n = 0}^\infty  {e^{ip_n (x - x') - iE_n t - \sigma E_n } }  \\ 
                                = \sum\limits_{n = 0}^\infty  {e^{i\displaystyle\frac{{2\pi n}}{L}(x - x')} (a\displaystyle\frac{{2\pi n}}{L} + 1)^{( - it - \sigma )/a}  = (\displaystyle\frac{L}{{2\pi a}})^{(it + \sigma )/a} \phi (\displaystyle\frac{{x - x'}}{L},} \displaystyle\frac{{it + \sigma }}{a},\displaystyle\frac{L}{{2\pi a}}) \\ 
 \end{array}
\end{equation}
where $\phi $
is the Lerch zeta function. Again the transition amplitude is proportional to the Lerch zeta function however this time it is evaluated at $s = i(t - i\sigma ) = it + \sigma $. The transition probability is then:
\begin{equation}
\begin{array}{l}
 P(x,t \to x')_\sigma   = \displaystyle\frac{{|\left\langle x \right.',\sigma /2|x,\sigma /2,\left. t \right\rangle |^2 }}{{\left\langle x \right.,\sigma /2|x,\sigma /\left. 2 \right\rangle \left\langle x \right.',\sigma /2|x',\sigma /\left. 2 \right\rangle }} \\ 
                       = \displaystyle\frac{{|\left\langle x \right.',\sigma /2|x,\sigma /2,\left. t \right\rangle |^2 }}{{z(\sigma )z(\sigma )}}  = \displaystyle\frac{{|\phi (\displaystyle\frac{{x - x'}}{L},\displaystyle\frac{{it + \sigma }}{a},\displaystyle\frac{L}{{2\pi a}})|^2 }}{{(\zeta _H (\displaystyle\frac{\sigma }{a},\displaystyle\frac{L}{{2\pi a}}))^2 }} \\ 
 \end{array}
\end{equation}

Now we can give a physical representation of the Riemann hypothesis by specializing to  $x - x' = \displaystyle\frac{L}{2} = \pi R$
and $\displaystyle\frac{L}{{2\pi a}} = \displaystyle\frac{R}{a} = 1$. In that case the transition probability becomes:
	\begin{equation}
\begin{array}{l}
 P(x,t \to x') = \displaystyle\frac{{|\phi (\displaystyle\frac{1}{2},\displaystyle\frac{{it + \sigma }}{a},1)|^2 }}{{(\zeta _H (\displaystyle\frac{\sigma }{a},1))^2 }} \\ 
  = \displaystyle\frac{{|\eta (\displaystyle\frac{{it + \sigma }}{a})|^2 }}{{(\zeta (\displaystyle\frac{\sigma }{a}))^2 }} = \displaystyle\frac{{|(1 - 2^{1 - (it + \sigma )/a} )\zeta (\displaystyle\frac{{it + \sigma }}{a})|^2 }}{{(\zeta (\displaystyle\frac{\sigma }{a}))^2 }} \\ 
 \end{array}
\end{equation}
and the Riemann hypothesis is the statement that:
\begin{equation}
P(x,t \to x')  = \displaystyle\frac{{|(1 - 2^{1 - (it + \sigma )/a} )\zeta (\displaystyle\frac{{it + \sigma }}{a})|^2 }}{{(\zeta (\displaystyle\frac{\sigma }{a}))^2 }} > 0
\end{equation}
for $\displaystyle\frac{1}{2} < \displaystyle\frac{\sigma }{a} < 1$ and ${x - x' = \displaystyle\frac{L}{2},\displaystyle\frac{L}{{2\pi }} = a}.$ The physical setup is illustrated in Figure 6. The particle is measured at initial position $x=0$. The probability that a particle will subsequently be measured at position $x=L/2$ is nonzero according to the Riemann hypothesis.

\begin{figure}[htbp]

\centerline{\hbox{
   \epsfxsize=2.0in
   \epsffile{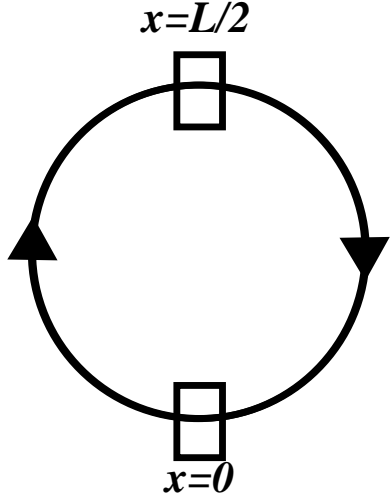}
     }
  }
  \caption{Physical setup for transition probability interpretation of the Riemann hypothesis. } 
          
  \label{fig6}

\end{figure}

One can also define the time averaged transition probability
	\begin{equation}
P(T) = \displaystyle\frac{1}{T}\displaystyle\frac{1}{{(z(\sigma ))^2 }}\int\limits_a^T {dt|\left\langle {x',\sigma /2|x,t,\left. {\sigma /2} \right\rangle |^2 } \right.}  = \displaystyle\frac{1}{{(\zeta (\sigma /a))^2 }}\displaystyle\frac{1}{T}\int\limits_a^T {dt} |\eta (\displaystyle\frac{{it + \sigma }}{a})|^2 
\end{equation}
The literature on zeta function theory contains a great deal of information on these averages for large $T$
as they are easier to study than the Riemann zeros.

A variant on the above analysis is to consider the transition between states:
\begin{equation}
\begin{array}{l}
 |\left. x \right\rangle  = \sum\limits_{n = 0}^\infty  {e^{2\pi i\displaystyle\frac{n}{R}x} |\left. {p_n } \right\rangle }  \\ 
 |\left. \sigma  \right\rangle  = \sum\limits_{n = 0}^\infty  {e^{ - \sigma E_n } |\left. {p_n } \right\rangle }  = \sum\limits_{n = 0}^\infty  {(a\displaystyle\frac{n}{R} + 1)^{ - \sigma /a} |\left. {p_n } \right\rangle }  \\ 
 \end{array}
\end{equation}

The $|\left. \sigma  \right\rangle$state is normalized by 
\begin{equation}\begin{array}{l}
 \left\langle {\sigma |\left. \sigma  \right\rangle } \right. = z(2\sigma ) = \sum\limits_{n = 0}^\infty  {e^{ - 2\sigma E_n } }  \\ 
  = \sum\limits_{n = 0}^\infty  {(a\displaystyle\frac{n}{R} + 1)^{ - 2\sigma /a}  = (\displaystyle\frac{R}{a}} )^{2\sigma /a} \zeta _H (\displaystyle\frac{{2\sigma }}{a},\displaystyle\frac{R}{a}) \\ 
 \end{array}\end{equation}
The $t$
evolution of the $|\left. x \right\rangle $
state is straightforward because of the simple evolution of$|\left. {p_n } \right\rangle $
by the phase $e^{ - iE_n t} $
and we have:
\begin{equation}
|\left. {x,t} \right\rangle  = \sum\limits_{n = 0}^\infty  {e^{2\pi i\displaystyle\frac{n}{R}x - iE_n t} |\left. {p_n } \right\rangle } 
\end{equation}
The transition amplitude is then:
	\begin{equation}
\begin{array}{l}
 \left\langle {\sigma |x,\left. t \right\rangle } \right. = \sum\limits_{n = 0}^\infty  {e^{ - \sigma E_n } e^{ - itE_n } e^{i\displaystyle\frac{n}{R}x} }  = \sum\limits_{n = 0}^\infty  {(a\displaystyle\frac{n}{R} + 1)^{( - \sigma  - it)/a} e^{i\displaystyle\frac{n}{R}x} }  \\ 
               = (\displaystyle\frac{R}{a})^{((\sigma  + it)/a} \phi (\displaystyle\frac{x}{{2\pi R}},\displaystyle\frac{{it + \sigma }}{a},\displaystyle\frac{R}{a}) \\ 
 \end{array}
\end{equation}
Taking the magnitude of the amplitude and dividing by the normalization of the $|\left. \sigma  \right\rangle $
states we have:
\begin{equation}
\displaystyle\frac{{|\left\langle {\sigma |x,\left. t \right\rangle } \right.|^2 }}{{\left\langle {\sigma |\left. \sigma  \right\rangle } \right.}} = \displaystyle\frac{{|\left\langle {\sigma |x,\left. t \right\rangle } \right.|^2 }}{{z(2\sigma )}} = \displaystyle\frac{{|\phi (\displaystyle\frac{x}{{2\pi R}},\displaystyle\frac{{it + \sigma }}{a},\displaystyle\frac{R}{a})|^2 }}{{\zeta _H (2\sigma /a,R/a)}}
\end{equation}
One can then use the asymptotic expression for the magnitude of the Lerch zeta function for $\displaystyle\frac{1}{2} < \sigma  < 1$
and $T \to \infty $ \cite{Laurincikas}
	\begin{equation}
\begin{array}{l}
 \int\limits_0^T {|\phi (x,t,\alpha )|^2 dt}  \\ 
  = \zeta_H (2\sigma ,\alpha )T + \displaystyle\frac{{(2\pi )^{2\sigma  - 1} }}{{2 - 2\sigma }}\zeta_H (2 - 2\sigma ,x)T^{2 - 2\sigma }  + BT^{1 - \sigma } \log T + BT^{\sigma /2}  \\ 
 \end{array}
\end{equation}
with $B$ a constant to obtain the asymptotic formula for the time averaged quantity:
\begin{equation}
\begin{array}{l}
 \displaystyle\frac{1}{T}\int\limits_0^T {dt} \displaystyle\frac{{|\left\langle {\sigma |x,\left. t \right\rangle } \right.|^2 }}{{\left\langle {\sigma |\left. \sigma  \right\rangle } \right.}} = \displaystyle\frac{1}{T}\int\limits_0^T {dt} \displaystyle\frac{{|\phi (\displaystyle\frac{x}{{2\pi R}},\displaystyle\frac{{it + \sigma }}{a},\displaystyle\frac{R}{a})|^2 }}{{\zeta _H (2\sigma /a,R/a)}}\\\;\; \approx  1 + \displaystyle\frac{1}{{2 - \displaystyle\frac{{2\sigma }}{a}}}\displaystyle\frac{{\zeta ((2a - 2\sigma )/a,x/2\pi R)}}{{\zeta (2\sigma /a,R/a)}}(\displaystyle\frac{T}{{2\pi a}})^{1 - 2\displaystyle\frac{\sigma }{a}}\\
\;\;\;\;  + \displaystyle\frac{{B(\displaystyle\frac{T}{a})^{ - \displaystyle\frac{\sigma }{a}} \log (T/a) + B(\displaystyle\frac{T}{a})^{\displaystyle\frac{\sigma }{{2a}} - 1} }}{{\zeta_H (2\sigma /a,R/a)}} \\ 
  \\ 
 \end{array}
\end{equation}
Specializing to the case relevant to the Riemann hypothesis we have $R = a$
and $x = \displaystyle\frac{L}{2} = \pi R = \pi a$
so that
\begin{equation}
\begin{array}{l}
 \displaystyle\frac{{|\left\langle {\sigma |x = \pi a,\left. t \right\rangle } \right.|^2 }}{{\left\langle {\sigma |\left. \sigma  \right\rangle } \right.}} = \displaystyle\frac{{|\left\langle {\sigma |x = \pi a,\left. t \right\rangle } \right.|^2 }}{{z(2\sigma )}} \\ 
  = \displaystyle\frac{{|\eta (\displaystyle\frac{{it + \sigma }}{a})|^2 }}{{\zeta (2\sigma /a)}} = \displaystyle\frac{{|(1 - 2^{1 - (it + \sigma )/a} )\zeta (\displaystyle\frac{{it + \sigma }}{a})|^2 }}{{\zeta (2\sigma /a)}} \\ 
 \end{array}
\end{equation}

The Riemann hypothesis is equivalent to the statement that the amplitude \begin{equation}
\left\langle {\sigma |x,\left. t \right\rangle } \right.
\end{equation}
is never zero in the range $\displaystyle\frac{1}{2} < \displaystyle\frac{\sigma }{a} < 1$
. The asymptotic formula of the time averaged quantity for $R = a$
,$x = \pi a$
,$\displaystyle\frac{1}{2} <  \displaystyle\frac{\sigma }{a} < 1$
, and $T \to \infty $
becomes:
\begin{equation}
\begin{array}{l}
 \displaystyle\frac{1}{T}\int\limits_0^T {dt} \displaystyle\frac{{|\left\langle {\sigma |x,\left. t \right\rangle } \right.|^2 }}{{\left\langle {\sigma |\left. \sigma  \right\rangle } \right.}} = \displaystyle\frac{1}{T}\int\limits_0^T {dt} \displaystyle\frac{{|\eta (\displaystyle\frac{{it + \sigma }}{a})|^2 }}{{\zeta (2\sigma /a)}} \approx  \\ 
   \;\;                                1 + \displaystyle\frac{1}{{2 - \displaystyle\frac{{2\sigma }}{a}}}\displaystyle\frac{{(2^{2 - 2\displaystyle\frac{\sigma }{a}}  - 1)\zeta ((2a - 2\sigma )/a)}}{{\zeta (2\sigma /a)}}(\displaystyle\frac{T}{{2\pi a}})^{1 - 2\displaystyle\frac{\sigma }{a}}\\
\;\;\;\;\;  + \displaystyle\frac{{B(\displaystyle\frac{T}{a})^{ - \displaystyle\frac{\sigma }{a}} \log (T/a) + B(\displaystyle\frac{T}{a})^{\displaystyle\frac{\sigma }{{2a}} - 1} }}{{\zeta (2\sigma /a)}} \\ 
  \\ 
 \end{array}
\end{equation}

\subsection{$\sigma $
in Partial sums and neutrino mixing analogy }

In this section we will develop an analogy of the transition probabilities of the previous section with the phenomenon of neutrino mixing. Partial sums for the Lerch zeta function are truncations of the series representation to $N$
 terms and are defined by:
	\begin{equation}
\phi _N (\displaystyle\frac{x}{L},\displaystyle\frac{{it + \sigma }}{a},\alpha ) = \sum\limits_{n = 1}^N {(n + \alpha )^{\displaystyle\frac{{ - it - \sigma }}{a}} e^{2\pi in\displaystyle\frac{x}{L}} } 
\end{equation}
One can formulate a sufficient condition for the Riemann hypothesis in terms of these partial sums as:
	\begin{equation}
\phi _N (x = \displaystyle\frac{1}{2},it + \sigma ,\alpha  = 1) =  - \eta _N (it + \sigma ) = \sum\limits_{n = 1}^N {( - 1)^n n^{ - it - \sigma } }  \ne 0
\end{equation}
for $0 < \sigma  < 1$
 (in this formula we have set $a = 1$
). Note that this a stronger condition than the usual Riemann hypothesis which only involves $N = \infty $. 

A nice feature of quantum mechanics is that it can be defined for finite dimensional Hilbert spaces so that one can also represent these partial sums as quantum transition amplitudes in a system with a finite number of states. Then quantum states are defined as  $|\left. {x,\sigma } \right\rangle  = \sum\limits_{n = 0}^N {e^{2\pi inx/L} e^{ - E_n \sigma } |\left. n \right\rangle } $
 with special states given by:
	\begin{equation}
\begin{array}{l}
 |\left. \nu  \right\rangle _\sigma   = |0,\left. {\sigma /2} \right\rangle  = \sum\limits_{n = 1}^N {e^{ - \sigma E_n /2} |\left. n \right\rangle }  \\ 
 |\left. {\nu '} \right\rangle _\sigma   = |\displaystyle\frac{L}{2},\left. {\sigma /2} \right\rangle  = \sum\limits_{n = 1}^N {( - 1)^n e^{ - \sigma E_n /2} |\left. n \right\rangle }  \\ 
 \end{array}
\end{equation}
The time dependence of the $|\left. \nu  \right\rangle _\sigma$ is easily determines from the evolution of the momentum basis states \begin{equation}
|\left. {\nu ,t} \right\rangle _\sigma   = \sum\limits_{n = 1}^N {e^{ - \sigma E_n /2} e^{ - iE_n t} |\left. n \right\rangle } 
\end{equation}
The normalization is defined by $\left\langle {\nu |\left. \nu  \right\rangle } \right. = \left\langle {\nu '|\left. {\nu '} \right\rangle } \right. = \sum\limits_{n = 0}^N {e^{ - \sigma E_n } }  = z_N (\sigma )$
So that the transition probability is:
\begin{equation}
\begin{array}{l}
 P_N (\nu  \to \nu ') = \displaystyle\frac{{|\left\langle {\nu '|\nu ,\left. t \right\rangle } \right.|^2 }}{{\left\langle {\nu |\left. \nu  \right\rangle \left\langle {\nu '|\left. {\nu '} \right\rangle } \right.} \right.}} = \displaystyle\frac{{|\sum\limits_{n = 1}^N {( - 1)^n e^{ - iE_n t - E_n \sigma } } |^2 }}{{z_N ^2 (\sigma )}} \\ 
  = \displaystyle\frac{{z_N (2\sigma ) - 2\sum\limits_{m \ne n}^N {e^{ - \sigma (E_m  + E_n )} \cos (t(E_m  - E_n )} }}{{z_N ^2 (\sigma )}} \\ 
 \end{array}
\end{equation}
Using in the dispersion relation $E_n  = \displaystyle\frac{1}{a}\log (a\displaystyle\frac{{2\pi n}}{L} + 1) = \displaystyle\frac{1}{a}\log (n + 1)$
 for $\displaystyle\frac{L}{{2\pi a}} = \displaystyle\frac{R}{a} = 1$
 we obtain a sufficient formulation of the Riemann hypothesis in terms of \begin{equation}
P(\nu  \to \nu ') = |\left\langle {\nu '|\nu ,\left. t \right\rangle } \right.|^2  > 0
\end{equation}
 where
\begin{equation}
P_N (\nu  \to \nu ') = \displaystyle\frac{{|\left\langle {\nu '|\nu ,\left. t \right\rangle } \right.|^2 }}{{|\left\langle {\nu |\left. \nu  \right\rangle } \right.|^2 |\left\langle {\nu '|\left. {\nu '} \right\rangle |^2 } \right.}} = \displaystyle\frac{{|\eta _N (it + \sigma )|^2 }}{{z_N^2 (\sigma )}} > 0
\end{equation}

To gain further insight into the mixing parameter let us consider truncating the momentum sum in (4.34) to two terms so that $N = 2$.

First define:
	\begin{equation}
\begin{array}{l}
 |\left. {x = \displaystyle\frac{L}{2}} \right\rangle  = \displaystyle\frac{1}{{2^{1/2} }}(\left. {|p = 0} \right\rangle  - |\left. {p = 1} \right\rangle ) \\ 
 |\left. {x = 0} \right\rangle  = \displaystyle\frac{1}{{2^{1/2} }}(\left. {|p = 0} \right\rangle  + |\left. {p = 1} \right\rangle ) \\ 
 \end{array}
\end{equation}
Then introduce the mixing parameter $\sigma $
by defining two states:
\begin{equation}
\begin{array}{l}
 |\left. \nu  \right\rangle _\sigma   = |0,\left. {\sigma /2} \right\rangle  = \sum\limits_{n = 1}^2 {e^{ - \sigma E_n /2} |\left. n \right\rangle }  = |\left. {p = 0} \right\rangle  + 2^{ - \sigma /2} |p = \left. 1 \right\rangle  \\ 
 |\left. {\nu '} \right\rangle _\sigma   = |\displaystyle\frac{L}{2},\left. {\sigma /2} \right\rangle  = \sum\limits_{n = 1}^N {( - 1)^n e^{ - \sigma E_n /2} |\left. n \right\rangle }  = |\left. {p = 0} \right\rangle  - 2^{ - \sigma /2} |p = \left. 1 \right\rangle  \\ 
 \end{array}
\end{equation}

The normalizing factor is $z(\sigma ) = 1 + 2^{ - \sigma } $
 so we set:
\begin{equation}
\begin{array}{l}
 |\left. \nu  \right\rangle _\sigma   = \cos \theta |\left. {p = 0} \right\rangle  + \sin \theta |p = \left. 1 \right\rangle  \\ 
 |\left. {\nu '} \right\rangle _\sigma   = \cos \theta |\left. {p = 0} \right\rangle  - \sin \theta |p = \left. 1 \right\rangle  \\ 
 \end{array}
\end{equation}
where we defined  $\theta  = \tan ^{ - 1} (2^{ - \sigma /2} )$.
The time dependence of the $\nu $
state is easily expressed as 
	\begin{equation}
\begin{array}{l}
 |\nu ,\left. t \right\rangle  = \cos \theta e^{ - iE_0 t} |p = \left. 0 \right\rangle  + \sin \theta e^{ - iE_1 t} |p = \left. 1 \right\rangle  \\ 
          = \cos \theta |p = \left. 0 \right\rangle  + \sin \theta 2^{ - it} |p = \left. 1 \right\rangle  \\ 
 \end{array}
\end{equation}
So that the transition probability is given by:
\begin{equation}
\begin{array}{l}
 P_{N = 2} (\nu  \to \nu ') = |\left\langle {\nu '|\nu ,\left. t \right\rangle } \right.|^2  = |(\cos \theta )^2  - 2^{ - it} (\sin \theta )^2 |^2  \\ 
                       = (\cos \theta )^4  + (\sin \theta )^4  - 2(\cos \theta \sin \theta )^2 \cos (t\log 2) \\ 
                       = 1 - 2(\cos \theta \sin \theta )^2 (1 + \cos (t\log 2)) \\ 
                       = 1 - (\sin 2\theta )^2 (\cos (t\log \sqrt 2 ))^2  \\ 
  \\ 
 \end{array}
\end{equation}
This is always greater than zero except for $\theta  = \displaystyle\frac{\pi }{4}$
or $\sigma  = 0$
which is outside the critical strip $0 < \sigma  < 1$. The transition probability for $N = 2$
 and $\sigma=.75$ is plotted as a function of time in Figure 7. It is identical with the two flavor neutrino mixing transition probability with $\Delta E = \log 2$. 

\begin{figure}[htbp]

 \centerline{\hbox{
   \epsfxsize=4.0in
   \epsffile{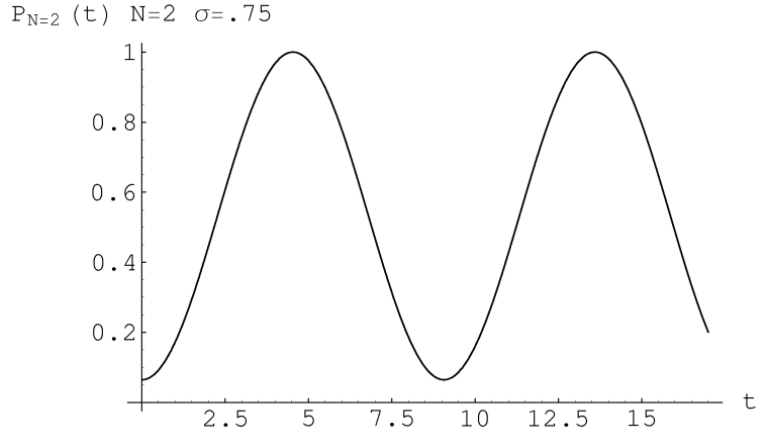}
     }
  }
  \caption{Transition probability for the $N=2$ partial sum for mixing parameter $\sigma=.75.$ } 
          
  \label{fig7}

\end{figure}

The $N$ partial sum transition probability is analogous to the $N$
flavor neutrino mixing transition probability. In figure 8 and figure 9 we plot the transition probability for $N = 4$ and $N = 16$. Unlike the $N = 2$
case we do not have periodicity of the probability amplitude. The case $N = \infty $
corresponds to the Riemann zeta function and is considered separately below.

\begin{figure}[htbp]

 \centerline{\hbox{
   \epsfxsize=4.0in
   \epsffile{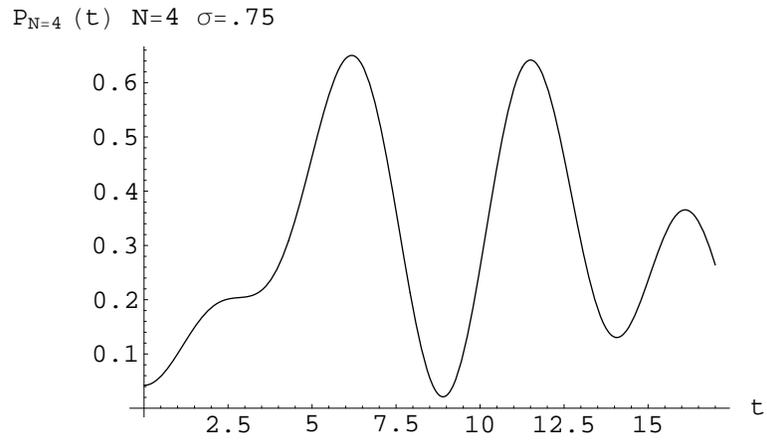}
     }
  }
  \caption{Transition probability for $N=4$ partial sum for mixing parameter $\sigma=.75$} 
          
  \label{fig8}

\end{figure}

\begin{figure}[htbp]

 \centerline{\hbox{
   \epsfxsize=4.0in
   \epsffile{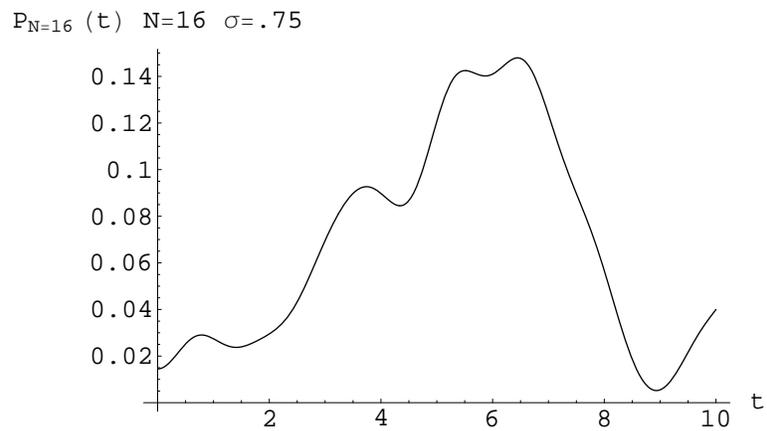}
     }
  }
  \caption{Transition probability for $N=16$ partial sum for mixing parameter $\sigma=.75$} 
          
  \label{fig9}

\end{figure}

The physical interpretation of these transition probabilities is as follows. One prepares an initial state weighted  by energy according to a Boltzmann distribution with mixing parameter $\sigma $
. One sets up a detector halfway around the universe at $\displaystyle\frac{L}{2}$
. One then counts particles received and records the time they were received. One then plots the number of particles received as a function of time and $\sigma $
. If there is time when no particles are received then the survival probability drops to zero. If the particle isn't at $\displaystyle\frac{L}{2}$
then the probability it is at another position is non zero.

We saw this at in the simple two state system above. In that case there were two positions $x = 0$
and $x = \displaystyle\frac{L}{2}$
(note that because of periodicity$x = L$
is the same point as $x = 0$
). The probability moves between the two states but the transition probability never vanished (except in the trivial case when the initial state was pure $\sigma  = 0$)
. This is exactly the behavior observed in the oscillation of massive neutrinos (or any other two state system that has two different energies and evolves from mixed superposition to a single state). Indeed the plot of the transition probability is familiar from such experimental neutrino studies.   

The $N = \infty $
transition probability discussed in the previous section also has a neutrino mixing analogy. In a Kaluza Klein description neutrino mixing involves the reduction from $5 \to 4$ dimensions. The five dimensional momentum $p_5  = m_n  = \displaystyle\frac{{2\pi n}}{L}$
 acts like a mass in four dimensions through the Dirac equation $(i\gamma ^\mu  \partial _\mu   + i\gamma ^5 \partial _5 )\psi  = 0$. Introducing a Kaluza Klein expansion $\psi (t,x_i ,x_5 ) = \sum\limits_{m =  - \infty }^\infty  {\psi _m (t,x_i )e{}^{2\pi i\displaystyle\frac{n}{L}x_5 }} $ the Dirac equation becomes $(i\gamma ^\mu  \partial _\mu   + \gamma _5 m_n )\psi _n  = 0$. The probability then moves between the different mass eigenstates with initial mixing dependent on the interaction with the boundary of the five dimensional space.

In our case the Riemann Zeta function condition $\alpha  = \displaystyle\frac{L}{{2\pi a}} = 1$
 means that we are essentially in a Kaluza Klein situation for any reasonably long time scales. Essentially the radius of the universe in fixed at a small scale proportional to the parameter $a$
. As in  Kaluza Klein theory  we can effect a reduction from $1 + 1 \to 0 + 1$
 dimensions through the expansion 
\begin{equation}\psi (t,x_1 ) = \sum\limits_{m =  - \infty }^\infty  {\psi _m (t)e{}^{2\pi i\displaystyle\frac{n}{L}x_1 }} \end{equation} 
with two special states defined by:
	\begin{equation}
\begin{array}{l}
 \left| {\nu ,t} \right\rangle  = \sum\limits_{n = 1}^\infty  {e^{ - iE_n t} e^{ - \sigma E_n /2} |\left. n \right\rangle }  \\ 
 \left| {\nu '} \right\rangle  = \sum\limits_{n = 1}^\infty  {( - 1)^n e^{ - \sigma E_n /2} |\left. n \right\rangle }  \\ 
 \end{array}
\end{equation}
So that the transition probability is:
	\begin{equation}
\begin{array}{l}
 P(\nu  \to \nu ') = \displaystyle\frac{{|\left\langle {\nu '|\nu ,\left. t \right\rangle } \right.|^2 }}{{|\left\langle {\nu '|\left. {\nu '} \right\rangle } \right.|^2 |\left\langle {\nu |\left. \nu  \right\rangle } \right.|^2 }} = \displaystyle\frac{{|\sum\limits_{n = 0}^\infty  {( - 1)^n e^{ - iE_n t - E_n \sigma } } |^2 }}{{z^2 (\sigma )}} \\ 
\;\;\;\;\;  = \displaystyle\frac{{z(2\sigma ) - 2\sum\limits_{m \ne n}^\infty  {e^{ - \sigma (E_m  + E_n )} \cos (t(E_m  - E_n )} }}{{z^2 (\sigma )}} \\ 
 \end{array}
\end{equation}
and the one particle partition function$z(\sigma ) = \sum\limits_{n = 0}^\infty  {e^{ - \sigma E_n } } $
is used to normalize the states.

The formula for the transition probability depends only on the dispersion relation. 
For case (2) $E_n  = \displaystyle\frac{{2\pi n}}{L}$
we have the $0 + 1$
 dimensional Dirac equation is written: $\left( {i\partial _t  + \displaystyle\frac{{2\pi n}}{L}} \right)\psi _n  = 0$
with solutions $\psi _n (t) = e^{ - im_n t}  = e^{ - i\displaystyle\frac{{2\pi n}}{L}t} $.
For case (4) $E_n  = \displaystyle\frac{1}{a}\log (a\displaystyle\frac{{2\pi n}}{L} + 1) = \displaystyle\frac{1}{a}\log (n + 1)$
these equations become:
\begin{equation}\begin{array}{l}
 \left( {\displaystyle\frac{1}{a}(e^{ai\partial _t }  - 1) + \displaystyle\frac{{2\pi n}}{L}} \right)\psi _n (t) = 0 \\ 
 \psi _n (t) = e^{ - im_n t}  = (\displaystyle\frac{{2\pi an}}{L} + 1)^{ - it/a}  = (n + 1)^{ - it/a}  \\ 
 \end{array}\end{equation}
The mixing in either case comes from a Boltzmann distributed initial state.

One may wonder where the particle goes so that the probability goes up and own as a function of time. As in the Kaluza Klein neutrino case the particle is lost to the bulk, in other words there is finite probability it will found at positions other than $x = 0$
or $x = \displaystyle\frac{L}{2}$. 

To compare with the Kaluza-Klein neutrino oscillation we note that
	\begin{equation}
|\left. \nu  \right\rangle  = \sum\limits_{n = 0}^\infty  {\displaystyle\frac{{(\displaystyle\frac{1}{a})^{\sigma /a} }}{{\sqrt {\displaystyle\frac{{(n + \displaystyle\frac{R}{a})^2 }}{{R^2 }}} ^{\sigma /a} }}} |n\left.\right\rangle 
\end{equation}
whereas the Kaluza Klein expansion is given by \cite{Lukas:2000wn}, \cite{Lukas:2000rg}, \cite{Ng:2003rv}, \cite{Dienes:2000ph}:
\begin{equation}
|\left. \nu  \right\rangle _{KK}  = \sum\limits_{n = 0}^\infty  {\displaystyle\frac{m}{{\sqrt {(\displaystyle\frac{{(n + R\mu _V )^2 }}{{R^2 }} + \mu _S^2 )} }}} |n\left.\right\rangle 
\end{equation}
So that the form is the same when $m = \displaystyle\frac{1}{a}$
, $\mu _V  = \displaystyle\frac{1}{a}$
, $\mu _S  = 0$
and $\sigma  = a.$ with $\mu _V$ and $\mu_S$ are vector and
scalar fermionic mass terms described in \cite{Lukas:2000rg}.
 However the time variation of this state is very different in our case because of the nontrivial logarithmic dispersion relation.

Using the methods of this section each of the cases 1-4 can be generalized to finite $\sigma $
with the transition probabilities where $\sigma $
is a mixing parameter.

\subsubsection*{Case (1)  $E = p$}

The $\nu $
 and $\nu '$
states are defined by:
\begin{equation}\begin{array}{l}
 |\left. {\nu ,t} \right\rangle  = \int {dpe^{i\pi ap} e^{ - itp - \sigma p/2} |\left. p \right\rangle }  \\ 
 |\left. {\nu '} \right\rangle  = \int {dpe^{ - \sigma p/2} |\left. p \right\rangle }  \\ 
 \end{array}\end{equation}
They are normalized by: $\left\langle {\nu |\left. \nu  \right\rangle } \right. = \left\langle {\nu '|\left. {\nu '} \right\rangle } \right. = \int\limits_0^\infty  {dpe^{ - \sigma p}  = \displaystyle\frac{1}{\sigma }} $.
The transition probability is given as:
\begin{equation}
\begin{array}{l}
 P_1 (\nu  \to \nu ',t) = \displaystyle\frac{{|\left\langle {\nu '|\nu ,\left. t \right\rangle } \right.|^2 }}{{|\left\langle {\nu '|\left. {\nu '} \right\rangle } \right.|^2 |\left\langle {\nu |\left. \nu  \right\rangle } \right.|^2 }} \\ 
  = \displaystyle\frac{{|g_1 (\pi a,it + \sigma )|^2 }}{{|\displaystyle\frac{1}{\sigma }|^2 }} = \displaystyle\frac{{|\displaystyle\frac{i}{{\pi a - (t - i\sigma )}}|^2 }}{{|\displaystyle\frac{1}{\sigma }|^2 }} = \displaystyle\frac{{\sigma ^2 }}{{(\pi a - t)^2  + \sigma ^2 }} \\ 
 \end{array}
\end{equation}
where we have used $g_1 (x,s) = \displaystyle\frac{i}{{x + is}}$
 computed in section 3.	
We plot the transition probability in figure 10 for $\sigma  = .75a$.

\begin{figure}[htbp]

 \centerline{\hbox{
   \epsfxsize=4.0in
   \epsffile{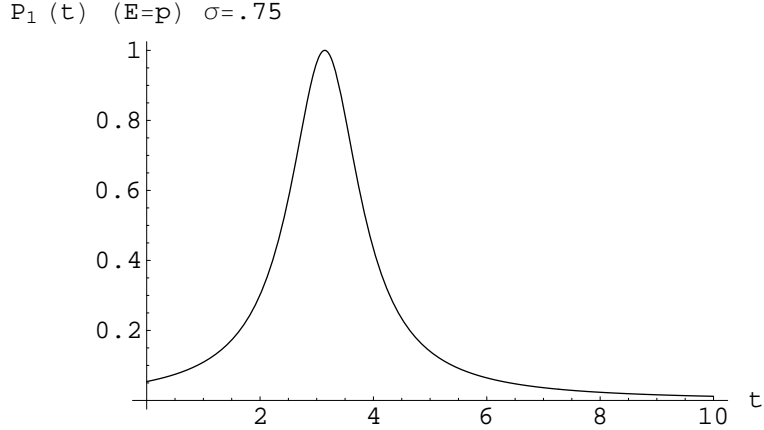}
     }
  }
  \caption{Transition probability for Case (1) $E=p.$ } 
          
  \label{fig10}

\end{figure}

\subsubsection*{Case (2) $E = \displaystyle\frac{n}{R}$with $R = a$}

The $\nu $
and $\nu '$
states are defined by:
	\begin{equation}
\begin{array}{l}
 |\left. {\nu ,t} \right\rangle  = \sum\limits_{n = 0}^\infty  {e^{ - itn/a} e^{ - \sigma n/2a} |\left. n \right\rangle }  \\ 
 |\left. {\nu '} \right\rangle  = \sum\limits_{n = 0}^\infty  {e^{i\pi n} e^{ - \sigma n/2a} |\left. n \right\rangle }  \\ 
 \end{array}
\end{equation}
These states are normalized by: $\left\langle {\nu |\left. \nu  \right\rangle } \right. = \left\langle {\nu '|\left. {\nu '} \right\rangle } \right. = \sum\limits_{n = 0}^\infty  {e^{ - n\sigma /a} }  = \displaystyle\frac{1}{{1 - e^{ - \sigma /a} }}$.
 Then the transition probability is:
\begin{equation}
\begin{array}{l}
 P_2 (\nu  \to \nu ',t) = \displaystyle\frac{{|\left\langle {\nu '|\nu ,\left. t \right\rangle } \right.|^2 }}{{\left\langle {\nu '|\left. {\nu '} \right\rangle } \right.\left\langle {\nu |\left. \nu  \right\rangle } \right.}} \\ 
  = \displaystyle\frac{{|g_2 (\pi a,it + \sigma )|^2 }}{{|\displaystyle\frac{1}{{1 - e^{ - \sigma /a} }}||^2 }} = \displaystyle\frac{{|\displaystyle\frac{1}{{1 - e^{i(x - t + i\sigma )/a} }}|^2 }}{{|\displaystyle\frac{1}{{1 - e^{ - \sigma /a} }}|^2 }} = |\displaystyle\frac{{1 - e^{ - \sigma /a} }}{{1 - e^{i(x - t + i\sigma )/a} }}|^2  \\ 
 \end{array}
\end{equation}
and we have used $g_2 (x,s) = \displaystyle\frac{1}{{1 - e^{i(x + is)/R} }}$
 from section 3. We plot the transition probability in figure 11 for $\sigma  = .75a$
 We show in section 8 that case (2) is the periodization of case (1) in the sense that $g_2 (x,s) = \sum\limits_{n =  - \infty }^\infty  {g_{1} (x + 2\pi Rn,s)} $
or
\begin{equation}G_2 (x,t,0,i\sigma ) = \sum\limits_{n =  - \infty }^\infty  {G_{1} (x + 2\pi Rn,t;0,i\sigma )} \end{equation}
This explains the multiple peaks in figure 11.

\begin{figure}[htbp]

 \centerline{\hbox{
   \epsfxsize=4.0in
   \epsffile{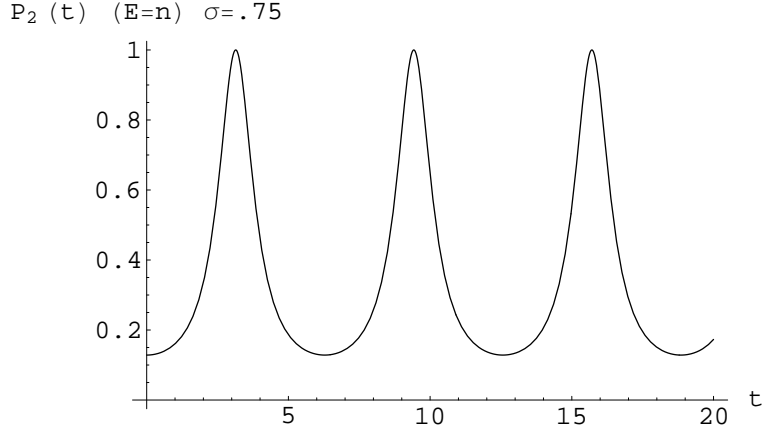}
     }
  }
  \caption{Transition probability for Case (2) $E = \displaystyle\frac{n}{R}$with $R = a$ and $\sigma=.75.$ } 
          
  \label{fig11}

\end{figure}

\subsubsection*{Case (3) $E = \displaystyle\frac{1}{a}\log (ap + 1)$}

The $\nu $
and $\nu '$
states are defined by:
\begin{equation}\begin{array}{l}
 |\left. {\nu ,t} \right\rangle  = \int {dp(ap + 1)^{ - it/a - \sigma /2a} |\left. p \right\rangle }  \\ 
 |\left. {\nu '} \right\rangle  = \int {dpe^{i\pi ap} (ap + 1)^{ - \sigma /2a} |\left. p \right\rangle }  \\ 
 \end{array}\end{equation}
The normalization is defined as $\left\langle {\nu |\left. \nu  \right\rangle } \right. = \left\langle {\nu '|\left. {\nu '} \right\rangle } \right. = \int\limits_0^\infty  {dp(ap + 1)^{ - \sigma /a}  = \displaystyle\frac{1}{{\sigma  - a}}} $.
The transition probability is:
\begin{equation}
\begin{array}{l}
 P_3 (\nu  \to \nu ',t) = \displaystyle\frac{{|\left\langle {\nu '|\nu ,\left. t \right\rangle } \right.|^2 }}{{|\left\langle {\nu '|\left. {\nu '} \right\rangle } \right.|^2 |\left\langle {\nu |\left. \nu  \right\rangle } \right.|^2 }} = \displaystyle\frac{{|g_3 (\pi a,it + \sigma )|^2 }}{{|\displaystyle\frac{1}{{\sigma  - a}}|^2 }} \\ 
  = |(\displaystyle\frac{{\sigma  - a}}{a})e^{( - i\displaystyle\frac{\pi }{2} + \log (\pi ))( - it + (1 - \sigma ))/a} \Gamma ( - i\pi ,1 - (it + \sigma )|^2  \\ 
 \end{array}
\end{equation}
where we have used $g_3 (x,s) = e^{ - ix/a} ( - ix)^{1 - s} \Gamma ( - ix,1 - s)$
for $s = (it + \sigma )/a$
derived in section 3. We plot the transition probability in figure 12 for $\sigma  = .75a$.

\begin{figure}[htbp]

 \centerline{\hbox{
   \epsfxsize=4.0in
   \epsffile{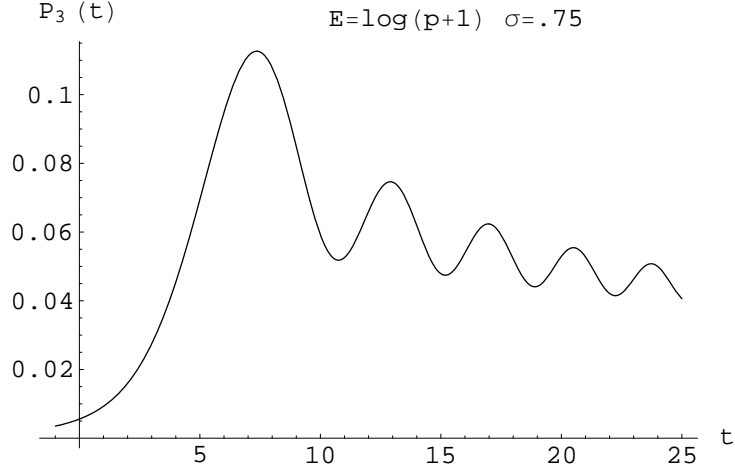}
     }
  }
  \caption{Transition probability for Case (3)  $E = \displaystyle\frac{1}{a}\log (ap + 1)$ for $\sigma=.75.$} 
          
  \label{fig12}

\end{figure}

\subsubsection{Case ($3'$) }

This is similar to the above case except $g_{3'} (x,s) = G_{3'} (x,t + ia - i\sigma ;0,0)$
is defined by the  $(E,p)$
integral:
\begin{equation}
\begin{array}{l}
 g_{3'} (x,s) = G_{3'} (x,t + ia - i\sigma ;0,0) \\ 
  = \displaystyle\frac{1}{{(2\pi )^2 }}\int {dEdp\displaystyle\frac{1}{{\displaystyle\frac{1}{a}(e^{aE}  - 1) - p + i\varepsilon }}e^{ - i(t + ia - i\sigma )E} e^{ipx} }  \\ 
  = \displaystyle\frac{1}{a}e^{ - ix/a} (\displaystyle\frac{{ - ix}}{a})^{ - 1 + s} \Gamma (1 - s) \\ 
 \end{array}
\end{equation}
The $\nu $
and $\nu '$
states are defined by
	\begin{equation}
\begin{array}{l}
 |\left. {\nu ,t} \right\rangle  = \displaystyle\frac{1}{{(2\pi )^2 }}\int {dEdp\displaystyle\frac{1}{{\displaystyle\frac{1}{a}(e^{aE}  - 1) - p + i\varepsilon }}e^{ - i(t + ia - i\sigma /2)E} |\left. p \right\rangle }  \\ 
 |\left. {\nu ',t'} \right\rangle  = \displaystyle\frac{1}{{(2\pi )^2 }}\int {dEdp\displaystyle\frac{1}{{\displaystyle\frac{1}{a}(e^{aE}  - 1) - p + i\varepsilon }}e^{ - i(t' + ia - i\sigma /2)E} e^{i\pi a/a} |\left. p \right\rangle }  \\ 
 \end{array}
\end{equation}
The transition probability is then given by:
\begin{equation}
\begin{array}{l}
 P_{3'} (\nu  \to \nu ') = \displaystyle\frac{{|\left\langle {\nu '|\nu ,\left. t \right\rangle } \right.|^2 }}{{|\left\langle {\nu '|\left. {\nu '} \right\rangle } \right.|^2 |\left\langle {\nu |\left. \nu  \right\rangle } \right.|^2 }} \\ 
  = \displaystyle\frac{{|g_{3'} (a\pi ,(it + \sigma )/a)|^2 }}{{|z_{3'} (\sigma )|^2 }} = |(e^{( - i\pi /2 + \log (\pi ))( - it + (1 - \sigma ))/a} )\Gamma (1 - (it + \sigma )/a)|^2  \\ 
 \end{array}
\end{equation}
In figure 13 we plot the transition probability for mixing parameter$\sigma  = .75a$

\begin{figure}[htbp]

 \centerline{\hbox{
   \epsfxsize=4.0in
   \epsffile{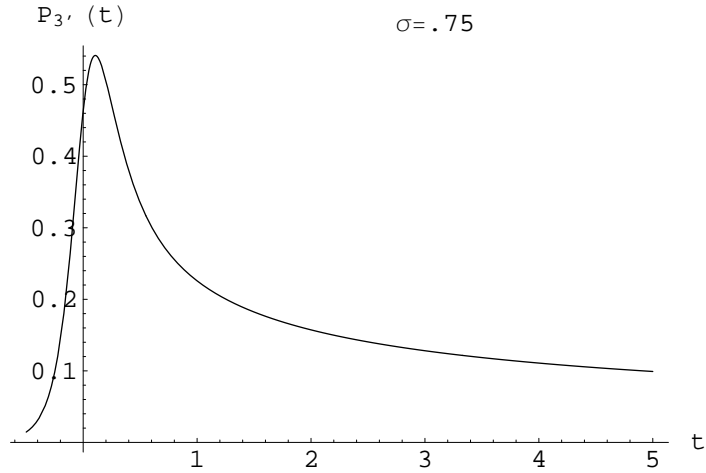}
     }
  }
  \caption{Transition probability for Case ($3'$) using $(E,p)$ representation for $\sigma=.75.$} 
          
  \label{fig13}

\end{figure}

\subsubsection*{Case (4) $E = \displaystyle\frac{1}{a}\log (a\displaystyle\frac{n}{R} + 1)$}

For $R = a$
this is the case relevant to the Riemann hypothesis. The $\nu $
and $\nu '$
 states are defined by:
\begin{equation}
\begin{array}{l}
 |\left. {\nu ,t} \right\rangle  = \sum\limits_{n = 1}^\infty  {n^{ - it/a} n^{ - \sigma /2a} |\left. n \right\rangle }  \\ 
 |\left. {\nu '} \right\rangle  = \sum\limits_{n = 0}^\infty  {e^{i\pi n} n^{ - \sigma /2a} |\left. n \right\rangle }  \\ 
 \end{array}
\end{equation}
The states are normalized by $\left\langle {\nu |\left. \nu  \right\rangle } \right. = \left\langle {\nu '|\left. {\nu '} \right\rangle } \right. = \sum\limits_{n = 1}^\infty  {n^{ - \sigma /a} }  = \zeta (\sigma /a)$
. The transition probability is given by:
\begin{equation}
\begin{array}{l}
 P_4 (\nu  \to \nu ',t) = \displaystyle\frac{{|\left\langle {\nu '|\nu ,\left. t \right\rangle } \right.|^2 }}{{\left\langle {\nu '|\left. {\nu '} \right\rangle } \right.\left\langle {\nu |\left. \nu  \right\rangle } \right.}} \\ 
                       = \displaystyle\frac{{|g_4 (\pi a,it + \sigma )|^2 }}{{|\zeta (\sigma /a)|^2 }} = \displaystyle\frac{{|\eta ((it + \sigma )/a)|^2 }}{{|\zeta (\sigma /a)|^2 }} = \displaystyle\frac{{|(1 - 2^{1 - (it + \sigma )} )\zeta ((it + \sigma )/a)|^2 }}{{|\zeta (\sigma /a)|^2 }} \\ 

 \end{array}
\end{equation}
We plot the transition probability for $\sigma  = .75$ in Figure 14.

\begin{figure}[htbp]

 \centerline{\hbox{
   \epsfxsize=4.0in
   \epsffile{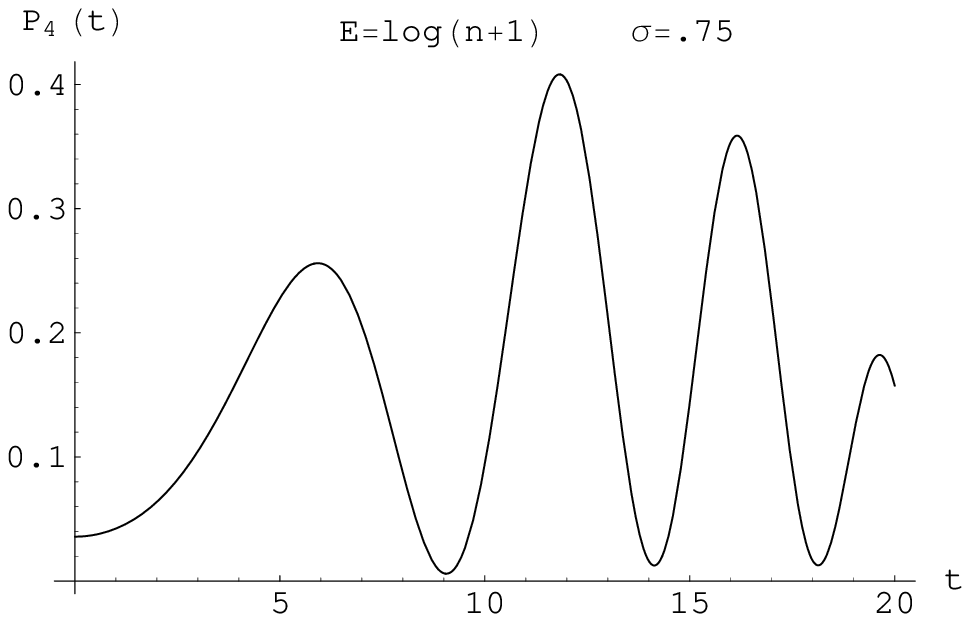}
     }
  }
  \caption{Transion probability for Case (4)  $E = \displaystyle\frac{1}{a}\log (a\displaystyle\frac{n}{R} + 1)$ with $R=a$ and $\sigma=.75.$} 
          
  \label{fig14}

\end{figure}

We show in section 8 that case (4) is the periodization of case ($3'$) in the sense that
$g_4 (x,s) = \sum\limits_{n =  - \infty }^\infty  {g_{3'} (x + 2\pi Rn,s)} $
or 
\begin{equation}G_4 (x,t,0,i\sigma ) = \sum\limits_{n =  - \infty }^\infty  {G_{3'} (x + 2\pi Rn,t;0,i\sigma )} \end{equation} 
This explains the multiple peaks in figure 14. Note that unlike the periodization from case (1) to case (2) one does not have periodization in $t$. This because in case $(3 - 4)$
the Green's function is not a function of $(x - t)$. The Riemann hypothesis in this context is the statement that that the transition probability from the state $\nu  \to \nu '$ is nonzero for mixing parameters in the range $\displaystyle\frac{1}{2}a < \sigma  < a$.

\subsection{$\sigma $
as a SUSY breaking parameter}

In supersymmetric quantum mechanics one considers Hamiltonians of the form \cite{Witten:1981nf}:
	\begin{equation}
H_1  = \displaystyle\frac{1}{2}(\displaystyle\frac{{p_x^2 }}{{m_1 }} + U^2 ) + \displaystyle\frac{{dU}}{{dx}}(b_1^\dag  b_1  - \displaystyle\frac{1}{2})
\end{equation}
Here $b_1^\dag  ,b_1 $
are fermionic creation and annihilation operators. In principle $U$
 can be an arbitrarily complicated function. Often its form it determined by symmetry properties. For example $U$
can be a modular function if $y$
represents a modulus or compactified scalar field \cite{Greene:1989ya}. If one takes the zeta function as $U$
then the symmetry property would be the functional equation together with the Dirichlet expansion which uniquely determines the form of the function up to a proportionality constant.

To be more specific we add to the above the Hamiltonian
\begin{equation}
H_2  = \displaystyle\frac{1}{2}(\displaystyle\frac{{p_y^2 }}{{m_2 }} + V^2 ) + \displaystyle\frac{{dV}}{{dx}}(b_2^\dag  b_2  - \displaystyle\frac{1}{2}))
\end{equation}
where $W = U + iV$
is an analytic function of $x + iy$. The total Hamiltonian becomes:
\begin{equation}\begin{array}{l}
 H = H_1  + H_2  \\ 
  = \displaystyle\frac{1}{2}(\displaystyle\frac{{p_x^2 }}{{m_1 }} + \displaystyle\frac{{p_y^2 }}{{m_2 }} + |W|^2 ) + \displaystyle\frac{{dU}}{{dx}}(b_1^\dag  b_1  - \displaystyle\frac{1}{2}) + \displaystyle\frac{{dV}}{{dy}}(b_2^\dag  b_2  - \displaystyle\frac{1}{2}) \\ 
 \end{array}\end{equation}
Now using the Cauchy-Riemann condition $\displaystyle\frac{{\partial U}}{{\partial x}} = \displaystyle\frac{{\partial V}}{{\partial y}}$ we have:
\begin{equation}H = \displaystyle\frac{1}{2}(\displaystyle\frac{{p_x^2 }}{{m_1 }} + \displaystyle\frac{{p_y^2 }}{{m_2 }} + |W(x + iy)|^2 ) + \displaystyle\frac{{dV}}{{dy}}(b_1^\dag  b_1  + b_2^\dag  b_2  - 1)\end{equation}
Setting $W(x + iy) = \zeta (x + iy)$
and relabeling $x$
by $\sigma $
we obtain the Hamiltonian:
\begin{equation}H = \displaystyle\frac{1}{2}(\displaystyle\frac{{p_\sigma ^2 }}{{m_1 }} + \displaystyle\frac{{p_y^2 }}{{m_2 }} + |\zeta (\sigma  + iy)|^2 ) + \displaystyle\frac{{dV}}{{dy}}(b_1^\dag  b_1  + b_2^\dag  b_2  - 1)\end{equation}
Finally taking $m_1  \to \infty $
 we have :
\begin{equation}H = \displaystyle\frac{1}{2}(\displaystyle\frac{{p_y^2 }}{{m_2 }} + |\zeta (\sigma  + iy)|^2 ) + \displaystyle\frac{{dV}}{{dy}}(b_1^\dag  b_1  + b_2^\dag  b_2  - 1)\end{equation}
In supersymmetric quantum mechanics the Hamiltonian is related to the supercharges by $H = \sum\limits_\alpha  {Q_\alpha ^2 } $
. This means supersymmetry will be broken unless the ground state has zero energy. Thus supersymmetry will be broken if the function $W$
does not intersect zero. In Figure 15 and 16 we plot $|W|^2  = |\zeta (iy + \displaystyle\frac{1}{2})|^2 $
 and $|W|^2  = |\zeta (iy + 0)|^2 $.

\begin{figure}[htbp]

 \centerline{\hbox{
   \epsfxsize=4.0in
   \epsffile{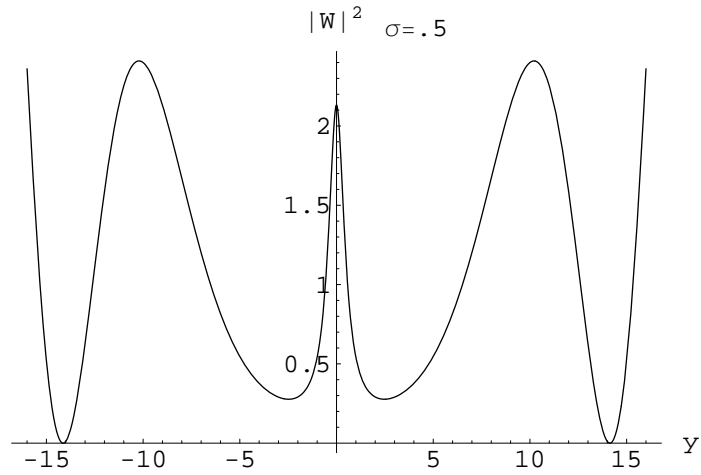}
     }
  }
  \caption{$|W|^2  = |\zeta (iy + \displaystyle\frac{1}{2})|^2.$ The presence of the first Riemann zero is consistent with unbroken supersymmetry.} 
          
  \label{fig15}

\end{figure}

\begin{figure}[htbp]

 \centerline{\hbox{
   \epsfxsize=4.0in
   \epsffile{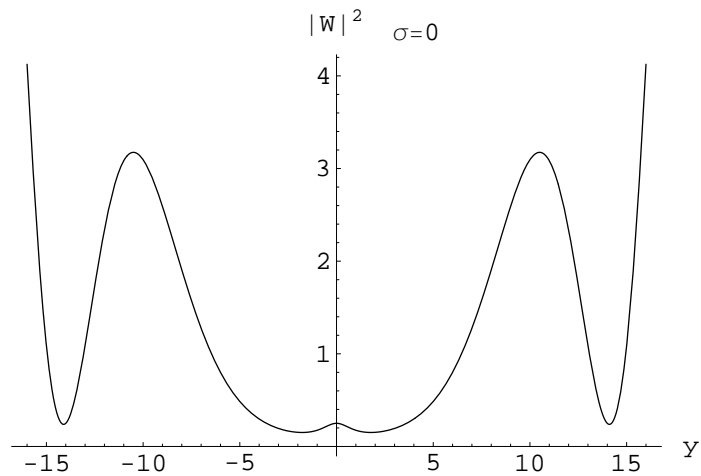}
     }
  }
  \caption{$|W|^2  = |\zeta (iy + 0)|^2.$ The absence of zeros according to the Riemann hypothesis leads to broken supersymmetry.} 
          
  \label{fig16}

\end{figure}

If the Riemann hypothesis is false then supersymmetry may remain as a symmetry of (4.70) assuming a zero occurs for large $y.$ To understand further how $\sigma $
manifests itself as a supersymmetry breaking parameter we need to understand how the potential is generated. For example finite temperature effects always break supersymmetry
because the bosons and fermions react differently to finite temperature. Bosons have integer momentum in the imaginary time formalism and fermions have half integer. This leads to potentials that have no zero for temperature $T > 0$
. In that case the supersymmetry breaking parameter is the temperature $T$
which is varied away from zero. In another example finite lattice effects can break supersymmetry because the translation symmetry is broken by the lattice and supersymmetry is related to space-time translations through the superpoincaire algebra. In that case the supersymmetry breaking parameter is the lattice spacing. In our  case the zeta function has both elements of discreteness from the Wick rotated Dirac equation (3.28) and momentum quantization from the periodicity of the spatial direction in (3.56).

\subsection{$(0 \to 0)$
String amplitudes and $(1 \to 1)$
amplitudes in Riemann dynamics}

\subsubsection*{$(0 \to 0)$String amplitudes}

One loop String vacuum amplitudes are defined by two dimensional sigma models of the form:
	\begin{equation}
\Lambda (0 \to 0) = \int {De_z^a } DX^M e^{ - \int {d^2 z(e(e_a^z e_a^{\bar z} e\partial _z X\partial _{\bar z} X)} } 
\end{equation}
The sigma model field $X^M (z,\bar z)$
 maps the two torus $T^2 $
to a target space of the form $S^1  \times M^{d - 1} $
 where the circle has radius $R$
. The two dimensional world sheet metric is related to the modular parameters of the torus through $e_z^a e_{\bar z}^a dzd\bar z = (d\sigma ^0 ,d\sigma ^1 )\displaystyle\frac{1}{{\tau _2 }}\left( {\begin{array}{*{20}c}
   {\tau _1^2  + \tau _2^2 } & {\tau _1 }  \\
   {\tau _1 } & 1  \\
\end{array}} \right)\left( \begin{array}{l}
 d\sigma ^0  \\ 
 d\sigma ^1  \\ 
 \end{array} \right)$
and $z = \displaystyle\frac{1}{2}(\sigma ^1  + i\sigma ^0 ),\bar z = \displaystyle\frac{1}{2}(\sigma ^1  - i\sigma ^0 )$
. On a torus the sigma model reduces to an integral over the modular parameter and its complex conjugate so that the path integral becomes:
 	\begin{equation}
\begin{array}{l}
 \Lambda (R) = \int\limits_F {d\tau } \theta (\tau ,\bar \tau ,R)P_L (\tau )P_R (\bar \tau ) \\ 
  = \int\limits_F {d\tau d\bar \tau } (\sum\limits_n {e^{ - \pi \tau (\displaystyle\frac{n}{R} + \displaystyle\frac{{mR}}{{\alpha '}})^2  - \pi \bar \tau (\displaystyle\frac{n}{R} - \displaystyle\frac{{mR}}{{\alpha '}})^2 } } )P_L (\tau )P_R (\bar \tau ) \\ 
 \end{array}
\end{equation}
where $P_{L,R} (\tau ) = tr_{L,R} (e^{ - \tau M^2 } )$
is a left or right moving string partition function and $R$
is a compactification radius. 

It is knownthat global anomalies can lead to zeros in one loop vacuum amplitudes of string models. In \cite{Moore:1987ue} models were found that had global anomalies in symmetries that were not required for the consistency of the theory so these string models still were sensible. For example zeros in one loop amplitudes were found in which the integrand of (4.71) had an additional modular symmetry of the form:
	\begin{equation}
\begin{array}{l}
 {\rm{Usual}}:         \tau  \to \tau  + 1;\tau  \to  - \displaystyle\frac{1}{\tau } \\ 
 {\rm{Additional}}:\tau  \to  - \displaystyle\frac{1}{{2\tau }} \\ 
 \end{array}
\end{equation}
The additional symmetry usually arises in the form of a discrete exchange of compactified spaces. This symmetry together with the modular properties of the integrand leads to the zeros.

Also zeros in the derivative of the vacuum energy of string models can also be obtained at special values of the compactified radii \cite{Nair:1986zn}. These are points of enhanced symmetry of the theory where the usual $U(1) \times U(1)$
of the compactified theory is enhanced to $SU(2)$
or $SU(3)$
and certain states becomes massless. These are also the location of thermal divergences if the compactified moduli are inverse temperature or chemical potentials in the imaginary time finite temperature string theory where the vacuum energy becomes the free energy.  

Other features of $(0 \to 0)$
string amplitudes are the existence of the fundamental region:
	\begin{equation}
F = \{ \tau :|\tau | > 1, - \displaystyle\frac{1}{2} < \tau _1  < \displaystyle\frac{1}{2}\} 
\end{equation}
and the duality symmetry under the transformation:
	\begin{equation}
R' = \displaystyle\frac{{\alpha '}}{R}
\end{equation}
These are very important properties of string theory which differentiate them from traditional point particle theories.

\subsubsection*{$(1 \to 1)$ Amplitudes in Riemann dynamics}
	
We have mainly used the mode expansion representation of the Green's function of the one particle to one particle  $(1 \to 1)$
amplitude. In section 8 we will derive several other representation of the Green's function. One of these is the point particle path integral representation: 
	\begin{equation}
A(1 \to 1) = G(x,t;0,  i\sigma )|_R  = \int\limits_{x(0) = 0,t(0) =   i\sigma }^{x(1) = x,t(1) = t} {De_0 D\chi _0 D\xi D\bar \xi DPDxDEDte^{ - I} } 
\end{equation}
The fields $(e_0 (\tau ),\chi _0 (\tau ))$
define a 0+1 dimensional supergravity associated with (super)reparametrization point particle action, $\xi $
and $\bar \xi $ are world line fermion fields which eventually become the target space gamma matrices. The point particle action is given by:
	\begin{equation}
\begin{array}{l}
 I = \int\limits_{\tau _1 }^{\tau _2 } {d\tau (P\dot x}  - E\dot t - \displaystyle\frac{{e_0 }}{2}(P^2  - \displaystyle\frac{1}{{a^2 }}(e^{aE}  - 1)^2 ) \\ 
\;\;\;\;\;  + \chi _0 \displaystyle\frac{1}{a}P\xi  + \chi _0 \displaystyle\frac{1}{a}(e^{aE}  - 1)\bar \xi  + i\bar \xi  \dot{ \bar \xi}  + i\xi \dot \xi ) \\ 
 \end{array}
\end{equation}
The fields $(x,t)$
map the interval $\tau  \in [\tau _1 ,\tau _2 ]$
to a cylindrical target space where the $S^1 $ factor
has the radius $R$
(Recall the string $0 \to 0$
amplitude amplitude mapped the two torus onto $S^1  \times M^{d - 1} $
target space). The periodicity of the target $x$
direction ensures that the momentum is quantized as $P = \displaystyle\frac{n}{R}$. The Tiechmuller parameter of the interval is: $C = \int\limits_{\tau _1 }^{\tau _2 } {e_0 (\tau )d\tau } $
We use the symbol $C$
to denote the Teichmuller parameter for the point particle to avoid confusion with parameterized time in the point particle action.

The point particle Path integral defining the $(1 \to 1)$
amplitude can be written as an integral over $C$
by:
\begin{equation}\begin{array}{l}
 A(1 \to 1) = (2^s  - 1)\pi ^{ - s/2} \Gamma (s/2)aG(x = \pi a,t;0, - i\sigma )|_{R = a}  \\ 
                          = (2^s  - 1)(2^{1 - s}  - 1)\pi ^{ - s/2} \Gamma (s/2)\zeta (s) = \int\limits_F {dCC^{s/2 - 1} (\theta _3 (C) - \theta _2 (C) - \theta _4 (C))}  \\ 
 \end{array}\end{equation}
The prefactor $\pi ^{ - s/2} (2^s  - 1)\Gamma (s/2)$
is introduced to allow simple transformation associated with the functional equation. We leave the details of the derivation to section 8. Some features can be recognized however. The theta functions occur because of the quantization of momentum associated with the periodicity of the target space and the form $e^{ - CP^2 } $
in the path integral. The unusual $C^{s/2} $
factor occurs because of the $e^{ - E(it + \sigma )} $
factor the change of variable $W = e^{aE} $
in the integral over $W$.

The usual modular transformation for a point particle on the interval world line is $C \to  - C$ associated with the discrete symmetry $e_0 (\tau ) \to  - e_0 ( - \tau )$
and the exchange of the endpoints $\tau _1 $
and $\tau _2 $. As shown in \cite{Gozzi:1988uj} it is this world line modular symmetry that leads to the discrete space-time reflection symmetry $CPT$
in the Target space. There is an additional modular symmetry $C \to \displaystyle\frac{1}{C}$
associated with the functional relation obeyed by (4.78). One can actually derive the functional equation by tracing the transformation properties of the theta functions under $C \to \displaystyle\frac{1}{C}$
and their effect on the $(1 \to 1)$
amplitude. Thus the modular transformations of (4.78) are given by:
	\begin{equation}
\begin{array}{l}
 {\rm{Usual}}:         C \to  - C \\ 
 {\rm{Additional}}:C \to \displaystyle\frac{1}{C} \\ 
 \end{array}
\end{equation}
The additional modular transformation is directly connected with the Riemann zeros. This is because the Riemann zeta function is uniquely determined up to a constant factor by its functional relation and the fact that it has a Dirichlet expansion. The functional relation is in turn derived from the action under the additional modular transformation. The Riemann zeta function is also uniquely determined as a product over the Riemann zeros as:
	\begin{equation}
\zeta (s) = \displaystyle\frac{{e^{bs} }}{{2(s - 1)\Gamma (1 + s/2)}}\prod\limits_\rho  {(1 - \displaystyle\frac{s}{\rho }} )e^{s/\rho } 
\end{equation}
where $b = \log (2\pi ) - 1 - \displaystyle\frac{1}{2}\gamma $
and $\rho $
are Riemann zeros. As both the position of the zeros and the functional relation uniquely determine the zeta function they must be related.
If one interprets $s$
as additional data for the $(1 \to 1)$
amplitude on can interpret the Riemann zeros as specific choices of this data that cause the amplitude to vanish. In the $(0 \to 0)$
string amplitude one has choices for the internal radii and boundary conditions. In the $(1 \to 1)$
point particle amplitude one has additional data namely the initial and final space-time positions and the mixing parameter $\sigma $.

The usual modular transformation $C \to  - C$
of the point particle action on the interval has important implications for the fundamental region $F$
. It allows one to choose the fundamental region of the point particle as $F = \{ C:C \ge 0\}  = [0,\infty ] $\cite{Giannakis:1988yh}. For example the energy momentum $(E,P)$
form of the Green's function 
	\begin{equation}
\begin{array}{l}
 \int\limits_F {dCdPdEe^{iC(P^2  - \displaystyle\frac{1}{{a^2 }}(e^{aE}  - 1)^2 } (P + \displaystyle\frac{1}{a}(e^{aE}  - 1))e^{ - iEt + iPx} }  \\ 
  = \int {dPdE\displaystyle\frac{1}{{P - \displaystyle\frac{1}{a}(e^{aE}  - 1)}}} e^{ - iEt + iPx}  \\ 
 \end{array}
\end{equation}
would not be obtained if one chose $F = [ - \infty ,\infty ].$ In that case one would have energy momentum delta functions instead of denominators. This would mean that point particles would not be able to propagate off shell. We discuss the $(E,P)$
form of the Green's function in more detail in section 8.

The additional modular transformation $C \to \displaystyle\frac{{a^4 }}{C}$
means that using the theta function representation (4.78) we can further restrict the fundamental region as
	\begin{equation}
F = \{ C:C \ge a^2 \}  = [a^2 ,\infty ]
\end{equation}
and we have restored the natural units of the parameter $C$. The fact that a point particle theory can have a fundamental region away from the ultraviolet divergence $C = 0$
is somewhat unexpected as such properties are thought to be exclusive properties of string models leading to their finiteness.

\subsection{Duality symmetry and the functional equation}

An important symmetry of string models is the Target duality $R' = \displaystyle\frac{{\alpha '}}{R}$
. Is there an analogy of this symmetry in the Green's function expressions (4.74) or (4.75)? 

From the relation of the Green's function to the Lerch zeta function 
	\begin{equation}
\begin{array}{l}
 2\pi G(x,t;0,i\sigma )|_R  = \displaystyle\frac{1}{R}\sum\limits_{n = 0}^\infty  {(a\displaystyle\frac{n}{R} + 1)^{ - s} e^{i\displaystyle\frac{n}{R}x} }  \\ 
  = \displaystyle\frac{1}{R}(\displaystyle\frac{R}{a})^s \sum\limits_{n = 0}^\infty  {(n + \displaystyle\frac{R}{a}} )^{ - s} e^{i2\pi n\displaystyle\frac{x}{{2\pi R}}}  = \displaystyle\frac{1}{R}(\displaystyle\frac{R}{a})^s \phi (\displaystyle\frac{x}{{2\pi R}},s,\displaystyle\frac{R}{a}) \\ 
 \end{array}
\end{equation}
and the corresponding functional relation (2.12) we have:
	\begin{equation}
\begin{array}{l}
 \sum\limits_n {\displaystyle\frac{1}{{(n + \displaystyle\frac{R}{a})^{1 - s} }}} e^{inx/R}  = (2\pi )^{ - s} \Gamma (s)e^{\pi is/2} e^{i\displaystyle\frac{x}{a}} \sum\limits_n {\displaystyle\frac{1}{{(n + \displaystyle\frac{x}{{2\pi R}})^s }}} e^{ - in\displaystyle\frac{2\pi R}{a}}  \\ 
  + (2\pi )^{ - s} \Gamma (s)e^{ - \pi is/2} e^{i2\pi \displaystyle\frac{R}{a}(1 - \displaystyle\frac{x}{{2\pi R}})} \sum\limits_n {\displaystyle\frac{1}{{(n + 1 - \displaystyle\frac{x}{{2\pi R}})^s }}} e^{in\displaystyle\frac{2 \pi R}{a}}  \\ 
  \\ 
 \end{array}
\end{equation}
These relations can be derived using a theta representation for the Green's function by considering the integral$\int {dC\sum\limits_{n = 0}^\infty  {e^{ - C(n + \displaystyle\frac{R}{a})^2 } } e^{\displaystyle\frac{{2\pi inx}}{{2\pi R}}} C^{s/2 - 1} } $. Now setting:
\begin{equation}
\begin{array}{l}
 \sum\limits_n {\displaystyle\frac{1}{{(n + \displaystyle\frac{{R'}}{a})^{1 - s} }}} e^{ - inx/R'}  = \sum\limits_n {\displaystyle\frac{1}{{(n + \displaystyle\frac{x}{{2\pi R}})^s }}} e^{ - in\displaystyle\frac{{2\pi R}}{a}}  \\ 
 \sum\limits_n {\displaystyle\frac{1}{{(n + \displaystyle\frac{{R''}}{a})^{1 - s} }}} e^{inx/R''}  = \sum\limits_n {\displaystyle\frac{1}{{(n + 1 - \displaystyle\frac{x}{{2\pi R}})^s }}} e^{in\displaystyle\frac{{2\pi R}}{a}}  \\ 
 \end{array}
\end{equation}
We obtain:
	\begin{equation}
\begin{array}{l}
 t' =  - t \\ 
 \sigma ' = \sigma '' = a - \sigma  \\ 
 x' = x \\ 
 R' = \displaystyle\frac{{ax}}{{2\pi }}\displaystyle\frac{1}{R} \\ 
 x'' = 2\pi R - x \\ 
 R'' = a - \displaystyle\frac{{ax}}{{2\pi }}\displaystyle\frac{1}{R} \\ 
 \end{array}
\end{equation}
For the case relevant to the zeta function $x = \pi a$
and 
\begin{equation}
\begin{array}{l}
 x' = \pi a \\ 
 R' = \displaystyle\frac{{a^2 }}{2}\displaystyle\frac{1}{R} \\ 
 x'' = 2\pi R - \pi a \\ 
 R'' = a - \displaystyle\frac{{a^2 }}{2}\displaystyle\frac{1}{R} \\ 
 \end{array}
\end{equation}
If one sets $R = a$
as in the previous section we have $x' = x'' = x = \pi a$
and $R' = R'' = R/2 = a/2$
. If instead one sets $R = \displaystyle\frac{1}{{\sqrt 2 }}a$
then $R = R' = \displaystyle\frac{a}{{\sqrt 2 }}$
, $x = x' = \pi a$
, $x'' = (\sqrt 2  - 1)\pi a$
and $R'' = (1 - \displaystyle\frac{1}{{\sqrt 2 }})a$
. Thus we find that $R = \displaystyle\frac{a}{{\sqrt 2 }}$
is a self dual point with respect to the transformation\begin{equation}
R' = \displaystyle\frac{{a^2 }}{2}\displaystyle\frac{1}{R}
\end{equation}
In string theory the duality relation says that string propagation amplitudes on a circle of radius $R$
can be written in terms of string propagation on $R' = \displaystyle\frac{{\alpha '}}{R}$. 
For the Riemann point particle the duality reflected in the functional equation (2.12) tells us that point particle fermionic propagator from $(x = 0,t = 0)$
to $(x,t - i\sigma  =  - isa)$
on a circle of radius $R$
can be written as a superposition of 
the propagator from $(x = 0,t = 0)$
to $(x' = x,t' - i\sigma ' =  - i(1 - s)a)$
on a circle of radius
 \begin{equation}
R' = \displaystyle\frac{{ax}}{{2\pi }}\displaystyle\frac{1}{R}
\end{equation}
and the propagator from $(x = 0,t = 0)$
to $(x'' = 2\pi R - x,t'' - i\sigma '' =  - i(1 - s)a)$
on a circle of radius 
\begin{equation}
R'' = a - \displaystyle\frac{{ax}}{{2\pi }}\displaystyle\frac{1}{R}
\end{equation}
Thus the radii, position and $\sigma $
values relevant to the Riemann zeta function $R = a,x = \pi a,\sigma  = \displaystyle\frac{a}{2}$
are special values with respect to the duality transformations.
 
\subsection{$\sigma  = \displaystyle\frac{1}{2}$
and path integral zeros}

Consider the integral representation of the zeta function where the integrand can be represented in terms of theta functions:
\begin{equation}
\zeta (s) = \displaystyle\frac{1}{{2^{1 - s}  - 1}}\displaystyle\frac{1}{{2^s  - 1}}\pi ^{\displaystyle\frac{s}{2}} \displaystyle\frac{1}{{\Gamma (\displaystyle\frac{s}{2})}}\int\limits_0^\infty  {\displaystyle\frac{{d\tau }}{\tau }} \tau ^{s/2} (\theta _4 (0|i\tau ) + \theta _2 (0|i\tau ) - \theta _3 (0|i\tau ))
\end{equation}
If one can equate this integral with the result of a path integral of a quantum mechanical system then any statement about Riemann zeros can be translated to a statement about path integral zeros.

Path integrals over gauge fields and fermions can develop zeros in certain situations \cite{Witten:1982fp}. An especially simple example is $0 + 1$
 dimensional gauge theory coupled to a single fermion with path integral \cite{Axenides:1993pn}, \cite{Kikukawa:1997qh}:
	\begin{equation}
Z = \int {DA_0 D\psi e^{\int\limits_0^1 {d\tau (i\psi ^* \partial _0 \psi  + A_0 \psi ^* \psi )} } } 
\end{equation}
The action has the gauge invariance $\psi '(\tau ) = e^{i\theta (\tau )} \psi (\tau ),A_0 '(\tau ) = A_0 (\tau ) + \partial _0 \theta (\tau )$. Note that this $A_0 (\tau )$
should not be confused with $A_1 $
parameter of section 3  as $A_0 (\tau )$
is a worldline gauge field and $A_1 $
is a target space parameter like $R$
. Similarly in this section $\psi (\tau )$
is a worldline fermion and $\psi (x,t)$
is a target space field. Fixing the gauge $A_0 (\tau ) = \alpha _0 $
for the $0 + 1$
dimensional gauge field then a global symmetry of the gauge fixed action is $\alpha _0  \to \alpha _0  + 2\pi $. The path integral after gauge fixing and integration over the fermion reduces to:
	\begin{equation}
Z = \int\limits_{ - \infty }^\infty  {d\alpha _0 \cos (\alpha _0 /2)} 
\end{equation}
The path integral is not invariant $\alpha _0  \to \alpha _0  + 2\pi $
. There is a global anomaly. Indicative of this is the fact that the path integral vanishes. What's more is that the path integral with the insertion of a gauge invariant function vanishes\cite{Witten:1982fp}.
That is :
\begin{equation}
Z = \int\limits_{ - \infty }^\infty  {d\alpha _0 \cos (\alpha _0 /2)f(\alpha _0 )}  = 0
\end{equation}
if $f(\alpha _0  + 2\pi ) = f(\alpha _0 )$
. However the path integral is invariant under $\alpha _0  \to  - \alpha _0 $
which is world line parity. There is a tension between parity and gauge invariance in this model. The gauge invariant form of the path integral is :
	\begin{equation}
\begin{array}{l}
 Z' = \int\limits_0^{2\pi } {d\alpha _0 \displaystyle\frac{1}{2}(e^{i\alpha _0 }  + 1)}  \\ 
  = \int\limits_0^{2\pi } {d\alpha _0 e^{i\alpha _0 /2} \cos (\alpha _0 /2)}  = \int {DA_0 D\psi e^{\int_0^1 {d\tau (\psi ^* (\dot \psi  + iA_0 \psi ) + iA_0 /2} } }  \\ 
 \end{array}
\end{equation}
This gauge invariant expression $Z'$
is invariant under $\alpha _0  \to \alpha _0  + 2\pi $
but not worldline parity. The parity invariant expression $Z$
is invariant under $\alpha _0  \to  - \alpha _0 $
but not gauge invariance.

In our case we have a point particle Green's function which can be defined by the path integral:
	\begin{equation}
G_{4'} (x,t;0,0)|_{R,A'}  = \int\limits_{x(0) = 0,t(0) = 0}^{x(1) = x,t(1) = t} {DpDEDxDtDe_0 D\xi D\bar \xi e^{ - I} } 
\end{equation}
Fixing the world line reprametrization symmetry by $e_0 (\tau ) = C$
we can use the methods of section 8.4 to express the path integral as:
\begin{equation}
\begin{array}{l}
 G_{4'} (x,t;0,0)|_{R,A'}  = \int\limits_0^\infty  {dCC^{s/2 - 1} \sum\limits_{n =  - \infty }^\infty  {e^{ - C(\displaystyle\frac{{n + A'}}{R})^2 } e^{i(\displaystyle\frac{{n + A'}}{R})x} } }  \\ 
\;\;\;\;  + \int\limits_0^\infty  {dCC^{(s + 1)/2 - 1} \sum\limits_{n =  - \infty }^\infty  {(\displaystyle\frac{{n + A'}}{R})e^{ - C(\displaystyle\frac{{n + A'}}{R})^2 } e^{i(\displaystyle\frac{{n + A'}}{R})x} } }  \\ 
 \end{array}
\end{equation}
Forming the superposition:
\begin{eqnarray}
 Z&= G_{4'} (x = 0,t - i\sigma ;0,0)|_{R,A' = 0} \nonumber \\ 
 &- \;G_{4'} (x = 0,t - i\sigma ;0,0)|_{R,A' = \displaystyle\frac{1}{2}}\nonumber  \\ 
&- \;G_{4'} (x = \pi R,t - i\sigma ;0,0)|_{R,A' = 0}  
 \end{eqnarray}
where we have introduced the $\sigma $
parameter through $t \to t - i\sigma $
 as in section 4. In forming this combination the second term in (4.97) drops out and we have the simplified expression:
\begin{eqnarray}
 Z &=& (2^s  - 1)(2^{1 - s}  - 1)\pi ^{ - s/2} \Gamma (s/2)\zeta (s)\nonumber \\ 
  &=& \int\limits_0^\infty  {dCC^{s/2 - 1} (\theta _3 (0|\displaystyle\frac{{iC}}{{R^2 }})}  - \theta _2 (0|\displaystyle\frac{{iC}}{{R^2 }}) - \theta _4 (0|\displaystyle\frac{{iC}}{{R^2 }})) 
 \end{eqnarray}
Introducing a Liouville type variable through $C = a^2 e^\varphi  $
 we have:
\begin{equation}
Z = \int\limits_0^\infty  {d\varphi e^{\varphi (s/2)} (\theta _3 (0|\displaystyle\frac{{a^2 e^\varphi  }}{{R^2 }})}  - \theta _2 (0|\displaystyle\frac{{a^2 e^\varphi  }}{{R^2 }}) - \theta _4 (0|\displaystyle\frac{{a^2 e^\varphi  }}{{R^2 }}))
\end{equation}
To see why we call this a Liouville type variable consider the first term in (4.100) which is of the form:
\begin{eqnarray}
 Z_1  &=& \int\limits_{ - \infty }^\infty  {d\varphi (\cos (t\varphi /2) + i\sin (t\varphi /2))e^{\varphi \sigma /2} \sum\limits_{n =  - \infty }^\infty  {e^{ - e^\varphi  \displaystyle\frac{{a^2 n^2 }}{{R^2 }}} } } \nonumber \\ 
  &=& \int {D\varphi DPDxD\psi e^{\int_0^1 {d\tau (P\dot x + \psi ^* \dot \psi  + \displaystyle\frac{\sigma }{2}\varphi  - e^\varphi  (a^2 P^2 ) + \psi ^* (it\varphi )\psi } } }   
 \end{eqnarray}
In writing this expression as a path integral we have formed a superposition of antiperiodic $(\psi (0) =  - \psi (1))$
and periodic $(\psi (0) = \psi (1))$
fermionic path integrals accounting for the cosine and sine terms above. The quantization of $P$
follows from the periodicity of the $x$
field. The Liouville interpretation follows from the fact that $P^2 $
acts like a cosmological constant in $0 + 1$
dimensions.

For the special case $\sigma  = \displaystyle\frac{1}{2}a$
and using the modular transformation properties of the theta functions $e^{\displaystyle\frac{1}{4}\varphi } (\theta _3 (0|e^\varphi  \displaystyle\frac{{a^2 }}{{R^2 }}) - \theta _2 (0|e^\varphi  \displaystyle\frac{{a^2 }}{{R^2 }}) - \theta _4 (0|e^\varphi  \displaystyle\frac{{a^2 }}{{R^2 }}))$
is an even function of $\varphi $. The integral with the sine term in (4.101) is then identically zero and we have:
\begin{equation}
\begin{array}{l}
 Z = \int\limits_{ - \infty }^\infty  {d\varphi (\cos (t\varphi /2)e^{\displaystyle\frac{1}{4}\varphi } (\theta _3 (0|e^\varphi  \displaystyle\frac{{a^2 }}{{R^2 }}) - \theta _2 (0|e^\varphi  \displaystyle\frac{{a^2 }}{{R^2 }}) - \theta _4 (0|e^\varphi  \displaystyle\frac{{a^2 }}{{R^2 }})) = }  \\ 
  = \int {D\varphi DPDxD\psi e^{\int_0^1 {d\tau (P\dot x + \psi ^* \dot \psi  + \displaystyle\frac{1}{4}\varphi  - e^\varphi  (a^2 P^2 ) + \psi ^* (it\varphi )\psi } } }  \\ 
 \end{array}
\end{equation}
At a zero of the zeta function $s_0  = t_0  + i\displaystyle\frac{1}{2}$
we have a path integral similar to the example of the beginning of this section and we have:
\begin{equation}
\begin{array}{l}
 0=Z(s_0) = (2^s  - 1)(2^{1 - s}  - 1)\pi ^{ - s/2} \Gamma (s/2)\zeta (s)|_{s = it_0 /a + 1/2}  \\ 
     = \int\limits_{ - \infty }^\infty  {d\varphi (\cos (t_0 \varphi /2)e^{\varphi /4} (\theta _3 (0|ie^\varphi  \displaystyle\frac{{a^2 }}{{R^2 }}) - \theta _2 (0|ie^\varphi  \displaystyle\frac{{a^2 }}{{R^2 }}) - \theta _4 (0|ie^\varphi  \displaystyle\frac{{a^2 }}{{R^2 }}))}  \\ 
     = \int\limits_{ - \infty }^\infty  {d\alpha _0 (\cos (\alpha _0 /2)e^{\alpha _0 /4t_0 } (\theta _3 (0|ie^{\alpha _0 /t_0 } \displaystyle\frac{{a^2 }}{{R^2 }}) - \theta _2 (0|ie^{\alpha _0 /t_0 } \displaystyle\frac{{a^2 }}{{R^2 }}) - \theta _4 (0|ie^{\alpha _0 /t_0 } \displaystyle\frac{{a^2 }}{{R^2 }}))}  \\ 
     = \int {DA_0 DPDxD\psi e^{\int_0^1 {d\tau (P\dot x + \psi ^* \dot \psi  + A_0 /4t_0  - e^{A_0 /t_0 } (a^2 P^2 ) + \psi ^* (iA_0 )\psi )} } } \\ 
 \end{array}
\end{equation}
Where we have defined $\alpha _0  = t_0 \varphi $. 

\begin{figure}[htbp]

 \centerline{\hbox{
   \epsfxsize=4.0in
   \epsffile{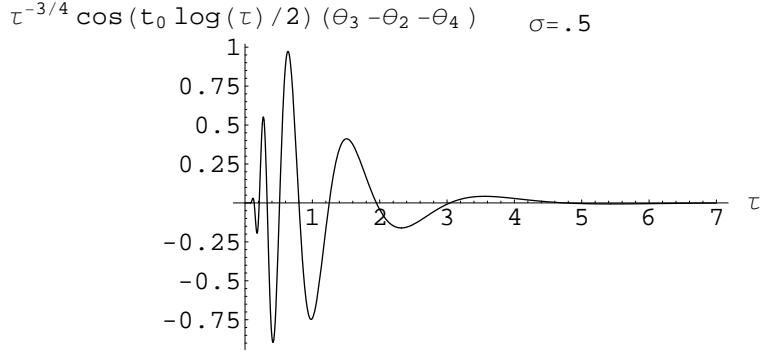}
     }
  }
  \caption{Integrand for the theta representation of the zeta function for the first Rieman zero}
  \label{fig17}

\end{figure}
\begin{figure}[htbp]

 \centerline{\hbox{
   \epsfxsize=3.0in
   \epsffile{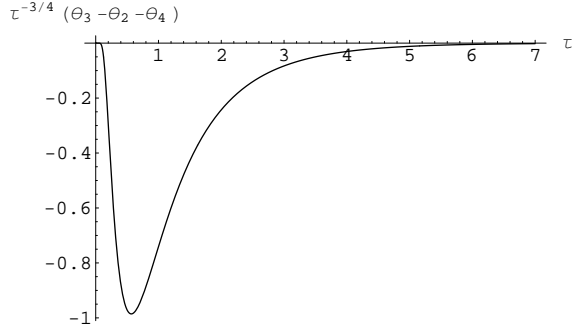}
     }
  }
  \caption{Integrand for the theta representation of the zeta function without the cosine factor.} 
          
  \label{fig18}

\end{figure}

In figure 17 and 18 we plot the real part of the integrand in (4.91) and its product with $\cos (t_0 \varphi /2)$
for the first nontrivial zero of the zeta function. 
Note like the gauge example at the beginning of this section it is the cosine factor which is crucial in obtaining a zero, the theta functions occur whenever the $x$ field is periodic. In terms of the additional modular symmetry $C \to \displaystyle\frac{1}{C}$, the variables $C = e^\varphi  $
transform as $\varphi  \to  - \varphi $. This is worldline parity in the transformed variables $A_0  \to  - A_0 $ . In terms of the original wordline variables $C = e_0 $
the usual modular transformation $C \to  - C$ yields worldline parity $e_0  \to  - e{}_0$
of the point particle einbein. Thus the relation of worldline parity to the usual and additional modular transformations is reversed upon the introduction of the Liouville type variable $\varphi $. 

One can also use the integral representation 
\begin{equation}Z = (1 - 2^{1 - s} )\Gamma (s)\zeta (s) = \int\limits_0^\infty  {dCC^{s - 1} \displaystyle\frac{1}{{e^C  + 1}}}
\end{equation}  
to obtain a path integral representation of the zeta zero condition. In that case:
	\begin{equation}
\begin{array}{l}
 0=Z(s_0) = (1 - 2^{1 - s} )\Gamma (s)\zeta (s) = \int\limits_0^\infty  {dCC^{s - 1} \displaystyle\frac{1}{{e^C  + 1}}}  \\ 
  = \sum\limits_{n = 0}^\infty  {\int\limits_0^\infty  {dCC^{it_0  - 1} C^{1/2} e^{ - C(n + 1) + i\pi n} } }\\
  = i\sum\limits_{n = 0}^\infty  {\int\limits_0^\infty  {dCC^{it_0  - 1} C^{1/2} e^{ - \displaystyle\frac{C}{2} - C(n + \displaystyle\frac{1}{2}) - i\pi (n + \displaystyle\frac{1}{2})} } }  \\ 
  = i\sum\limits_{n = 0}^\infty  {\int\limits_{ - \infty }^\infty  {d\varphi \cos (t_0 \varphi /2)e^{\varphi /4} e^{ - \displaystyle\frac{{e^{\varphi /2} }}{2} - e^{\varphi /2} (n + \displaystyle\frac{1}{2}) - i\pi (n + \displaystyle\frac{1}{2})} } }  \\ 
  = i\int {D\varphi DpDqD\psi } e^{\int_0^1 {(p\dot q + \psi ^* \dot \psi  + \displaystyle\frac{\varphi }{4} - \displaystyle\frac{{e^{\varphi /2} }}{2} - (i\pi  + e^{\varphi /2} )\displaystyle\frac{1}{2}(p^2  + q^2 ) - it_0 \varphi \psi ^* \psi } }  \\ 
  = i\int {DA_0 DpDqD\psi } e^{\int_0^1 {p\dot q + \psi ^* \dot \psi  + A_0 /4t_0  - iA_0 \psi ^* \psi - \displaystyle\frac{{e^{A_0 /2t_0 } }}{2} - (i\pi  + e^{A_0 /2t_0 } )\displaystyle\frac{1}{2}(p^2  + q^2 )  } }\\ 
 \end{array}
\end{equation}
Here we have defined the Liouville type field by $C = e^{\varphi /2}  = e^{A_0 /2t_0 } $. The quantization comes from the harmonic oscillator Hamiltonian $\displaystyle\frac{1}{2}(p^2  + q^2 )$ with eigenvalues $(n + \displaystyle\frac{1}{2})$
instead of from the periodicity condition.

\subsection{$\sigma $
as an extra dimension}

The transformation $s' = 1 - s$
or $\sigma ' = a - \sigma $
could be understood as a reflection or parity symmetry if $\sigma $
could be considered as an extra dimension. Also the spectral properties of the Dirac operator can be investigated from the point of view of extra dimensions especially with respect to its chiral properties. Examples of this include fermionic field theory in the presence of a defect, overlap and domain wall approaches to lattice fermions \cite{Callan:1984sa}, \cite{Kaplan:1992bt}, \cite{Neuberger:2003nu}.

To begin the investigation of $\sigma $
as an extra coordinate we write the Dirac equation equation (3.54) as: 
\begin{equation}
(\sigma ^0 \bar \partial _t  + \sigma ^1 \partial _x )\psi (t,x) = (\sigma ^0 \displaystyle\frac{1}{{ai}}(e^{ia\partial _t }  - 1) + \sigma ^1 \partial _x )\psi (t,x) = 0
\end{equation}
Where we have defined $\bar \partial _t  = \displaystyle\frac{1}{{ia}}(e^{ia\partial _t }  - 1)$. Writing this equation in matrix form we have:
	\begin{equation}
\left( {\begin{array}{*{20}c}
   0 & D  \\
   { - \bar D} & 0  \\
\end{array}} \right)\left( \begin{array}{l}
 \psi _1  \\ 
 \psi _2  \\ 
 \end{array} \right) = \left( {\begin{array}{*{20}c}
   0 & {\bar \partial _t  + \partial _x }  \\
   {\bar \partial _t  - \partial _x } & 0  \\
\end{array}} \right)\left( \begin{array}{l}
 \psi _1  \\ 
 \psi _2  \\ 
 \end{array} \right) = 0
\end{equation}
Now we can modify the above equation by replacing the zero diagonal components by operators $M$
and $\bar M$
so that the equation becomes
\begin{equation}
\left( {\begin{array}{*{20}c}
   {\bar M} & D  \\
   { - \bar D} & M  \\
\end{array}} \right)\left( \begin{array}{l}
 \psi _1  \\ 
 \psi _2  \\ 
 \end{array} \right) = \left( {\begin{array}{*{20}c}
   {\bar M} & {\bar \partial _t  + \partial _x }  \\
   {\bar \partial _t  - \partial _x } & M  \\
\end{array}} \right)\left( \begin{array}{l}
 \psi _1  \\ 
 \psi _2  \\ 
 \end{array} \right) = 0
\end{equation}
The usual choice for $M = mI$
leads to the massive Dirac equation whereas $M = \partial _\sigma   - m{\mathop{\rm sgn}} (\sigma )$
leads to the overlap equation with a mass acting nonuniformly in flavor space \cite{Neuberger:2003nu}. Neither of these choices leads to the zeta function as a Green's function. Instead we choose $\bar M = \bar \partial _\sigma   + i\tilde \partial _t $
and $\bar M = \bar \partial _\sigma   - i\tilde \partial _t $
 where $\tilde \partial _t  = \displaystyle\frac{1}{{ - a}}(e^{ - a\partial _t }  - 1)$
and $\bar \partial _\sigma   = \displaystyle\frac{1}{{ - ai}}(e^{ - ai\partial _\sigma  }  - 1)$. Then we have:
\begin{equation}
\left( {\begin{array}{*{20}c}
   {\bar M} & D  \\
   { - \bar D} & M  \\
\end{array}} \right)\left( \begin{array}{l}
 \psi _1  \\ 
 \psi _2  \\ 
 \end{array} \right)\\
 = \left( {\begin{array}{*{20}c}
   {\bar \partial _\sigma   - i\tilde \partial _t } & {\bar \partial _t  + \partial _x }  \\
   {\bar \partial _t  - \partial _x } & {\bar \partial _\sigma   - i\tilde \partial _t }  \\
\end{array}} \right)\left( \begin{array}{l}
 \psi _1  \\ 
 \psi _2  \\ 
 \end{array} \right) = 0
\end{equation}
After Fourier transforming the above equation using the mode decomposition\\
 $e^{ - iEt + ipx + ik\sigma } $ the above equation becomes 
\begin{equation}
  \left( {\begin{array}{*{20}c}
   {(e^{ak}  - 1) - (e^{iaE}  - 1)} & {(e^{aE}  - 1) + pa}  \\
   {(e^{aE}  - 1) - pa} &{ -(e^{ak}  - 1) - (e^{iaE}  - 1)}  \\
\end{array}} \right)\left( \begin{array}{l}
 \psi _1  \\ 
 \psi _2  \\ 
 \end{array} \right) = 0 \\ 
\end{equation}
Searching for right moving solutions as above we set
\begin{equation}
\left( \begin{array}{l}
 \psi _1  \\ 
 \psi _2  \\ 
 \end{array} \right) = \left( \begin{array}{l}
 \psi _1  \\ 
 0 \\ 
 \end{array} \right)
\end{equation}
 and obtain the equations:
\begin{equation}
\begin{array}{l}
 (\displaystyle\frac{1}{a}(e^{ak}  - 1) - \displaystyle\frac{1}{a}(e^{iaE}  - 1))\psi _1  = 0 \\ 
 (\displaystyle\frac{1}{a}(e^{aE}  - 1) - \displaystyle\frac{n}{R})\psi _1  = 0 \\ 
 \end{array}
\end{equation}
Solving this equation we have $E = \displaystyle\frac{1}{a}\log (a\displaystyle\frac{n}{R} + 1)$
and $k = iE$
so that the mode solutions are of the form 
\begin{equation}\psi _n (x,t,\sigma ) = (\displaystyle\frac{a}{R})^{ - (it + \sigma )} (n + \displaystyle\frac{R}{a})^{ - it/a} (n + \displaystyle\frac{R}{a})^{ - \sigma /a} e^{in\displaystyle\frac{x}{R}} \end{equation}
and the Green's function is given by:
	\begin{equation}
\begin{array}{l}
 G(x,t,\sigma ;x',t',\sigma ') = \sum\limits_n {\psi _n (x,t,\sigma )\psi _n^* (x',t',\sigma ')}  \\ 
  = (\displaystyle\frac{a}{R})^{ - i(t - t') - (\sigma  - \sigma ')} \sum\limits_n {(n + \displaystyle\frac{R}{a}} )^{ - i(t - t')/a - (\sigma  - \sigma ')/a} e^{in\displaystyle\frac{{(x - x')}}{R}}  \\ 
 \end{array}
\end{equation}
which again is proportional to the Lerch zeta function.

In deriving the mode expansion one can also start with the 4D Dirac equation for massless fermion:
	\begin{equation}
(\sigma ^0 \bar \partial _t  + \sigma ^1 \partial _x  + \sigma ^2 \bar \partial _\sigma   + \sigma ^3 \tilde \partial _z )\psi (t,x,\sigma ,z) = 0
\end{equation}
where we have used the Weyl representation
\begin{equation}
\begin{array}{l}
 \gamma ^\mu   = \left( {\begin{array}{*{20}c}
   0 & {\sigma ^\mu  }  \\
   {\bar \sigma ^\mu  } & 0  \\
\end{array}} \right) \\ 
 \psi _\alpha   = \left( \begin{array}{l}
 \psi  \\ 
 0 \\ 
 \end{array} \right) \\ 
 \end{array}
\end{equation}
In component form the corresponding equation is given by:
\begin{equation}
\left( {\begin{array}{*{20}c}
   {\bar M} & D  \\
   { - \bar D} & M  \\
\end{array}} \right)\left( \begin{array}{l}
 \psi _1  \\ 
 \psi _2  \\ 
 \end{array} \right) = \left( {\begin{array}{*{20}c}
   {\bar \partial _\sigma   - i\tilde \partial _z } & {\bar \partial _t  + \partial _x }  \\
   {\bar \partial _t  - \partial _x } & {\bar \partial _\sigma   + i\tilde \partial _z }  \\
\end{array}} \right)\left( \begin{array}{l}
 \psi _1  \\ 
 \psi _2  \\ 
 \end{array} \right) = 0
\end{equation}
If one identifies $z$
coordinate with $t$
we obtain an equation in the form (4.110). 

\subsection{$\sigma $ and Space-time Reflections} 

The transformation $s' = 1 - s,s'' = s^ *  $
,$s''' = (1 - s)^* $
of the Riemann zeta function have interpretations as discrete space-time reflections when $\sigma $
is interpreted as an extra dimension. Using $s = \displaystyle\frac{{it + \sigma }}{a}$
 we have
\begin{equation}\begin{array}{l}
 it' + \sigma ' = it + (a - \sigma ) \\ 
 it'' + \sigma '' =  - it + \sigma  \\ 
 it''' + \sigma ''' =  - it + (a - \sigma ) \\ 
 \end{array}\end{equation}

and thus the time reversal $T$
and parity $P$
transformations are given by :
	\begin{equation}
\begin{array}{l}
 T:        t \to  - t ,\;\;\;\;   \sigma  \to \sigma  \\ 
 PT:     t \to  - t ,\;\;\;\;     \sigma  \to a - \sigma  \\ 
 P:      t \to t ,\;\;\;\;        \sigma  \to a - \sigma  \\ 
 \end{array}
\end{equation}
Note that the parity transformation includes a small translation by $a$
in its definition. This reduces to the usual parity transformation at large distances $\sigma  \gg a$.

The transformation properties of the Riemann zeta function under $s \to s^ *  $
and  $s \to 1 - s$
 have well known applications to the Riemann zeros. If $s$
is a zero than so is $s^ *  $
,$1 - s$
and 
$1 - s^* $
. This means that if a zero in the critical region exists there would actually be four zeros. Thus from the transformations (4.120) for $P,T$
and $PT$
one can restrict the search for zeros to the region $\displaystyle\frac{1}{2} < \displaystyle\frac{\sigma }{a} < 1$
 and $t > 0$. In the interpretation of $\sigma $
as an extra dimension the zeros have the interpretation as locations $(x,t,\sigma )$
that thatfor which the fermion has zero probability amplitude to evolve to, essentially forbidden points of transition. 

\section{Statistical mechanics of the fermionic theory} 

\subsection{Partition function}

It is known that the zeta function behavior at $s = 1$
can be interpreted as a thermal divergence \cite{Spector:1988nn}, \cite{Julia:1993pz}, \cite{Bowick:1990mx}. Also the mixing parameter $\sigma $
parameterizes a Boltzmann distributed initial state so it is interesting to study the Weyl fermion theory at finite temperature to see how this phenomena is represented. In this section we compute the finite temperature partition  function associated with cases (2) and (4) with Hamiltonians
  \begin{equation}
H = \int {dxi\psi \partial _x \psi } 
\end{equation}
and 
\begin{equation}
H = \int {dx} \psi \displaystyle\frac{1}{a}\log (ai\partial _x  + 1)\psi 
\end{equation}
respectively. The partition function is defined by:
	\begin{equation}
Z(\beta ) = tr(e^{ - \beta H_{\rm{2nd}} } )
\end{equation}
with $H_{\rm{2nd}} $
the second quantized Hamiltonian. Because we use the second quantized Hamiltonian we will compute the grand canonical partition function which contains arbitrary numbers of particles. Nevertheless we shall show that it is the one particle partition function $z_{\rm{one}} (\beta )$
which controls the behavior of the partition function near the thermal divergence.
The general formula we shall need is that for a noninteracting fermionic field theory the logarithm of the partition function is given by:
	\begin{equation}
\log Z = \sum\limits_{n = 0}^\infty  {\log (1 + e^{ - \beta (\varepsilon _n  - \mu )} )} 
\end{equation}
where $\beta $
is the inverse temperature and $\mu $
is the chemical potential.
	
\subsubsection*{Case (2) $\epsilon_n=2\pi n/L$ }

Again we begin by reviewing the well studied case of a free right moving fermion on a cylinder of circumference $L = 2\pi R$
with Hamiltonian
	\begin{equation}
H = \int {dxi\psi \partial _x \psi } 
\end{equation}
In terms of Fourier modes it can be written:
	\begin{equation}
H = \sum\limits_{n = 1}^\infty  {\varepsilon _n b_{ - n} } b_n  = \displaystyle\frac{{2\pi }}{L}\sum\limits_{n = 1}^\infty  {nb_{ - n} } b_n 
\end{equation}
So that:
	\begin{equation}
Z(\beta ) = \prod\limits_{n = 1}^\infty  {(1 + e^{2\pi n\displaystyle\frac{\beta }{L}} } )
\end{equation}
If we had studied a scalar particle the partition function would be:
\begin{equation}
Z_{{\rm{scalar}}} (\beta ) = \prod\limits_{n = 1}^\infty  {(1 - e^{2\pi n\displaystyle\frac{\beta }{L}} } )^{ - 1} 
\end{equation}
In appendix A we discuss the treatment of the scalar particle associated with the nonstandard dispersion relation (3.31).

For the fermionic theory the number of states at energy $E$
is given by the inverse Laplace transform:
	\begin{equation}
\sigma (E) = \int\limits_{ - i\infty }^{i\infty } {d\beta } e^{\beta E} Z(\beta )
\end{equation}
As $E = \displaystyle\frac{1}{R}\sum\limits_{n = 1}^\infty  {nN_n } $
then $\sigma (E)$
gives the number of ways of writing $E$
as a sum of positive integers irrespective of order with no repeats. It is this property of the partition function that plays a role in additive number theory where one seeks to enumerate the ways of writing a positive integer as a sum.

\subsubsection*{Case (4) $a\epsilon_n=\log(1+a2\pi n/L)$}

In this case we have $H = \int {dx\bar \psi } \displaystyle\frac{1}{a}\log (ia\partial _x  + 1)\psi $
with periodic boundary conditions $\psi (x + 2\pi R) = \psi (x)$. Fourier transforming $\psi (x) = \sum\limits_{n = 0}^\infty  {b_n e^{ - i\varepsilon _n t + in\displaystyle\frac{x}{R}} } $
and using the dispersion relation for the one particle energies $\varepsilon _n  = \displaystyle\frac{1}{a}\log (a\displaystyle\frac{n}{R} + 1)$ we have:
\begin{equation}
H = \sum\limits_{n = 1}^\infty  {\displaystyle\frac{1}{a}\log (a\displaystyle\frac{{2\pi n}}{L}n}  + 1)b_{ - n} b_n 
\end{equation}
Then from the formula for Fermi-Dirac partition functions we obtain:
	\begin{equation}
Z(\beta ) = \prod\limits_{n = 1}^\infty  {(1 + e^{ - \beta \varepsilon _n } )}  = \prod\limits_{n = 1}^\infty  {(1 + (a\displaystyle\frac{n}{R} + 1)^{ - \beta /a} )} 
\end{equation}
A general result on infinite products states that an infinite product such as\\ $\prod\limits_{n = 0}^\infty  {(1 + a_n )} $
diverges whenever $\sum\limits_{n = 0}^\infty  {a_n } $
diverges \cite{Andrews}. Applying this to our case we have\\  $\prod\limits_{n = 0}^\infty  {(1 + e^{ - \beta \varepsilon _n } } )$
diverges whenever the one particle partition function $z_{\rm{\rm{one}}} (\beta ) = \sum\limits_{n = 0}^\infty  {e^{ - \beta \varepsilon _n } } $
diverges. Thus the grand partition function diverges whenever the one particle partition function diverges. For the dispersion relation (3.31) the one particle partition function is 
\begin{equation}z_{\rm{one}} (\beta ) = (\displaystyle\frac{R}{a})^{\beta /a} \sum\limits_{n = 0}^\infty  {\displaystyle\frac{1}{{(n + \displaystyle\frac{R}{a})^{\beta /a} }}}  = (\displaystyle\frac{R}{a})^{\beta /a} \zeta _H (\displaystyle\frac{\beta }{a},\displaystyle\frac{R}{a})\end{equation}
which has a simple pole divergence only at $\beta  = a$. Thus we see that the grand partition function of the nonstandard fermion also has a divergence at $\beta  = a.$ Studying the Yang-Lee zeros of the one particle partition function yields another physical representation of the Riemann hypothesis. We discuss the one particle partition function in Appendix B. 

Now consider the fermionic partition function associated with the special case $R = a$
. The formula for the total energy is then $E_{\rm{total}}  = \displaystyle\frac{1}{a}\sum\limits_{n = 2}^\infty  {(\log n)N_n } $ and the partition function becomes:
\begin{equation}Z(\beta ,L = 2\pi a) = \prod\limits_{n = 2}^\infty  {(1 + n^{ - \beta /a} } )\end{equation}
From the definition of the partition function the number of states at a given energy is given by:
\begin{equation}
\sigma (E) = \int\limits_{ - i\infty }^{i\infty } {d\beta e^{\beta E} Z(\beta )} 
\end{equation}
The usual situation is that the number of states grows with the energy. However because of the definition of energy $\sigma _{\rm{mult}} (E)$
gives the number of factors of $E$
without regards to order and with no repeats. For example $E = \log(120)a^{ - 1} $
the states without regards to order with no repeats are of the form $\prod\limits_{N_n  = 0,1;n = 2, \ldots ,exp(aE) }^{} {b_{ - n}^{N_n } } \left| 0 \right\rangle $ and can be listed as:
\begin{equation}
\begin{array}{l}
 120 = 5! =  \\ 
 5 \cdot 4 \cdot 3 \cdot 2;b_{ - 5} b_{ - 4} b_{ - 3} b_{ - 2} |\left. 0 \right\rangle  \\ 
 20 \cdot 3 \cdot 2;b_{ - 20} b_{ - 3} b_{ - 2} |\left. 0 \right\rangle  \\ 
 15 \cdot 4 \cdot 2;b_{ - 15} b_{ - 4} b_{ - 2} |\left. 0 \right\rangle  \\ 
 12 \cdot 5 \cdot 2;b_{ - 12} b_{ - 5} b_{ - 2} |\left. 0 \right\rangle  \\ 
 10 \cdot 4 \cdot 3;b_{ - 10} b_{ - 4} b_{ - 3} |\left. 0 \right\rangle  \\ 
 8 \cdot 5 \cdot 3;b_{ - 8} b_{ - 5} b_{ - 3} |\left. 0 \right\rangle  \\ 
 6 \cdot 5 \cdot 4;b_{ - 6} b_{ - 5} b_{ - 4} |\left. 0 \right\rangle  \\ 
 60 \cdot 2;b_{ - 60} b_{ - 2} |\left. 0 \right\rangle  \\ 
 40 \cdot 3;b_{ - 40} b_{ - 3} |\left. 0 \right\rangle  \\ 
 30 \cdot 4;b_{ - 30} b_{ - 4} |\left. 0 \right\rangle  \\ 
 24 \cdot 5;b_{ - 24} b_{ - 5} |\left. 0 \right\rangle  \\ 
 20 \cdot 6;b_{ - 20} b_{ - 6} |\left. 0 \right\rangle  \\ 
15 \cdot 8;b_{ - 15} b_{ - 8} |\left. 0 \right\rangle  \\ 
12 \cdot 10;b_{ - 12} b_{ - 10} |\left. 0 \right\rangle  \\ 
 120;b_{ - 120} |\left. 0 \right\rangle  \\ 
 \end{array}
\end{equation}
There $\sigma (\displaystyle\frac{1}{a}\log (120)) = 15$
 states without regard to order with no repeats. These consist of 1 four-particle state, 6 three-particle states 7 two-particle states and 1 one-particle state.
A dramatic effect occurs when $E$
is the logarithm of a prime number in units of a. Then there is only one state and $\sigma _{mult} (E) = 1$. This behavior leads to an inhomogeneous density of states and a potential  loss of equilibrium. 

Returning to the general case of arbitrary $R = L/2\pi $
the partition function can be written:
\begin{equation}
 Z(\beta ,L) = \prod\limits_{n = 1}^\infty  {(1 + (a\displaystyle\frac{{2\pi n}}{L} + 1)^{ - \beta /a} } )
 \end{equation}
For $L \gg a$
we obtain the additive partition function (5.7). For $L = 2\pi a$
we obtain the multiplicative partition function with $\sigma _{mult} (E=\log(\rm{Prime})) = 1$ . For the additive density of states the number of states grows rapidly with energy $\sigma _{\rm{add}} (E) \sim e^{2\sqrt {\pi LE/24} }  = e^{\pi \sqrt {RE/3} } $. The general expression  $\sigma (E,L)$
interpolates between these two extremes.

\subsection{Series representation of the free energy}

From the representation (5.4) specialized to $\varepsilon _n  = \displaystyle\frac{1}{a}\log (a\displaystyle\frac{n}{R} + 1)$
we have:
	\begin{equation}
\log Z = \sum\limits_{n = 0}^\infty  {\log (1 + (\displaystyle\frac{n}{R}}  + 1)^{ - \beta /a} e^{\beta \mu } )
\end{equation}
Now expanding the logarithm in the sum using $\log (1 + y) =  - \sum\limits_{m = 1}^\infty  {\displaystyle\frac{{( - y)^m }}{m}} $
we obtain:
	\begin{equation}
\begin{array}{l}
 \log Z = \sum\limits_{m = 1}^\infty  {\sum\limits_{n = 0}^\infty  { - ( -1 )^m \displaystyle\frac{1}{m}(a\displaystyle\frac{n}{R} + 1)^{ - \beta m/a} e^{\beta \mu m} )} }  \\ 
  = \sum\limits_{m = 1}^\infty  { - ( -1 )^m \displaystyle\frac{1}{m}(\displaystyle\frac{R}{a})^{m\beta /a} e^{\beta \mu m} \zeta _H (\displaystyle\frac{{\beta m}}{a},\displaystyle\frac{R}{a})}  \\ 
 \end{array}
\end{equation}
For the special case of $R = a$
and $\mu  = 0$
this becomes:
	\begin{equation}
\log Z(R = a,\mu  = 0) = \sum\limits_{m = 1}^\infty  { - ( -1 )^m \displaystyle\frac{1}{m}} \zeta (\displaystyle\frac{{\beta m}}{a})
\end{equation}
In this form we see that the divergence at $\beta  = a$
is contained in the first term of the series.

\subsection{Thermodynamics quantities}

Various thermodynamic quantities can be obtained from the partition function. For example the free energy $f$
, average energy $U$
, average pressure $P$
and average occupation number $N$
are given by:
\begin{eqnarray}
 f(\beta ,L)& =& \displaystyle\frac{1}{\beta }\log Z(\beta ) = \displaystyle\frac{1}{\beta }\sum\limits_{n = 1}^\infty  {\log (1 + (a\displaystyle\frac{n}{R} + 1)^{ - \displaystyle\frac{\beta }{a}} } ) \sim \displaystyle\frac{1}{\beta }\sum {n^{ - \displaystyle\frac{\beta }{a}} } \nonumber \\ 
 U &=&  - \displaystyle\frac{\partial }{{\partial \beta }}\log Z = \displaystyle\frac{1}{a}\sum\limits_{n = 1}^\infty  {\displaystyle\frac{{\log (a\displaystyle\frac{n}{R} + 1)}}{{1 + (a\displaystyle\frac{n}{R} + 1)^{\displaystyle\frac{\beta }{a}} }}}  \sim \displaystyle\frac{1}{a}\sum {n^{ - \displaystyle\frac{\beta }{a}} \log n} \nonumber \\ 
 P& =& \displaystyle\frac{\partial }{{\partial V}}U = \displaystyle\frac{1}{{2\pi }}\displaystyle\frac{\partial }{{\partial R}}U\nonumber\\
 &=& \displaystyle\frac{1}{{2\pi }}\displaystyle\frac{1}{{R^2 }}\sum\limits_{n = 1}^\infty  {\displaystyle\frac{n}{{(a\displaystyle\frac{n}{R} + 1)}}\displaystyle\frac{1}{{1 + (a\displaystyle\frac{n}{R} + 1)^{\displaystyle\frac{\beta }{a}} }}} \left( {\displaystyle\frac{{\log (a\displaystyle\frac{n}{R} + 1)}}{{1 + (a\displaystyle\frac{n}{R} + 1)^{ - \displaystyle\frac{\beta }{a}} }}\displaystyle\frac{\beta }{a} - 1} \right)\nonumber \\ 
              & \sim& \displaystyle\frac{1}{{2\pi }}\displaystyle\frac{1}{{R^2 }}\sum\limits_n^\infty  {n^{ - \beta /a} \log n}  \nonumber\\ 
 N& =&  - \displaystyle\frac{\partial }{{\partial \alpha }}\sum\limits_{n = 0}^\infty  {\log (1 + (a\displaystyle\frac{n}{R} + 1)^{ - \beta /a} e^{ - \alpha } ) = \sum\limits_{n = 0}^\infty  {\displaystyle\frac{1}{{e^\alpha   + (a\displaystyle\frac{n}{R} + 1)^{\displaystyle\frac{\beta }{a}} }}} }\nonumber \\
& \sim& \sum {n^{ - \beta /a} } 
 \end{eqnarray}
All these quantities diverge at the critical inverse temperature  $\beta _c  = a$.

\subsection{Thermal Green's function}

The thermal Green's function for a fermionic field theory is defined by:
	\begin{equation}
G_\beta  (x,t;0,0) = tr(e^{ - \beta H} \psi (x,t)\bar \psi (0)) = \sum\limits_{m =  - \infty }^\infty  {( - 1)^m G(x,t + i\beta m;0,0)} 
\end{equation}
where $G(x,t;0,0)$
is the zero temperature fermionic Green's function. From the definition we see that the thermal Green's function is antiperiodic in imaginary time with period $2\pi \beta $
. Now using the formula for the zero temperature Green's function associated with case (4):
\begin{eqnarray}
 G(x,t;0,0)& =& \sum\limits_{n = 0}^\infty  {\displaystyle\frac{1}{{(a\displaystyle\frac{n}{R} + 1)^{it/a} }}} e^{i\displaystyle\frac{n}{R}x} \nonumber \\ 
 & =& (\displaystyle\frac{R}{a})^{it/a} \sum\limits_{n = 0}^\infty  {\displaystyle\frac{1}{{(n + \displaystyle\frac{R}{a})^{it/a} }}} e^{i\displaystyle\frac{n}{R}x} \nonumber\\ 
&=& (\displaystyle\frac{R}{a})^{it/a} \phi (\displaystyle\frac{x}{{2\pi R}},\displaystyle\frac{{it}}{a},\displaystyle\frac{R}{a}) 
 \end{eqnarray}
the thermal Green's function is given by:
\begin{equation}
G_\beta  (x,t;0,0) = \sum\limits_{m =  - \infty }^\infty  {( - 1)^m (\displaystyle\frac{R}{a})^{ - m\beta /a + it/a} \phi (\displaystyle\frac{x}{{2\pi R}},\displaystyle\frac{{it}}{a} - \displaystyle\frac{{m\beta }}{a},\displaystyle\frac{R}{a})} 
\end{equation}
For the special case relevant to the Riemann zeta function we have $R = a,x = 0,t = 0$
so that
	\begin{equation}
G_\beta  (0,0;0,0) = \sum\limits_{m =  - \infty }^\infty  {( - 1)^m \zeta (m\displaystyle\frac{\beta }{a})} 
\end{equation}
As the Riemann zeta function $\zeta (z)$
has a pole at $z = 1$
we see that the terms of the sum for the thermal Green's function develop a divergence whenever $\beta  = \displaystyle\frac{a}{m}$
 and temperature $kT_c  = \displaystyle\frac{m}{a}$.

The crucial difference between the thermal Green's function and the mixing Green's function used to describe the Riemann hypothesis is that a trace is taken in (5.21). It is simple matter to introduce the mixing parameter $\sigma $
into the thermal Green's function and one  has:
\begin{equation}
G_\beta  (x,t;0,i\sigma ) = \sum\limits_{m =  - \infty }^\infty  {( - 1)^m (\displaystyle\frac{R}{a})^{(it - m\beta )/a} \phi (\displaystyle\frac{x}{{2\pi R}},\displaystyle\frac{{i(t + im\beta  - i\sigma )}}{a},\displaystyle\frac{R}{a})} 
\end{equation}
Again specializing to the case relevant tot the Riemann zeta function $R = a,x = 0,t = 0$
we obtain:
\begin{equation}
G_\beta  (0,0;0,i\sigma) = \sum\limits_{m =  - \infty }^\infty  {( - 1)^m \zeta (m\displaystyle\frac{\beta }{a} + \displaystyle\frac{\sigma }{a})} 
\end{equation}
In the presence of the mixing parameter $\sigma $
the terms of the sum develop a thermal divergence at $\beta  = \displaystyle\frac{{a - \sigma }}{m}$
 thus the temperature of the thermal divergence is modified to $kT_c  = \displaystyle\frac{m}{{a - \sigma }}$.

\section{Harmonic oscillator representation of the zeta function, 2D fermionic string theory and Matrix models.}

\subsection{Logarithmic oscillator and fermionic string theory}

In the previous sections we obtained the zeta function as a Green's function from a fermion theory compactified on a circle $R = a$
 and described by the actions:
	\begin{equation}
I = \int {dtdx(i\psi ^* } \partial _t \psi  + \psi ^* \displaystyle\frac{1}{a}\log (ia\partial _x  + 1)\psi )
\end{equation}
or
\begin{equation}
I = \int {dtdx(\psi ^* \displaystyle\frac{1}{a}(} e^{ai\partial _t }  - 1)\partial _t \psi  + \psi ^* i\partial _x \psi )
\end{equation}
The sum over integers in the mode sum for zeta arose because of the compactified momentum condition $p = \displaystyle\frac{n}{R}$
and the nonstandard energy momentum dispersion relation
$E = \displaystyle\frac{1}{a}\log (ap + 1)$
or $p = \displaystyle\frac{1}{a}(e^{aE}  - 1)$. 

Another method of introducing the integer $n$ into the mode expansion is to use the quantization of a nonstandard oscillator with energy 
\begin{equation}E = \displaystyle\frac{1}{a}\log (a\displaystyle\frac{1}{2}(p^2  + \omega ^2 \lambda ^2 ) + 1)\end{equation}
Then the Schrodinger equation for this system is:
	\begin{equation}
i\partial _t \psi  + \displaystyle\frac{1}{a}\log (a\displaystyle\frac{1}{2}( - \partial _\lambda ^2  + \omega ^2 \lambda ^2 ) + 1)\psi  = 0
\end{equation}
with energy quantization condition $E = \displaystyle\frac{1}{a}\log (a\omega (n + \displaystyle\frac{1}{2}) + 1) = \displaystyle\frac{1}{a}\log (a\omega n + \displaystyle\frac{1}{2}a\omega  + 1)$. The relation with the Lerch zeta function follows immediately upon forming the one particle partition function
\begin{equation}\begin{array}{l}
 z_{\log-{\rm{osc}}} (\beta ,\mu ) = \sum\limits_{n = 0} {e^{ - \beta E_n  - n\mu } }  = \sum\limits_{n = 0} {e^{ - \beta \displaystyle\frac{1}{a}\log (a\omega n + \displaystyle\frac{1}{2}a\omega  + 1) - n\mu } }  \\ 
                  = (a\omega )^{ - \displaystyle\frac{\beta }{a}} \sum\limits_{n = 0} {\displaystyle\frac{1}{{(n + \displaystyle\frac{1}{{a\omega }} + \displaystyle\frac{1}{2})^{\beta /a} }}} e^{ - \mu n}  = (a\omega )^{ - \displaystyle\frac{\beta }{a}} \phi (i\mu ,\displaystyle\frac{\beta }{a},\displaystyle\frac{1}{{a\omega }} + \displaystyle\frac{1}{2}) \\ 
 \end{array}\end{equation}
Here we have introduced a chemical potential associated with oscillator number. $z_{\log-\rm{osc}} (\beta ,\mu )$
is proportional to the Riemann zeta function for $a\omega  = 2$
and $\mu  = 0$
and the Dirchlet eta function for $a\omega  = 2$
and $\mu  = i\pi $. The condition $a\omega  = 2$
takes the place of the condition $R = a$
of the previous section. The Riemann hypothesis can be formulated in by saying that the one particle partition function only has zeros for ${\mathop{\rm Im}\nolimits} \beta  = \displaystyle\frac{a}{2}$
when $\mu  = i\pi $
 or that there are no phase transitions in the form of Yang-Lee zeros in the nonstandard quantum oscillator for $\displaystyle\frac{1}{2}a < {\mathop{\rm Im}\nolimits} \beta  < \displaystyle\frac{1}{a}$.

The dynamics of the classical logarithmic nonstandard oscillator are given by the equations where $(q = \lambda )$:
\begin{equation}
\begin{array}{l}
 {\rm{Velocity: }}\;\;                  \dot q = \displaystyle\frac{{\partial E}}{{\partial p}} = \displaystyle\frac{p}{m}\displaystyle\frac{1}{{a(\displaystyle\frac{{p^2 }}{{2m}} + \displaystyle\frac{1}{2}m\omega ^2 q^2 ) + 1}} = \displaystyle\frac{p}{m}e^{ - aE}  \\ 
 {\rm{Force: }}\;\;                     \dot p =  - \displaystyle\frac{{\partial E}}{{\partial q}} =  - m\omega ^2 q\displaystyle\frac{1}{{a(\displaystyle\frac{{p^2 }}{{2m}} + \displaystyle\frac{1}{2}m\omega ^2 q^2 ) + 1}} =  - m\omega ^2 qe^{ - aE}  \\ 
 {\rm{Position: }}\;\;                 q(t) = q_0 \cos (\omega te^{ - aE} ) + \displaystyle\frac{{p_0 }}{{m\omega }}\sin (\omega te^{ - aE} ) \\ 
 {\rm{Period: }}\;\;                   T = \displaystyle\frac{{2\pi }}{\omega }e^{aE}  \\ 
 {\rm{Dispersion:}}\;\; e^{aE}  = a(\displaystyle\frac{{p_0^2 }}{{2m}} + \displaystyle\frac{{m\omega ^2 }}{2}q_0^2 ) + 1 \\ 
 \end{array}
\end{equation}
For initial momentum $p_0  = 0$
we see that the period \begin{equation}
T = \displaystyle\frac{{2\pi }}{\omega }e^{aE}  = \displaystyle\frac{{2\pi }}{\omega }(a\displaystyle\frac{{m\omega ^2 }}{2}q_0^2  + 1)
\end{equation}
So unlike the standard oscillator the period is amplitude dependent.

Returning to the Schrodinger equation (6.4) we see that this equation follows from the two dimensional field theory:
\begin{equation}I = \int {dtd\lambda (i\psi ^* \partial _t } \psi  + \psi ^* \displaystyle\frac{1}{a}\log (a\displaystyle\frac{1}{2}( - \partial _\lambda ^2  + \omega ^2 \lambda ^2 ) + 1)\psi \end{equation}
In the limit of $a\omega  \to 0$
this reduces to the fermionic formulation of the c=1 matrix model \cite{Moore:1991sf}:

\begin{equation}I = \int {dtd\lambda (i\psi ^* \partial _t } \psi  + \psi ^* \displaystyle\frac{1}{2}( - \partial _\lambda ^2  + \displaystyle\frac{1}{{\alpha '}}\lambda ^2 )\psi )\end{equation}
for $\displaystyle\frac{1}{{\alpha '}} = \omega ^2 $. In the c=1 matrix model one can go from the right side up to inverted harmonic oscillator potential by sending $\alpha ' \to  - \alpha '$ \cite{Berenstein:2004kk}. When one does so one changes the background for the model from as well as the vacuum energy. One also replaces the Hermite polynomials by parabolic cylinder functions and the physical description is quite different in the two cases.

The Green's function associated to the action is given by:
	\begin{equation}
G(t,\lambda ;t',\lambda ') = \int {dpe^{ipt} \int {dse^{ - sp + i\mu s} \left\langle {\lambda _1 |e^{ish} \left| {\lambda _2 } \right\rangle } \right.} } 
\end{equation}
with
\begin{equation}
\left\langle {\lambda _1 |e^{i\tau H} \left| {\lambda _2 } \right\rangle } \right. = \sum\limits_n {e^{ - iE_n \tau } H_n (\lambda )H_n (\lambda ') = } \sum\limits_n {(a\omega (n + \displaystyle\frac{1}{2}) + 1)^{ - i\tau /a} H_n (\lambda )H_n (\lambda ')} 
\end{equation}
Unlike the case (4) of the previous section the Green's function does not yield the zeta function. However the one particle partition function defined by $z_{\rm{osc}} (\beta ) = \int {d\lambda G(\lambda ,\lambda ,\beta )} $
does yield the zeta function as we saw above.

One of the many fascinating aspects of the usual c=1 fermionic theory is that it is quadratic in fields yet reproduces the genus expansion for 2D interacting c=1 noncritical string theory through the formula \cite{Gross:1990st}:
\begin{equation}\begin{array}{l}
 \partial _\mu  \rho  = \displaystyle\frac{1}{\pi }\displaystyle\frac{\partial }{{\partial \mu }}{\mathop{\rm Im}\nolimits} \int {d\lambda } \left\langle {\lambda |} \right.\displaystyle\frac{1}{{h_0  + \mu }}|\left. \lambda  \right\rangle  = \displaystyle\frac{1}{{2\pi }}{\mathop{\rm Im}\nolimits} \int\limits_0^\infty  {d\tau e^{ - i\mu \tau } \tau \int\limits_{ - \infty }^\infty  {d\lambda G(\lambda ,\lambda ;\tau )} }  =  \\ 
  = \displaystyle\frac{1}{{2\pi }}{\mathop{\rm Im}\nolimits} \int\limits_0^\infty  {d\tau e^{ - i\mu \tau } \tau z_{\rm{osc}} (\tau )}  = \displaystyle\frac{1}{{2\pi }}{\mathop{\rm Im}\nolimits} \int\limits_0^\infty  {d\tau e^{ - i\mu \tau } \tau \displaystyle\frac{1}{{\sinh (\omega \tau /2)}}}  \\ 
  = \partial _\mu  (\displaystyle\frac{1}{{2\pi }}{\mathop{\rm Re}\nolimits} (\sum\limits_n {\displaystyle\frac{1}{{(n + \displaystyle\frac{1}{2})\omega  + i\mu }}} ) = \displaystyle\frac{1}{\omega }\partial _\mu  (\displaystyle\frac{1}{{2\pi }}{\mathop{\rm Re}\nolimits} \Psi (\displaystyle\frac{1}{2} + i\displaystyle\frac{\mu }{\omega })) \\ 
 \end{array}\end{equation}
The genus expansion results from expanding the above expression in powers of $\mu ^{2g - 2} $
where $g$
is the genus. 

Here we study the modification to the above formula resulting from using the logarithmic fermionic oscillator theory (6.8) . From the structure of (6.12) we see that we have:
\begin{equation}\begin{array}{l}
 \partial _\mu  \rho  = \displaystyle\frac{1}{{2\pi }}{\mathop{\rm Im}\nolimits} \int\limits_0^\infty  {d\tau e^{ - i\mu \tau } \tau \int\limits_{ - \infty }^\infty  {d\lambda G(\lambda ,\lambda ;\tau )} }  =  \\ 
  = \displaystyle\frac{1}{{2\pi }}{\mathop{\rm Im}\nolimits} \int\limits_0^\infty  {d\tau e^{ - i\mu \tau } \tau z_{\log  -\rm{ osc}} (a,\tau )}  = \displaystyle\frac{1}{{2\pi }}{\mathop{\rm Im}\nolimits} \int\limits_0^\infty  {d\tau e^{ - i\mu \tau } \tau (a\omega )^{ - \tau /a} \zeta _H (\displaystyle\frac{\tau }{a},\displaystyle\frac{1}{{a\omega }} + \displaystyle\frac{1}{2})}  \\ 
  = \partial _\mu  (\displaystyle\frac{1}{{2\pi }}{\mathop{\rm Re}\nolimits} (\sum\limits_n {\displaystyle\frac{1}{{\displaystyle\frac{1}{a}\log (a(n + \displaystyle\frac{1}{2})\omega  + 1) + i\mu }}} )\\ 
 \end{array}\end{equation}
Also for the special value $a\omega  = 2$
 the integrand becomes:
\begin{equation}\begin{array}{l}
 \partial _\mu  \rho (\mu ) = \sum\limits_m {\displaystyle\frac{1}{{2\pi }}{\mathop{\rm Im}\nolimits} \int\limits_0^\infty  {d\tau e^{ - i\mu \tau } \tau (a\omega )^{ - \tau /a} \zeta (\displaystyle\frac{\tau }{a})} }  \\ 
  = \sum\limits_m {\displaystyle\frac{1}{{2\pi }}{\mathop{\rm Im}\nolimits} \int\limits_0^1 {d\tau e^{ - i\mu \tau } \tau (a\omega )^{ - \tau /a} \zeta (\displaystyle\frac{\tau }{a})} }\\
  + \sum\limits_m {\displaystyle\frac{1}{{2\pi }}{\mathop{\rm Im}\nolimits} \int\limits_1^\infty  {d\tau e^{ - i\mu \tau } \tau (a\omega )^{ - \tau /a} \prod\limits_{p\varepsilon primes} {(1 - p^{ - \tau /a} )^{ - 1} } } }  \\ 
 \end{array}\end{equation}
where the first integral is over the critical strip and second integrand involves the Euler product over prime numbers. It would be interesting to investigate if this occurrence of prime numbers has a connection with the adelic 2D string theory studied in \cite{Freund:1991md}.

Another manifestation of the $c=1$ 2D string model is the compactification of the $t$
direction with radius $R$
. In the usual case we have \cite{Klebanov:1991qa}:
\begin{equation}I = \int\limits_0^{2\pi R} {dt\int {d\lambda (i\psi ^* \partial _t } \psi  + \psi ^* \displaystyle\frac{1}{2}( - \partial _\lambda ^2  + \omega ^2 \lambda ^2 )\psi )} \end{equation}
with the result:
\begin{equation}\partial _\mu  \rho (\mu ,\omega ,R) = \partial _\mu  (\displaystyle\frac{1}{{2\pi }}{\mathop{\rm Re}\nolimits} (\sum\limits_{mn} {\displaystyle\frac{1}{{(n + \displaystyle\frac{1}{2})\omega  + 1 + (m + \displaystyle\frac{1}{2})\displaystyle\frac{1}{R} + i\mu }}} )\end{equation}
We compute the modifications the take place with the logarithmic oscillator action (6.8)
\begin{equation}I = \int\limits_0^{2\pi R} {dt\int {d\lambda (i\psi ^* \partial _t } \psi  + \psi ^* \displaystyle\frac{1}{a}\log (a\displaystyle\frac{1}{2}( - \partial _\lambda ^2  + \omega ^2 \lambda ^2 ) + 1)\psi )}\end{equation}
As in the usual case the propagator is antiperiodic \\
 $G_R (\lambda ,\lambda ';\tau ) = \sum\limits_n {( - 1)^n G(\lambda ,\lambda ';\tau  + 2\pi Rn)} $ and we have
\begin{equation}\begin{array}{l}
 \partial _\mu  \rho  = \displaystyle\frac{1}{{2\pi }}{\mathop{\rm Im}\nolimits} \int\limits_0^\infty  {d\tau e^{ - i\mu \tau } \tau \int\limits_{ - \infty }^\infty  {d\lambda G_R (\lambda ,\lambda ;\tau )} }  \\ 
  = \sum\limits_m {\displaystyle\frac{1}{{2\pi }}{\mathop{\rm Im}\nolimits} \int\limits_0^\infty  {d\tau e^{ - i\mu \tau } \tau z_{\log-\rm{osc}} (a,\tau  + 2\pi mR)} } ( - 1)^m  \\ 
  = \sum\limits_m {\displaystyle\frac{1}{{2\pi }}{\mathop{\rm Im}\nolimits} \int\limits_0^\infty  {d\tau e^{ - i\mu \tau } \displaystyle\frac{{\tau /2R}}{{\sinh (\tau /2R)}}\tau (a\omega )^{ - \tau /a} \zeta _H (\displaystyle\frac{\tau }{a},\displaystyle\frac{1}{{a\omega }} + \displaystyle\frac{1}{2})} }  \\ 
  = \partial _\mu  (\displaystyle\frac{1}{{2\pi }}{\mathop{\rm Re}\nolimits} (\sum\limits_{mn} {\displaystyle\frac{1}{{\displaystyle\frac{1}{a}\log (a(n + \displaystyle\frac{1}{2})\omega  + 1) + (m + \displaystyle\frac{1}{2})\displaystyle\frac{1}{R} + i\mu }}} ) \\ 
 \end{array}\end{equation}
In computing the above we have used the fact that the finite $R$
dependence can be implemented by inserting the factor $\displaystyle\frac{{\tau /2R}}{{\sinh (\tau /2R)}}$
into the integrand as in \cite{Klebanov:1991qa}. This expression agrees with the usual vacuum energy  (6.16) in the limit $a \to 0$
. Note that the duality symmetry $R \to \displaystyle\frac{{\alpha '}}{R} = \displaystyle\frac{1}{{\omega ^2 R}}$
of (6.16) is violated by the $a$
dependence in this Harmonic oscillator representation. Previously we found  a generalization of the duality symmetry using the Green's function representation in section 4.

 \subsection{Matrix Model}

It is well known that the fermionic action (6.9) has an interpretation as a matrix model \cite{Klebanov:1991qa}. Does the nonstandard  fermionic action (6.8) also lead to a random Matrix interpretation? If we write the matrix $M(t)$
in terms of it is eigenvalues
\begin{equation}M(t) = \left( {\begin{array}{*{20}c}
   {\lambda _1 (t)} &  \ldots  & 0  \\
    \vdots  &  \ddots  &  \vdots   \\
   0 &  \cdots  & {\lambda _N (t)}  \\
\end{array}} \right)\end{equation}
The fermionic action is associated with the Matrix integral
\begin{equation}
Z = \int {DMD\Pi e^{\int {dt(\rm{tr}(\dot M(t)\Pi (t) - \displaystyle\frac{1}{{\sqrt {\alpha '} }}\rm{tr}\log \left( {\sqrt {\alpha '} (\Pi (t)^2  + \displaystyle\frac{1}{{\alpha '^2 }}M(t))^2+1)} \right)} } } 
\end{equation}
where $\Pi (t)$
is a conjugate auxiliary field. In the eigenvalue basis the integral becomes:
\begin{equation}
Z = \int {D\lambda Dp\Delta (\lambda )e^{\int {dt(\dot \lambda p - \displaystyle\frac{1}{a}\log (a\displaystyle\frac{{p^2  + \omega ^2 \lambda ^2 }}{2} + 1)} } } 
\end{equation}
with Vandermonde determinant $\Delta (\lambda ) = \prod\limits_{t < t'} {(\lambda (t) - \lambda (t'))} $
 arises from the measure of hermitean matrices . It is the antisymmetry of the Vandermond determinant which connects the matrix integral with the fermionic theory \cite{Moore:1991sf}.

The eigenvalue action is somewhat more complicated than is usually considered for Matrix models. After eliminating the auxiliary field $p$
we find the eigenvalue action:
\begin{equation}S = \int {dt} (Q(\lambda ,\dot \lambda ) - V(\lambda ))\end{equation}
with
\begin{equation}\begin{array}{l}
 Q(\lambda ,\dot \lambda ) = \dot \lambda ^2 \displaystyle\frac{2}{{1 + \sqrt {1 - 2a\dot \lambda ^2 (1 + \displaystyle\frac{{a\omega ^2 \lambda ^2 }}{2})} }}(1 + \displaystyle\frac{{a\omega ^2 \lambda ^2 }}{2})\\
  \;\;\;\;\;\; - \displaystyle\frac{1}{a}\log (\displaystyle\frac{2}{{1 + \sqrt {1 - 2a\dot \lambda ^2 (1 + \displaystyle\frac{{a\omega ^2 \lambda ^2 }}{2})} }}) \\ 
 V(\lambda ) = \displaystyle\frac{1}{a}\log (1 + \displaystyle\frac{{a\omega ^2 \lambda ^2 }}{2}) \\ 
 \end{array}\end{equation}
This reduces to the usual matrix eigenvalue action $\int {dt(} \displaystyle\frac{{\dot \lambda ^2 }}{2} - \omega ^2 \displaystyle\frac{{\lambda ^2 }}{2})$
in the limit of $a\omega  \to 0$
. In figure 20 we plot the potential $V$
for the Matrix model. One can add an inverted oscillator potential to this matrix action:
\begin{equation}V(\lambda ) =  - \displaystyle\frac{{\omega '^2 }}{2}\lambda ^2  + \displaystyle\frac{1}{a}\log (1 + a\displaystyle\frac{{\omega ^2 }}{2}\lambda ^2 )\end{equation}
This form of the potential has been shown to survive large N scaling and has been used to describe a $D_0 $
brane contribution to type $0B$
string models in two dimensions \cite{Takayanagi:2004jz} . Shown in figure 19 and 20 is the potential energy used in that model. Taken with the analysis above  Dirichlet series may play a fundamental role in the short distance dynamics of string models.

\begin{figure}[htbp]

\centerline{\hbox{
   \epsfxsize=2.0in
   \epsffile{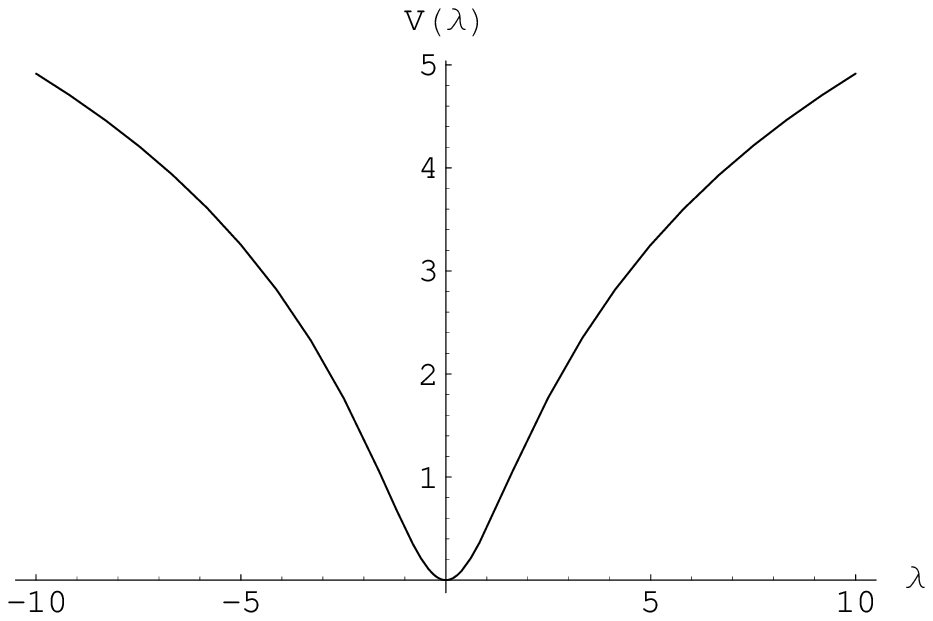}
     }
  }
  \caption{Logarithmic harmonic oscillator potential $V(\lambda ) = \displaystyle\frac{1}{a}\log (1 + \displaystyle\frac{{a\omega ^2 \lambda ^2 }}{2}).$} 
          
  \label{fig19}

\end{figure}

\begin{figure}[htbp]

  \centerline{\hbox{
   \epsfxsize=2.0in
   \epsffile{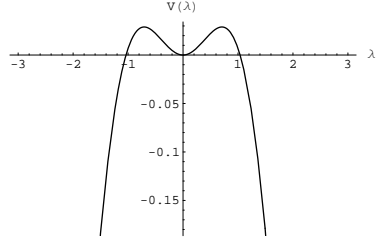}
     }
  }
  \caption{Sum of inverted and logarithmic potential. The same potential was used to describe $0B$ string theory in two dimensions in [37]} 
          
  \label{fig20}

\end{figure}

Setting $\beta  = N\varepsilon $
one can write the path integral representation for the ordinary oscillator as \cite{Creutz:1980gp}:
	\begin{equation}
\begin{array}{l}
 Z(\beta ) = \int {\prod\limits_{n = 1}^\infty   \ldots  dq_n } dp_n e^{ - \varepsilon (p_n^2  + \displaystyle\frac{{\omega ^2 }}{2}q_n^2 ) - ip_n q_n  + ip_n q_{n + 1} } dq_{n - 1}  \ldots   \\ 
\;\;\;= \int {\prod\limits_{n = 1}^\infty   \ldots  dq_n } dp_n \left\langle {p_{n - 1} |\left. {q_n } \right\rangle } \right.\left\langle {q_n |T|\left. {p_n } \right\rangle } \right.\left\langle {p_n |\left. {q_{n + 1} } \right\rangle } \right. \ldots   \\ 
\;\;\;= \int {\prod\limits_{n = 1}^\infty   \ldots  dq_n } \left\langle {q_{n - 1} |T|\left. {q_n } \right\rangle } \right.\left\langle {q_n |T} \right.|\left. {q_{n + 1} } \right\rangle  \ldots  = Tr(T^N ) \\ 
 \end{array}
\end{equation}
where the transfer operator in the mixed basis is given by: 
\begin{equation}
\left\langle {q|T|} \right.\left. p \right\rangle  = e^{ - ipq - \varepsilon \displaystyle\frac{1}{2}(p^2  + \omega ^2 q^2 )} 
\end{equation}
The partition function for the ordinary oscillator is:
\begin{equation}z_{{\rm{osc}}} (\beta ) = \sum\limits_{n = 0}^\infty  {e^{ - \beta (n + 1/2)\omega } }  = e^{ - \displaystyle\frac{1}{2}\beta \omega } \displaystyle\frac{1}{{1 - e^{ - \beta \omega } }} = \displaystyle\frac{1}{{2\sinh (\beta \omega /2)}}\end{equation}
For small $\beta $
this is $z_{\rm{osc}} (\beta ) = \displaystyle\frac{1}{{\beta \omega }}$.

For the logarithmic oscillator one can proceed in a similar way as:
\begin{equation}
\begin{array}{l}
 Z(\beta ) = \int {\prod\limits_{n = 1}^\infty   \ldots  dq_n } dp_n e^{ - \varepsilon \log (a(p_n^2  + \displaystyle\frac{{\omega ^2 }}{2}q_n^2 ) + 1) - ip_n q_n  + ip_n q_{n + 1} } dq_{n - 1}  \ldots  \\ 
\;\;= \int {\prod\limits_{n = 1}^\infty   \ldots  dq_n } dp_n \left\langle {p_{n - 1} |\left. {q_n } \right\rangle } \right.\left\langle {q_n |T|\left. {p_n } \right\rangle } \right.\left\langle {p_n |\left. {q_{n + 1} } \right\rangle } \right. \ldots  \\ 
\;\;= \int {\prod\limits_{n = 1}^\infty   \ldots  dq_n } \left\langle {q_{n - 1} |T|\left. {q_n } \right\rangle } \right.\left\langle {q_n |T} \right.|\left. {q_{n + 1} } \right\rangle  \ldots  = Tr(T^N ) \\ 
 \end{array}
\end{equation}
So that in the mixed basis the transfer operator is:

\begin{equation}
\left\langle {q|T|} \right.\left. p \right\rangle  = e^{ - ipq - \varepsilon \displaystyle\frac{1}{a}\log (a\displaystyle\frac{1}{2}(p^2  + \omega ^2 q^2 ) + 1)}  = e^{ - ipq} (a\displaystyle\frac{1}{2}(p^2  + \omega ^2 q^2 ) + 1)^{ - \varepsilon /a} 
\end{equation}
The partition function for the logarithmic oscillator is then:
\begin{equation}z_{{\rm{log - osc}}} (\beta ) = \sum\limits_{n = 0}^\infty  {(a(n + 1/2)\omega  + 1)^{ - \beta /a}  = (a\omega )^{ - \beta /a} \zeta _H (\displaystyle\frac{\beta }{a},\displaystyle\frac{1}{2}}  + \displaystyle\frac{1}{{a\omega }})\end{equation}
Near $\beta _c  = a$
this is $z_{\log  -\rm{ osc}} (\beta ) = (a\omega )^{ - \beta /a} \displaystyle\frac{a}{{\beta  - a}}$.

\subsection{Zeta function in the Topological Matrix Model}
Another matrix model related to 2d string theory is the Penner or topological matrix model used to compute the Euler characteristic of the moduli space of Riemann surfaces \cite{Imbimbo:1995yv}, \cite{Imbimbo:1995ns}. The partition function in this case is given by the single matrix integral:
	\begin{equation}
Z(A) = \int {dMe^{ - \nu tr(MA) + (\nu  - N)tr\log (M)} }  = \int {dM(\det M)^{\nu  - N} e^{ - \nu tr(MA)} } 
\end{equation}
This matrix integral is a matrix analog of the Gamma function. As the contour representation of the Gamma function is so similar to the contour representation of the zeta function it is worthwhile investigating whether there is a Matrix representation of the zeta function as well. Indeed one can introduce the matrix integral:
\begin{equation}
Z'(A) = \int {dM} (\det (M))^{\nu  - N} \displaystyle\frac{1}{{e^{\nu trMA}  - 1}} = \sum\limits_{m = 1} {\int {dM} (\det (M))^{\nu  - N} e^{ - m\nu trMA} } \end{equation}
Now if we scale $M$
by $m^{ - 1} $
and take into account the scale transformation properties of the measure together with the Vandermonde determinant we have:
\begin{equation}
\begin{array}{l}
Z'(A) = \sum\limits_{m = 1} {\displaystyle\frac{1}{{m^{\nu N - N^2  + N^2  + \displaystyle\frac{{N(N - 1)}}{2}} }}} \int {dM} (\det (M))^{\nu  - N} e^{ - m\nu trMA}\\
\;\;\;\; = \zeta (\nu N + \displaystyle\frac{{N(N - 1)}}{2})Z(A)
\end{array}
\end{equation}
Thus we can express the zeta function as a Matrix integral through
	\begin{equation}
\zeta (s) = \displaystyle\frac{1}{{Z(A)}}\int {dM} (\det (M))^{\nu  - N} \displaystyle\frac{1}{{e^{\nu trMA}  - 1}}
\end{equation}
where $s = \nu N + \displaystyle\frac{{N(N - 1)}}{2}$.

\subsection{$\displaystyle\frac{\sigma }{a} = \displaystyle\frac{1}{2}$
 inverted oscillator and Riemann zeros}

Recall the relation between the inverted oscillator and it's density of states \cite{Hyun:2005fq}.
The Hamiltonian of the inverted oscillator is:
	\begin{equation}
H = \displaystyle\frac{1}{2}(p^2  - \omega ^2 q^2 ) =  - \displaystyle\frac{1}{2}(x_ +  x_ -   + x_ -  x_ +  )
\end{equation}
where $x_ \pm   = (\omega q \pm p)/\sqrt 2 $
. A simple set of wave functions are given by \cite{Hyun:2005fq}:
	\begin{equation}
\psi _E (x_ +  ) = \int\limits_0^\infty  {dk(k/\omega )^{ - iE/\omega  - \sigma /a} e^{ikx_ +  } }  = (\displaystyle\frac{{x_ +  }}{i})^{iE/\omega  + \sigma /a - 1} \Gamma (1 - \displaystyle\frac{\sigma }{a} - i\displaystyle\frac{E}{\omega })
\end{equation}
These are energy eigenstates for $\displaystyle\frac{\sigma }{a} = \displaystyle\frac{1}{2}$
so that:
\begin{equation}
\psi _E (x_ +  ) = (\displaystyle\frac{{x_ +  }}{i})^{iE/\omega  - \displaystyle\frac{1}{2}} \Gamma (\displaystyle\frac{1}{2} - i\displaystyle\frac{E}{\omega })
\end{equation}
Placing a zero condition at $q = \Lambda ^{ - 1} $
 enforces a quantization condition:
	\begin{equation}
\vartheta (E,x_ +  ) = {\mathop{\rm Im}\nolimits} \log (\psi _E (x_ +  )) = (n + \displaystyle\frac{1}{2})\pi 
\end{equation}
The density of states is given by $\rho (E) = \displaystyle\frac{{dn}}{{dE}} = \displaystyle\frac{1}{{2\pi }}(\displaystyle\frac{{d\vartheta }}{{dE}} - \log \Lambda )$. 

Here we note the similarity between (6.36) and case $3''$ defined by the dispersion relation $E = \displaystyle\frac{1}{a}\log (p/a)$
with wave function:
	\begin{equation}
\psi _t (x) = \int\limits_0^\infty  {dp(pa)^{ - it/a - \sigma /a} e^{ipx} }  = (\displaystyle\frac{x}{{ai}})^{it/a + \sigma /a - 1} \Gamma (1 - it/a + \sigma /a)
\end{equation}
The correspondence is:
	\begin{equation}
\begin{array}{l}
 \sigma /a \to 1/2 \\ 
 t/a \to E/\omega  \\ 
 x \to x{}_ +  \\ 
 \end{array}
\end{equation}
In section 8 we show that one can go from expressions like (6.39) to the zeta function by the process of periodization. To see how this works in this case form:
\begin{equation}
\psi _t (x)|_R  = \sum\limits_{m =  - \infty }^\infty  {(\displaystyle\frac{{x - 2\pi Rm}}{{ai}})^{it/a + \sigma /a - 1} \Gamma (1 - it/a + \sigma /a)} 
\end{equation}
Using the definition of the Hurwitz zeta function this can be expressed as:
\begin{equation}
\begin{array}{l}
 \psi _t (x)|_R  = (\displaystyle\frac{{2\pi R}}{a})^{s - 1} (\zeta _H (1 - s,\displaystyle\frac{x}{{2\pi R}}) \\ 
  + ( - 1)^{s - 1} \zeta _H (1 - s,1 - \displaystyle\frac{x}{{2\pi R}}))i^{1 - s} \Gamma (1 - s) \\ 
 \end{array}
\end{equation}
with $s = (it + \sigma )/a$
. If we choose $x = \pi R$
and $R = a$
 and use $(2^{1 - s}  - 1)\zeta (1 - s) = \zeta _H (1-s,1/2)$
we have:
\begin{equation}
\psi _t (x = \pi a)|_{R = a}  = (2\pi )^{s - 1} \zeta (1 - s)(1 + ( - 1)^{s - 1} )(2^{1 - s}  - 1)i^{1 - s} \Gamma (1 - s)
\end{equation}
Now using the functional equation:$(2\pi )^{s - 1} \Gamma (1 - s)\zeta (1 - s) = \displaystyle\frac{{\zeta (s)}}{{2\sin (s\pi /2)}}$
 this becomes:
\begin{equation}
\begin{array}{l}
 \psi _t (x = \pi a)|_{R = a}  =  - \zeta (s)(i^s  - i^{ - s} )^{ - 1} (1 + ( - 1)^{s - 1} )(2^{1 - s}  - 1)i^{ - s}  \\ 
  =  - \zeta (s)(( - 1)^s  - 1)^{ - 1} (1 - ( - 1)^s )(2^{1 - s}  - 1) = \zeta (s)(2^{1 - s}  - 1) =  - \eta (s) \\ 
 \end{array}
\end{equation}

The formula that relates the phase of the zeta function at $\sigma /a = 1/2$
to the number of zeros on the positive  $t$
axis to a value of $T$ is
given by \cite{Edwards}:

	\begin{equation}
\begin{array}{l}
 N(T) = \displaystyle\frac{1}{\pi }\vartheta (T) + 1 + \displaystyle\frac{1}{\pi }{\mathop{\rm Im}\nolimits} \int_C {(\log \zeta (s))'ds}  \\ 
  = \displaystyle\frac{1}{\pi }\vartheta (T) + 1 + \displaystyle\frac{1}{\pi }{\mathop{\rm Im}\nolimits} (\log \zeta (\displaystyle\frac{1}{2} + iT)) \\ 
 \end{array}
\end{equation}
where $\vartheta (t) = {\mathop{\rm Im}\nolimits} (\log \Gamma (\displaystyle\frac{{it}}{2} + \displaystyle\frac{1}{4})) - \displaystyle\frac{t}{2}\pi $
and the contour $C$
is the boundary of the rectangle defined by$ - \varepsilon  \le {\mathop{\rm Re}\nolimits} s \le 1 + \varepsilon ,0 \le {\mathop{\rm Im}\nolimits} s \le T.$ The formula is not modified by the $2^{1 - s}  - 1$
factor in (6.44) as both $\zeta (s)$
and $\eta (s)$
have the same number of zeros in the critical strip. Formula (6.45) means that the process of periodization and the correspondence (6.40) to the quantum inverted oscillator leads to a condition on the number of zeros in place of the quantization condition (6.38). If we define:
	\begin{equation}
\psi _E (x_ +   = \pi ) = \Gamma (i\displaystyle\frac{E}{2} + \displaystyle\frac{1}{4})i^{ - E} \zeta (\displaystyle\frac{1}{2} + iE)
\end{equation}
Then the zero condition after periodization becomes:
	\begin{equation}
\pi N(E) = {\mathop{\rm Im}\nolimits} (\log (\psi _E (x_ +   = \pi ))
\end{equation}
Here $N(E)$
is the number of zeros in the function  $f(\epsilon)=\psi _{\epsilon} (x_ +   = \pi )$
up to energy $E$ which corresponds with the Riemann zeros. It is somewhat surprising that periodization through (6.41) leads to such a relation as the usual derivation of (6.38) uses the asymptotic form of the wave function at $\infty $
and periodicity does not permit such a region.

Various other permutations of the dynamical variables can be considered besides $t/a \to E/\omega $
and these are considered in appendix A. Often the physical picture changes including the introduction of a nontrivial background field due to the modification of the equations of motion under the permutation. For example, if instead of $x \to x_ +  $
one chooses $t/a \to E/\omega $
and $x \to p$
then the relation $t = \displaystyle\frac{a}{2}(xp + px)$
becomes $E = \displaystyle\frac{\omega }{2}(xp + px)$. This is the  Hamiltonian of Berry and Keating \cite{Berry} as well as the dilatation operator. If one uses the original variables $t = \displaystyle\frac{a}{2}(xp + px)$ and $H = \displaystyle\frac{1}{a}\log (ap)$ then $t$
has the intepretaion of the time of arrival operator. Usually one has difficulty in defining a time  operator in quantum mechanics because time takes its values on the real line whereas energies are positive. It is for this reason that the momentum-position uncertainty relations and energy-time uncertaity relations have very different derivations. However in this case because of the the logarithmic dispersion relation the energy takes its values on the real line the treatment of time as a quantum operator is less problematical.

\section{Ground rings, algebraic curves, modular forms and Dirchlet L-series}

There is a remarkable connection between the inverted oscillator with $t$
periodic with period $\beta _c  = \sqrt {\alpha '}  = \displaystyle\frac{1}{\omega }$
and topological string theories on non compact Calabi-Yau manifolds \cite{Ghoshal:1995wm} which are determined by algebraic equations. If this connection generalizes to the above process of periodization of $x_+$
with period $R_c  = a$
it would represent a bridge between the zeta function, 2d gravity (quantum surfaces) and the algebraic geometry of complex manifolds. We investigate some of these issues in  this section. 

\subsection{Elliptic curves}
\indent Projective elliptic curves $E$
are defined as the points $(X,Y)$
satisfying the equation \cite{Milne}:
	\begin{equation}
Y^2 Z = X^3  + aXZ^2  + bZ^3 
\end{equation}
The affine version of the elliptic curve is obtained by setting $Z = 1$. Elliptic curves are characterized by two quantities. The discriminant which is given by:
	\begin{equation}
\Delta  =  - 16(4a^3  + 27b^2 )
\end{equation}
and the $j$
 invariant which is defined by:
	\begin{equation}
j = \displaystyle\frac{{1728 \cdot 4a^3 }}{\Delta }
\end{equation}
When there is a term on the right hand side of the elliptic equation of the form $cX^2 Z$
these formulas are modified by $(a \to a - c^2 /3)$
and $(b \to b + \displaystyle\frac{2}{{27}}c^3  - \displaystyle\frac{1}{3}ac)$
. The discriminant may also be written as $\Delta  = 16(r_1  - r_2 )^2(r_2  - r_3 )^2(r_1  - r_3 )^2$
where $r_1 ,r_2 $
and $r_3 $
are the three roots of the right hand side of the cubic equation.

The most famous elliptic curve doesn't actually exist \cite{Ribet}. Consider the curve:
	\begin{equation}
Y^2 Z = X^3  - (S + T)X^2 Z + (ST)XZ^2 
\end{equation}
If $S = r^n $ and $T = k^n $ the discrimant of the curve is 
$\Delta  = 16S^2T^2(S-T)^2=16r^{2n} (k^n  - r^n )^2 k^{2n} $. If $k^n  - r^n  = i^n $ for some integer $i$
then there is no modular form associated with elliptic curve in contradiction with the Taniyama-Shimura theorem \cite{Shimura}. Thus $i^n  + r^n  \ne k^n $
which is Fermat's theorem. An algebraic equation of the form (7.4) is called a Frey curve.

Recall the Boltzmann probabilities $P_k  = \displaystyle\frac{{e^{ - \sigma E_k } }}{{z(\sigma )}} = (k^{ - \sigma /a} )\displaystyle\frac{1}{{\zeta (\sigma /a)}}$
 considered in section 4. If $\displaystyle\frac{\sigma }{a}$
is an integer $n > 2$
then  besides obeying  the usual identities  $P_k  < 1,$ $P_i  + P_r  < 1,$ $P_i P_j  < 1$ and $\sum\limits_{r = 1}^\infty  {P_r }  = 1$
 the probabilities also obey the inequality $\displaystyle\frac{1}{{P_k }} \ne \displaystyle\frac{1}{{P_i }} + \displaystyle\frac{1}{{P_r }}$
because of Fermat $i^n  + r^n  \ne k^n $. Rewritting this inequality as \begin{equation}
2P_k (P_i  + P_r ) \ne 2P_i P_r 
\end{equation}
This expression has the interpretation in terms of logical operations as $P(k \wedge (i \vee r)) \ne P(i \wedge r)$. Thus the probability of choosing a pair of numbers $a$
and $b$ from a Boltzman distribution of the form above is never equal to the probability of choosing a number $c$ together with ($a$ or $b$) . To illustrate this consider the simple situation of three objects $a$, $b$ and $c$ with probabilities all equal to $\displaystyle\frac{1}{3}$. The 9 possible ways of choosing the two objects at random from $\{ a,b,c\} $
are
\begin{equation}
\left( {\begin{array}{*{20}c}
   {aa} & {ab} & {ac}  \\
   {ba} & {bb} & {bc}  \\
   {cc} & {cb} & {cc}  \\
\end{array}} \right)
\end{equation}
 As we are assuming uniform probabilities in this simple example there are two ways of choosing $a$
and $b$
namely $(ab,ba)$
so the probability $P(a \wedge b)$
is $\displaystyle\frac{2}{9} = 2\displaystyle\frac{1}{3} \cdot \displaystyle\frac{1}{3}$
. There are four ways choosing $c$
and ($a$
or $b$
) namely \begin{equation}
(ac,ca,bc,cb)
\end{equation}
so the probability $P(c \wedge (a \vee b))$
is $\displaystyle\frac{4}{9} = 2\displaystyle\frac{1}{3} \cdot (\displaystyle\frac{1}{3} + \displaystyle\frac{1}{3})$
thus the inequality is obeyed in this case.

Returning to the probabilities $P_k  = \displaystyle\frac{{e^{ - \sigma E_k } }}{z} = e^{ - \displaystyle\frac{\sigma }{a}\log (a\displaystyle\frac{k}{R} + A)} \displaystyle\frac{1}{z} = k^{ - n} \displaystyle\frac{1}{{\zeta (n)}}$
we see from the discussion of section 3 that the fact that the $k$
is an integer follows from quantum mechanics and the fact that wave function obey $\psi (x + 2\pi R) = \psi (x)$
on a periodic space, so that momentum is quantized $p = \displaystyle\frac{k}{R}$
. The special nature of $\displaystyle\frac{R}{a} = 1$
and $\displaystyle\frac{\sigma }{a} = n$
follow from the duality symmetry (4.88) and the thermal partition function (5.11). For algebraic curves this means that if one considers an elliptic curve over a set of operators (like $e^{ - \sigma H} $
considered above) which are quantized, than coefficients of the curve which would normally takes values in the complex numbers $C$ or reals $R$ become curves over the rationals $Q$. It is elliptic curves over $Q$ which are in one to one correspondence with modular forms and Dirichlet L-series which in turn have a  form of the Riemann hypothesis \cite{Milne} associated to them.

\subsection{Ground rings}
\indent In the Liouville theory of noncritical $(c = 1)$
strings ground rings are defined from the operators \cite{Douglas:2003up}:
	\begin{equation}
\begin{array}{l}
 x \equiv O_{\displaystyle\frac{1}{2},\displaystyle\frac{1}{2}} (z) = (cb + i\partial X^ -  )e^{iX^ +  }  \\ 
 y \equiv O_{\displaystyle\frac{1}{2}, - \displaystyle\frac{1}{2}} (z) = (cb - i\partial X^ +  )e^{ - iX^ -  }  \\ 
 \end{array}
\end{equation}
Where $X^ \pm   = \displaystyle\frac{1}{{\sqrt 2 }}(X \pm i\phi )$
an $c$
and $b$
are ghost fields. Defining the operators $X = x\bar x,  Y = y\bar y,  S = x\bar y,  T = y\bar x$ yields four generators of the ground ring with one relation between them:
	\begin{equation}
XY - ST = 0
\end{equation}
It is the relation (7.9) that corresponds to an algebraic surface, in this case the quadratic equation describing the conifold.

In the matrix formulation of noncritical $c = 1$
string theory the ground ring is defined by the operators \cite{Douglas:2003up}:
	\begin{equation}
\begin{array}{l}
 O_{1,2}  = (q\omega  + p)e^{ - t\omega }  \\ 
 O_{2,1}  = (q\omega  - p)e^{t\omega }  \\ 
 \end{array}
\end{equation}
With one relation between them $O_{1,2} O_{2,1}  = \omega ^2 q^2  - p^2  =  - 2H$
where $H$
is the Hamiltonian of the inverted oscillator.

For $c = 1$
noncritical string theory defined at $A - D - E$
points at special radii $R = n/(\omega \sqrt 2 )$
 much more complicated algebraic equations are derived. For example for the $D_2 $
 $(n = 2)$
string theory the chiral ground ring operators where determined in \cite{Ghoshal:1992kx} and are given by:
	\begin{equation}
T = \displaystyle\frac{1}{2}(x^4  + y^4 ),\;\;Y = \displaystyle\frac{1}{2}xy(x^4  - y^4 ),\;\;X = (xy)^2 
\end{equation}
There is one relation between the ground ring operators which is:
	\begin{equation}
Y^2 Z  = X^3  - (T^2 )XZ^2 
\end{equation}
Here we interpret $T$
as a Neron parameter associated to the elliptic curve $E$
, $Z$
is a parameter of the projective curve,. Again one can set $Z = 1$
to obtain the affine curve.

The discriminant and $j$
invariant of this curve are
\begin{equation}\begin{array}{l}
 \Delta  =  - 16(4a^3  + 27b^2 ) =  - 16(4( - T^6 ) = 64T^6  \\ 
 j = \displaystyle\frac{{1728 \cdot 16a^3 }}{\Delta } = 1728 \\ 
 \end{array}\end{equation}

As another example the non chiral $D_2 $
ground ring of $c = 1$
 string theory at radius $R = \sqrt {2\alpha '} $ which
 was also computed in \cite{Ghoshal:1992kx}. The ground ring is given by the operators:
\begin{equation}
\begin{array}{l}
 X = x\bar xy\bar y \\ 
 Y = \displaystyle\frac{1}{4}(x\bar x - y\bar y)((x\bar y)^2  - (y\bar x)^2 ) \\ 
 T = \displaystyle\frac{1}{2}((x\bar y)^2  + (y\bar x)^2 ) \\ 
 S = \displaystyle\frac{1}{2}(x\bar x + y\bar y) \\ 
 \end{array}
\end{equation}
Again there is one relation between these operators and it is given by:
	\begin{equation}
Y^2 Z = X^3  - X(T^2  + S^2 )Z^2  + T^2 S^2 Z^3 
\end{equation}
Again we shall interpret this relation as an elliptic curve with Neron parameters $S$
and $T$
. The discriminant and $j$
 invariant for the curve are determined to be:
	\begin{equation}
\begin{array}{l}
 \Delta  =  - 16( - 4(S^2  + T^2 )^3  + 27T^4 S^4 ) \\ 
 j =  - 16( - 4 \cdot 1728(S^2  + T^2 )^3 /\Delta  \\ 
 \end{array}
\end{equation}
These examples suffice to illustrate the close relationship between the ground rings of $(c = 1)$
string theory and algebraic curves.

\subsection{Modular forms and elliptic curves}

\indent Consider the ordinary fermionic propagator in two Euclidean dimensions given by $g_1 (z) = 2\pi \left\langle {0|\psi (z)\psi (0)|\left. 0 \right\rangle } \right. = \displaystyle\frac{1}{z}$
. This function satisfies the simple relation $(\partial ^2 g_1 )^2  = 4(\partial g_1 )^3 $
 or 
\begin{equation}
(2\pi \partial _z T_{zz} )^2  = 4(2\pi T_{zz} )^3 
\end{equation}
with $2\pi T_{zz}  = \partial g_1  = \displaystyle\frac{1}{{z^2 }}$. On a two torus the relation is more complicated and is given by:
	\begin{equation}
(2\pi \partial _z T_{zz} )^2  = 4(2\pi T_{zz} )^3  - 60G_4 (\Lambda )2\pi T_{zz}  - 120G_6 (\Lambda )
\end{equation}
This equation is in one to one correspondence with the elliptic equation \cite{Milne}
	\begin{equation}
Y^2 Z = 4X^3  - G_4 (\Lambda )XZ^2  - G_6 (\Lambda )Z^3 
\end{equation}
Setting $Z = 4$ we obtain an elliptic equation with $a =  - 4G_4 (\Lambda )$
and $b =  - 16G_6 (\Lambda )$. The correspondence is through $(X,Y) \leftrightarrow (T_{zz} ,\partial _z T_{zz} )$
. On the two torus the stress energy tensor can be expressed in terms of the Weierstrass elliptic functions through
	\begin{equation}
2\pi T_{zz} (z) = \displaystyle\frac{1}{{z^2 }} + \sum\limits_{\omega \varepsilon \Lambda } {(\displaystyle\frac{1}{{(z - \omega )^2 }}}  - \displaystyle\frac{1}{{\omega ^2 }})
\end{equation}
with the lattice functions 
	\begin{equation}
\begin{array}{l}
 G_4 (\Lambda ) = 60\sum\limits_{\omega  \in \Lambda } {\displaystyle\frac{1}{{\omega ^4 }}}  \\ 
 G_6 (\Lambda ) = 140\sum\limits_{\omega  \in \Lambda } {\displaystyle\frac{1}{{\omega ^6 }}}  \\ 
 \end{array}
\end{equation}
Using the modular parameter of the torus the lattice vectors are of the form $\omega  = 2m\omega _1  + 2n\omega _2  = 2\pi R_1 m + 2\pi (\tau _1  + iR_2 )n = 2\pi R_1 (m + n\tau )$
.and the above expressions define Eisenstein modular forms. The discriminant and $j$ invariant also define modular forms and are given by:
	\begin{equation}
\Delta  = G_4 (\Lambda )^3  - 27G_6 (\Lambda )^2  = q\prod\limits_{n = 1}^\infty  {(1 - q^n )^{24} }  = 2^{ - 8} \theta _1^{'8}  = 2^{ - 8} \theta _2^8 \theta _3^8 \theta _4^8 
\end{equation}
\begin{equation}
j = \displaystyle\frac{{1728(4a^3 )}}{{4a^3  + 27b^2 }} = \displaystyle\frac{{1728G_2 (\Lambda )^3 }}{{G_4 (\Lambda )^3  - 27G_6 (\Lambda )^2 }}
\end{equation}
The $j$
function is a modular form of weight $0$ \cite{Milne} and can be written as:

	\begin{equation}
j = \displaystyle\frac{{(\theta _3^{8}  +\theta _4^8  + \theta _2^8 )^3 }}{8{\Delta }}
\end{equation}
This review is enough to illustrate the close connection between modular forms and elliptic equations.

\subsection{Dirichlet series, modular forms and elliptic equations}

\indent The correspondence between theta and modular functions and Dirichlet series is transparent through the Mellin transform. Consider the formula for the Riemann and Lerch zeta functions:
\begin{equation}
\zeta (s) = \displaystyle\frac{1}{{2^{1 - s}  - 1}}\displaystyle\frac{1}{{2^s  - 1}}\pi ^{\displaystyle\frac{s}{2}} \displaystyle\frac{1}{{\Gamma (\displaystyle\frac{s}{2})}}\int\limits_0^\infty  {\displaystyle\frac{{d\tau }}{\tau }} \tau ^{s/2} (\theta _4 (0|i\tau ) + \theta _2 (0|i\tau ) - \theta _3 (0|i\tau ))
\end{equation}
\begin{equation}
\begin{array}{l}
 \phi (x,s,\alpha ) + e^{ - 2\pi ix} \phi ( - x,s,1 - \alpha ) = \displaystyle\frac{{\pi ^{s/2} }}{{\Gamma (s/2)}}\int\limits_0^\infty  {d\tau   \tau ^{s/2 - 1/2} \sum\limits_{n\varepsilon Z} {e^{ - \pi (n + \alpha )^2 \tau } e^{2\pi inx} } }  \\ 
  \\ 
 \end{array}
\end{equation}
A further example is the Ramanujan zeta function which in this case is the Mellin transform of  the modular $\Delta $
function:
	\begin{equation}
\begin{array}{l}
F(s) = \sum\limits_{n = 1}^\infty  {\tau (n)n^{ - s}} \\
 = \pi ^{s/2} \displaystyle\frac{1}{{\Gamma (s/2)}}\int\limits_0^\infty  {dCC^{s/2 - 1} \Delta (C)} 
  \end{array}
\end{equation}
Where the $\tau (n)$ coefficients are defined from the series expansion\\
 $\Delta  = \sum\limits_{n = 0}^\infty  {\tau (n)e^{ - 2\pi nC} }.$ We treat this and other zeta functions more fully in the next section when we discuss string modifications to the fermionic Green's function.

Another way of relating algebraic equations to modular functions is through elliptic genera. In this case the algebraic equation defines a manifold and the path integral of a sigma model with that manifold as a target  space defines an invariant of that space. Computing the torus partition function then yields the relation to modular functions . An example of this type of analysis for singular $K_3 $ and Calabi-Yau
manifolds is given in \cite{Naka:2000uy}.

Finally when Dirichlet series have Euler products representations  and are associated with modular forms there is a direct relationship between elliptic curves $E$
over the rationals and Dirichlet series. In this case the Euler product is directly associated with the elliptic curve by forming \cite{Milne}:
	\begin{equation}
F(E,s) = \prod\limits_p {\displaystyle\frac{1}{{1 - a{}_pp^{ - s}  + p^{1 - s} }}}  \cdot \prod\limits_p {\displaystyle\frac{1}{{1 - a{}_pp^{ - s} }}} 
\end{equation}
with
	\begin{equation}
a_p  = p + 1 - N_p ,1, - 1,0
\end{equation}
if $p$
is good, split nodal, nonsplit nodal, cuspidal respectively and where $N_p $
is the number of points $(x,y)$
satisfying the elliptic equation $E$
 modulo a prime $p$. An example is the Ramanujan zeta function defined above which has the Euler product representation:
	\begin{equation}
F(s) = \prod\limits_p {(1 - \tau (p)p^{ - s}  + p^{11 - 2s} )^{ - 1} }:      (\sigma  > \displaystyle\frac{{13}}{2})
\end{equation}
This can be compared with the relatively simple Euler product for the Riemann zeta function
	\begin{equation}
\zeta (s) = \prod\limits_p {\displaystyle\frac{1}{{1 - p^{ - s} }}}  = Z_f Z_b 
\end{equation}
In the above we can express the product over primes as two factors \begin{equation}
Z_b  = \prod\limits_{n = 2}^\infty  {\displaystyle\frac{1}{{1 - n^{ - s} }}} 
\end{equation}
and \begin{equation}
Z_f  = \prod\limits_{m \le r = 2}^\infty  {(1 - (mr)^{ - s} )} 
\end{equation}
 which have the interpretation of bonic and fermion partition functions with Hamiltonian
	\begin{equation}
H = \sum\limits_{m \le r = 2}^\infty  {(\log (m) + \log (r))b_{ - m, - r} b_{m,r} }  + \sum\limits_{n = 2}^\infty  {(\log (n)a_{ - n} a_n } 
\end{equation}
And single particle energies 
\begin{equation}
\begin{array}{l}
 \varepsilon _{m,r}  = \log (m) + \log (r) \\ 
 \varepsilon _n  = \log (n) \\ 
 \end{array}
\end{equation}
The reason for writing the zeta function this way is that if $n = mr$
for some $(m,r)$
than the factors in the numerator and denominator cancel and one is left with the product over primes.

\section{Other representations of the fermionic Green's function}

\indent The basic representation of the fermionic Green's function we have been using is the mode expansion representation given by:
	\begin{equation}
G(x,t;x',t') = \left\langle {0|\hat \psi (x,t) \bar \psi (x',t')|\left. 0 \right\rangle } \right. = \sum\limits_n {\psi _n (x,t)\psi _n^* (x',t')} 
\end{equation}
where $\psi _n (x,t)$
solves the generalized Dirac equation. However many other representations of the fermionic Green's function can be derived including the energy-momentum representation, point particle path integral representation \cite{Brink:1976sz}, \cite{Brink:1976uf}, \cite{Fainberg:1987jr}, \cite{Fradkin:1991ci}, random walk representation \cite{Ambjorn:1989ba},  \cite{Mikovic:1990dz}, spin model representation \cite{Jacobson:1983xt}, 1D Ising model representation \cite{Feynman:1965}, \cite{Gersch:1981fv}, \cite{Jacobson:1983qi}, 2D Ising model representation \cite{Polyakov:1987ez} etc. As we have shown the zeta function can be interpreted as a Green's function so we should be able to obtain many of the representations for the zeta function using these methods. In this section we explore some of the ways of representing the fermionic Green's function associated with the Riemann, Lerch and Ramanujan zeta functions.

\subsection{ (E,p) Energy momentum integral representation}

\subsubsection*{Usual case: linear dispersion $E=p$}
\indent Recall that the usual massless 2d Dirac equation is given by
\begin{equation}(\gamma ^0 \partial _t  + \gamma ^1 \partial _x )\psi (x,t) = 0\end{equation}
The Fourier transform of this equation is then
	\begin{equation}
(\gamma ^0 E + \gamma ^1 p)\psi (p,E) = 0
\end{equation}
The Green's function $G$
is then the inverse of the differential operator in (8.2) and is expressed as
	\begin{equation}
G(x,t;0,0) = \displaystyle\frac{1}{{(2\pi )^2 }}\int {dpdE\displaystyle\frac{{\gamma ^0 E + \gamma ^1 p}}{{p^2  - E^2  + i\varepsilon }}} e^{ - iEt + ipx} 
\end{equation}
Writing out the gamma matrices the Green's function is a 2x2 matrix given by:
	\begin{equation}
G(x,t;0,0) = \left( {\begin{array}{*{20}c}
   0 & {\displaystyle\frac{1}{{(2\pi )^2 }}\int {dpdE\displaystyle\frac{e^{ - iEt + ipx}}{{p - E + i\varepsilon }} } }  \\
   {\displaystyle\frac{1}{{(2\pi )^2 }}\int {dpdE\displaystyle\frac{e^{ - iEt + ipx}}{{p + E + i\varepsilon }} } } & 0  \\
\end{array}} \right)
\end{equation}
We shall concentrate on the upper right part of the matrix associated with the propagation of a right moving particle. Then the integral becomes
\begin{equation}
G_{12} (x,t;0,0) = \displaystyle\frac{1}{{(2\pi )^2 }}\int {dpdE\displaystyle\frac{1}{{p - E + i\varepsilon }}e^{ - iEt + ipx} }  = \displaystyle\frac{1}{{2\pi }}\displaystyle\frac{i}{{x - t}}
\end{equation}
in agreement with (3.5).

\subsubsection*{Nonstandard case : Logarithmic dispersion $aE=\log(ap+1)$}

\indent Now applying this representation to the Dirac equation:
\begin{equation}(\gamma ^0 \partial _t  + \gamma ^1 \displaystyle\frac{1}{{ia}}\log (ai\partial _x  + 1)\psi (x,t) = 0\end{equation}
The Fourier transform form of the equation is 
	\begin{equation}
(\gamma ^0 E + \gamma ^1 \displaystyle\frac{1}{a}\log (ap + 1)\psi (p,E) = 0
\end{equation}
Inverting this operator and again picking out the right moving portion the modified form of (8.8) becomes
\begin{equation}
G_{12} (x,t;0,0) = \displaystyle\frac{1}{{(2\pi )^2 }}\int {dpdE\displaystyle\frac{1}{{\displaystyle\frac{1}{a}\log (ap + 1) - E + i\varepsilon }}e^{ - iEt + ipx} } 
\end{equation}
If $x$
is periodic with radius $R$
the momentum is quantized so the $p$
integral becomes a sum and we have:
\begin{equation}
G_{12} (x,t;0,0) = \displaystyle\frac{1}{{(2\pi )^2 }}\displaystyle\frac{1}{R}\sum\limits_n {\int {dE\displaystyle\frac{1}{{\displaystyle\frac{1}{a}\log (a\displaystyle\frac{n}{R} + 1) - E + i\varepsilon }}e^{ - iEt + inx/R} } } 
\end{equation}
Then if we assume $R = a$
which is the case relevant to the Riemann zeta function we have:

\begin{equation}
G_{12} (x,t;0,0) = \displaystyle\frac{1}{{(2\pi )^2 }}\displaystyle\frac{1}{a}\sum\limits_{n = 1} {\int {dE\displaystyle\frac{1}{{\displaystyle\frac{1}{a}\log (n) - E + i\varepsilon }}e^{ - iEt + inx/a} } } 
\end{equation}
Performing the contour integral over $E$
we obtain the $\varphi $
 function 
\begin{equation}G(x,t;0,0)|_{R = a}  = \varphi (\displaystyle\frac{x}{{2\pi a}},\displaystyle\frac{t}{a}) = \sum\limits_{n = 1}^\infty  {\displaystyle\frac{1}{{n^{it/a} }}} e^{2\pi ix/2\pi a} \end{equation}
For $x = 0$
we obtain the Riemann zeta function:
	\begin{equation}
G_{12} (x = 0,t;0,0)|_{R = a}  = \displaystyle\frac{1}{a}\zeta (\displaystyle\frac{{it}}{a})
\end{equation}
in agreement with the mode expansion representation.

\subsubsection*{Nonstandard case: exponential dispersion $ap+1=e^{aE}$}

\indent On the other hand if we start with the second form of the modified Dirac equation:
\begin{equation}
(\gamma ^0 \displaystyle\frac{1}{{ai}}(e^{ia\partial _t }  - 1) + \gamma ^1 \partial _x )\psi (x,t) = 0
\end{equation}
The Fourier transform form of the equation is 
	\begin{equation}
(\gamma ^0 \displaystyle\frac{1}{a}(e^{aE}  - 1) + \gamma ^1 p)\psi (p,E) = 0
\end{equation}
Then the modified form of (8.6) becomes
\begin{equation}
\begin{array}{l}
 \displaystyle\frac{1}{{2\pi }}g_{3'} (x,s) = G_{12} (x,t + ia;0,0) \\ 
  = \displaystyle\frac{1}{{(2\pi )^2 }}\int {dpdE\displaystyle\frac{1}{{p - \displaystyle\frac{1}{a}(e^{aE}  - 1) + i\varepsilon }}e^{ - iE(t + ia) + ipx} }  \\ 
 \end{array}
\end{equation}
Again If $x$
is periodic with radius $R$
the momentum is quantized so the $p$
integral becomes a sum and we have:
\begin{equation}
\begin{array}{l}
 g_{4'} (x,s) = G_{4'} (x,t + ia;0,0) \\ 
  = \displaystyle\frac{1}{{(2\pi )^2 }}\displaystyle\frac{1}{R}\sum\limits_n {\int {dE\displaystyle\frac{1}{{\displaystyle\frac{n}{R} - \displaystyle\frac{1}{a}(e^{aE}  - 1) + i\varepsilon }}e^{ - iE(t + ia) + inx/R} } }  \\ 
 \end{array}
\end{equation}

Now we can use the formula $\sum\limits_{n =  - \infty }^\infty  {\displaystyle\frac{{e^{in\displaystyle\frac{x}{R}} }}{{u - n}}}  = \displaystyle\frac{{\pi e^{ - i\pi u} e^{iux/R} }}{{\sin (\pi u)}}$
 with $u = \displaystyle\frac{R}{a}(e^{aE}  - 1)$
and $W = e^{aE} $
to write:
                          \begin{equation}
\begin{array}{l}
 g_{4'} (x,s) = G_{4'} (x,t + ia;0,0) \\ 
  = \displaystyle\frac{1}{{2\pi }}\displaystyle\frac{1}{a}\int {\displaystyle\frac{{dW}}{W}} \displaystyle\frac{{\pi e^{ - i\pi \displaystyle\frac{R}{a}(W - 1)} }}{{\sin (\pi \displaystyle\frac{R}{a}(W - 1))}}e^{i\displaystyle\frac{R}{a}(W - 1)\displaystyle\frac{x}{R}} W^{(\displaystyle\frac{{ - it}}{a} + 1)}  \\ 
 \end{array}
\end{equation}
Switching variables to $u = 2i\pi \displaystyle\frac{R}{a}W$
we write this as:
\begin{equation}
\begin{array}{l}
 g_{4'} (x,s) = G_{4'} (x,t + ia;0,0) \\ 
  = \displaystyle\frac{1}{a}e^{2\pi i\displaystyle\frac{R}{a}} e^{ - ix/a} (\displaystyle\frac{R}{a})^{s - 1} (2\pi )^{s - 1} e^{i\pi (s - 1)/2} \int {\displaystyle\frac{{du}}{u}} \displaystyle\frac{ u^{ - s + 1}}{{e^u  - e^{2\pi iR/a} }}e^{ - u( - \displaystyle\frac{x}{{2\pi R}})}  \\ 
  = \displaystyle\frac{1}{a}e^{2\pi i\displaystyle\frac{R}{a}} e^{ - ix/a} (\displaystyle\frac{R}{a})^{s - 1} (2\pi )^{s - 1} e^{i\pi (s - 1)/2} \Gamma (1 - s)\phi (\displaystyle\frac{R}{a},1 - s,1 - \displaystyle\frac{x}{{2\pi R}}) \\ 
 \end{array}
\end{equation}
For the case relevant to the Riemann zeta function $R = a$
and $x = \pi a$
this simplifies considerably to:
\begin{equation}
\begin{array}{l}
 g_{4'} (x = \pi a,s) = G_{4'} (x = \pi a,t + ia;0,0)|_{R = a}  \\ 
  = \displaystyle\frac{1}{a}(2\pi )^{s - 1} e^{i\pi (s - 1)/2} \Gamma (1 - s)(2^{1 - s}  - 1)\zeta (1 - s) \\ 
 \end{array}
\end{equation}
Using the functional equation for the zeta function this can be written as:
\begin{equation}
\begin{array}{l}
 g_{4'} (x = \pi a,s) = G_{4'} (x = \pi a,t + ia;0,0)|_{R = a}  \\ 
  = \displaystyle\frac{1}{a}\displaystyle\frac{1}{2}\displaystyle\frac{1}{{1 - e^{i\pi s} }}(2^{1 - s}  - 1)\zeta (s) \\ 
 \end{array}
\end{equation}
Thus the integral $(E,p)$
form of the Green's function leads to the similar conclusion as the mode sum: The physical transition probability of the fermion leaving from $(x = 0,t = 0)$
and arriving to $(x = \pi a,t)$
vanishes at the Riemann zeros. 

\subsection{Symmetry properties of the Green's function}

\indent The usual Dirac fermion is two dimensions has a number of symmetries including translation, Lorentz rotation, scaling and especially conformal transformations. To determine the symmetries of the fermionic Green's function that leads to the Riemann zeta function we perform a Wick rotation to Euclidean space which changes the Green's function to:
\begin{equation}
\begin{array}{l}
 G_{\alpha \beta } (x,\bar t;0,0) = \int {dpdE\displaystyle\frac{{\gamma ^0 \displaystyle\frac{1}{a}(e^{ia\bar E}  - 1) + \gamma ^1 p}}{{p^2  - \displaystyle\frac{1}{{a^2 }}(e^{ia\bar E}  - 1)^2 }}} e^{ - i\bar E\bar t + ipx}  \\ 
 \end{array}
\end{equation}
Upon Wick rotation the nonstandard Dirac equation becomes the difference equation:
	\begin{equation}
\begin{array}{l}
 (\gamma ^0 \displaystyle\frac{1}{a}(e^{a\partial _t }  - 1) + \gamma ^1 \partial _x )\psi (x,\bar t) \\ 
  = (\gamma ^0 (\psi (x,\bar t + a) - \psi (x,\bar t)) + \gamma ^1 \partial _x )\psi (x,\bar t) = 0 \\ 
 \end{array}
\end{equation}
which is a field equation from the discrete field theory:
	\begin{equation}
I = \int {d\bar tdx(\bar \psi (x,\bar t)(\gamma ^0 (\psi (x,\bar t + a) - \psi (x,\bar t)) + \gamma ^1 \partial _x )\psi (x,\bar t)} 
\end{equation}
The Green's function is modified under scale transformation as:
\begin{equation}
\begin{array}{l}
 G(\lambda x,\lambda \bar t;0,0;\lambda L,\lambda a) = \lambda ^{ - 1} G(x,t;0,0;L,a) \\ 
  = \lambda ^{ - 1} \sum\limits_{p = \displaystyle\frac{{2\pi n}}{L}} {\int {dE\displaystyle\frac{{\gamma ^0 \displaystyle\frac{1}{a}(e^{ai\bar E}  - 1) + \gamma ^1 p}}{{p^2  + \displaystyle\frac{1}{{a^2 }}(e^{ai\bar E}  - 1)^2 }}} e^{ - iE\bar t + ipx} }  \\ 
 \end{array}
\end{equation}
with $\lambda $
an integer. . The representation of discrete rotation and scale invariance can be found be studying the transformation properties of the denominator in (8.22) which results from applying the discrete Dirac operator twice. Setting $K(p,\bar E) = p^2  - \displaystyle\frac{1}{{a^2 }}(e^{ai\bar E}  - 1)^2  = k^2  - \displaystyle\frac{1}{{a^2 }}(e^{ia\omega }  - 1)^2 $
the generators of discrete rotation invariance are found to be $\ell _0  - \bar \ell _0  = \displaystyle\frac{1}{{ai}}(e^{ia\omega }  - 1)\partial _k  - k\partial _\omega  $
and dilatation\\
 $\ell _0  + \bar \ell _0  = \partial _k k + \partial _\omega  \displaystyle\frac{1}{{ai}}(e^{ia\omega }  - 1).$ In terms of position coordinates these are given by\\
 $\ell _0  - \bar \ell _0  = x\displaystyle\frac{1}{a}(e^{a\partial _t }  - 1)\partial _k  - \bar t\partial _x $
 and $\ell _0  + \bar \ell _0  = x\partial _x  + t\displaystyle\frac{1}{a}(e^{a\partial _t }  - 1).$ Further understanding of the symmetry properties can be found by using discrete conformal field theory as in \cite{Henkel:1998zj} or by using methods of stress energy tensors for higher derivative theories. We hope to return to these issues in future work.

\subsection{Green's function from periodization}

In the usual case $(E = p)$
we can form the Green's function on a cylinder from the Green's function on a plane by periodization. That is beginning with  \\
$2\pi G_1 (x,t;0,0) = \displaystyle\frac{i}{{x - t}}$  we form $G_2 (x,t;0,0)$ through:
	\begin{equation}
2\pi G_2 (x,t;0,0) = \sum\limits_{n =  - \infty }^\infty  {2\pi G_1 (x - 2\pi Rn,t;0,0) = } \sum\limits_{n =  - \infty }^\infty  {\displaystyle\frac{i}{{x - 2\pi nR - t}}} 
\end{equation}
Now using the formula $\sum\limits_{n =  - \infty }^\infty  {\displaystyle\frac{1}{{u - n}}}  = \displaystyle\frac{{\pi e^{ - i\pi u} }}{{\sin (\pi u)}} = \displaystyle\frac{{ - 2i\pi }}{{1 - e^{2\pi iu} }}$
we have:
	\begin{equation}
2\pi G_2 (x,t;0,0) = \displaystyle\frac{1}{R}\displaystyle\frac{1}{{1 - e^{i(x - t)/R} }}
\end{equation}
in agreement with (3.16).

We can use a similar method to construct the nonstandard Green's function $G_4 (x,t;0,0)$
. Beginning with the Green's function $g_{3'} (x,s) = 2\pi G_{3'} (x,t - i\sigma ;0,0) = \displaystyle\frac{1}{a}( - ix/a)^{s - 1} e^{ - ix/a} \Gamma (1 - s)$
associated with the dispersion relation $E = \displaystyle\frac{1}{a}\log (ap + 1)$
 and with $s = it + \sigma $
we form:
\begin{equation}
\begin{array}{l}
 g_4 (x,s) = 2\pi G_4 (x,t - i\sigma ;0,0) = \sum\limits_{n =  - \infty }^\infty  {g_{3'} (x - 2\pi Rn,s)}  \\ 
               = \Gamma \left( {1 - s} \right)\displaystyle\frac{1}{a}\sum\limits_{n =  - \infty }^\infty  {( - i(x - 2\pi nR)/a)^{s - 1} e^{ - i(x - 2\pi nR)/a} }  \\ 
              = \Gamma \left( {1 - s} \right)\displaystyle\frac{1}{a}(\displaystyle\frac{R}{a})^{s - 1} \sum\limits_{n =  - \infty }^\infty  {( - i\displaystyle\frac{x}{R} - 2\pi n)^{s - 1} e^{ - i(\displaystyle\frac{x}{R} - 2\pi n)\displaystyle\frac{R}{a}} }  \\ 
              = \displaystyle\frac{1}{R}(\displaystyle\frac{R}{a})^s \phi (\displaystyle\frac{x}{{2\pi R}},s,\displaystyle\frac{R}{a}) \\ 
 \end{array}
\end{equation}
in agreement with (3.52).

\subsection{Point particle path integral representation}

\indent Another way of representing the Dirac Green's function is as a point particle path integral from point $(0,0)$
to $(x,t)$
. This is given by:
\begin{equation}
G(x,t;0,0) = \int {De_0 D\chi _0 DxDtDpDED\lambda e^{iI} } 
\end{equation}
In the path integral $(e_0 ,\chi _0 )$
represent a (0+1) dimensional gravitational and Rarita Schwinger field, $(\xi ,\tilde \xi )$
represent worldline fermions which eventually become the space-time gamma matrices, $(x,t,p,E)$
are the position, time, momentum and energy associated with the point particle.

For the usual case of the Dirac particle the (0+1)  point particle action is:
\begin{equation}
I = \int {d\tau (p\dot x - E\dot t}  - \displaystyle\frac{{e_0 }}{2}(p^2  - E^2 ) - i\xi \dot \xi  - i\tilde \xi \dot{ \tilde \xi}  - ip\chi _0 \xi  - iE\chi _0 \tilde \xi )
\end{equation}

This point particle action has (super) reparametrization symmetry associated with the $(0 + 1)$
dimensional supergravity. As discussed in several references \cite{Brink:1976sz}, \cite{Brink:1976uf}, \cite{Fainberg:1987jr}, \cite{Fradkin:1991ci} the Green's function reduces to an integral over a single Teichmuller parameter $C = \int\limits_{\tau _1 }^{\tau _2 } {e_0 (\tau )d\tau } $
 and the zero modes of $p$
and $E$
. In \cite{Gozzi:1988uj} the modular group was described by by the $Z_2 $
transform $e_0 (\tau ) =  - e_0 ( - \tau )$
. Under this transformation the modular parameter changes sign and one restricts it to a fundamental region $F = [0,\infty ]$
. Upon Wick rotation the Euclidean Green's function is given by:
	\begin{equation}
G(x,\bar t;0,0) = \int\limits_F {dCdpdE}e^{ - Cp^2  - C\bar E^2 } (\xi p + \tilde \xi \bar E)e^{ipx - i\bar E\bar t}  
\end{equation}
After integrating over $C$
and identifying the zero modes of $(\xi ,\tilde \xi )$
with the gamma matrices we obtain the usual form of the Dirac Green's function.

In our case we can follow essentially the same procedure except our point particle action is more complicated, also the spatial direction is periodic or antiperiodic which causes the momentum to be quantized. The point particle action is:
\begin{equation}
\begin{array}{l}
 I = \int\limits_{\tau _1 }^{\tau _2 } {d\tau (p\dot x}  - E\dot t - \displaystyle\frac{{e_0 }}{2}(p^2  - \displaystyle\frac{1}{{a^2 }}(e^{aE}  - 1)^2 ) \\ 
\;\;\;\;\;  + \chi _0 \displaystyle\frac{1}{a}p\xi  + \chi _0 \displaystyle\frac{1}{a}(e^{aE}  - 1)\tilde \xi  + i\bar \xi  \dot{ \tilde \xi}  + i\xi \dot \xi ) \\ 
 \end{array}
\end{equation}
The path integral reduces to modular integral as before and we have:
\begin{equation}
\begin{array}{l}
 2\pi G(x,\bar t;0,0) \\ 
  = \displaystyle\frac{1}{{2\pi }}\int\limits_F {dCdpdEe^{ - Cp^2  + C\displaystyle\frac{1}{{a^2 }}(e^{ia\bar E}  - 1)^2 } (\xi p + \tilde \xi \displaystyle\frac{1}{a}(e^{ia\bar E}  - 1))e^{ipx - i\bar E\bar t} }  \\ 
 \end{array}
\end{equation}
The zero modes of the $\xi ,\tilde \xi $
fields become the gamma matrices $\gamma ^1 ,\gamma ^0 $
so after integrating over $C$
we obtain the Green's function (8.22). To obtain the theta function representation of the Green's function we send $p = \displaystyle\frac{n}{R}$
, $R = ma$
and $w = e^{ia\bar E} $
the Green's function reduces to:
\begin{equation}
\begin{array}{l}
 G(x,\bar t;0,0) \\ 
  = \displaystyle\frac{1}{{2\pi }}\displaystyle\frac{1}{a}\int\limits_F {dC\sum\limits_n {e^{ - C\displaystyle\frac{{n^2 }}{{R^2 }} + \displaystyle\frac{{inx}}{R}} } dWe^{ - C\displaystyle\frac{1}{{a^2 }}(W-1)^2 } (\xi \displaystyle\frac{{an}}{R} -  + \tilde \xi (W - 1)W^{ - \bar t/a - 1} }  \\ 
 \end{array}
\end{equation}
For simplicity consider the case $x = 0$
and $R = a$
which is relevant to the Riemann zeta function.
	\begin{equation}
\begin{array}{l}
 g_{4'} (0,s) = 2\pi G_{4'} (0,t + ia;0,0) \\ 
  = \int\limits_F {dC\displaystyle\frac{{dW}}{W}e^{ - Cn^2 } e^{ - C(W - 1)^2 } (n - 1 + W)W^{ - s + 1} }  \\ 
 \end{array}
\end{equation}
Now use the fact that
	\begin{equation}
\begin{array}{l}
 \sum\limits_{n =  - \infty }^\infty  {\int {dCe^{ - Cn^2 } e^{ - C(W - 1)^2 } (n + W - 1)} }  \\ 
  = \sum\limits_{n =  - \infty }^\infty  {\int {dCe^{ - C(n + 1)^2 } e^{ - CW^2 } (n + W + 1) = } } \sum\limits_{n =  - \infty }^\infty  {\int {dCe^{ - Cn^2 } e^{ - CW^2 } W} }  \\ 
 \end{array}
\end{equation}
We have:
\begin{equation}
g_{4'} (0,s) = 2\pi G_{4'} (0,t + ia;0,0) = \int\limits_F {dC\displaystyle\frac{{dW}}{W}\sum\limits_{n =  - \infty }^\infty  {e^{ - Cn^2 } } e^{ - CW^2 } W^{ - s + 2} } 
\end{equation}
Performing the integral over $W$
we have the theta representation. In terms of the zeta function this becomes:
\begin{equation}
g_{4'} (0,s) = 2\pi G_{4'} (0,t + ia;0,0) = \displaystyle\frac{1}{a}\displaystyle\frac{1}{{\sin (\pi s/2)}}\zeta (s)
\end{equation}

\subsection{String modified Green's function}

The representation of the zeta function using theta functions represents a convenient way to study string modifications to the Green's function. This is because of the form of the expression :
\begin{equation}2\pi G(x = 0,t;0,0) = \displaystyle\frac{1}{\Gamma(s/2)}\int\limits_F {dC(\sum\limits_{n = 0}^\infty  {e^{ - C\displaystyle\frac{{(n + A)^2 }}{{R^2 }}} )} e^{ - CM^2 } C^{ - it/a} } \end{equation}
where the modified dispersion relation including a mass $M$ is
	\begin{equation}
(\displaystyle\frac{1}{a}e^{aE} )^2  = (\displaystyle\frac{{n + A}}{R})^2  + M^2 
\end{equation}
or $E = \displaystyle\frac{1}{{2a}}\log (a((\displaystyle\frac{{n + A}}{R})^2  + M^2 ))$. For the choice $A = \displaystyle\frac{R}{a}$
this becomes: $E = \displaystyle\frac{1}{{2a}}\log ((a\displaystyle\frac{n}{R} + 1)^2  + a^2 M^2 ))$
which reduces to the usual dispersion relation$E = \displaystyle\frac{1}{a}\log (a\displaystyle\frac{n}{R} + 1)$
 if $M = 0$. To study the string modification to the Green's function from a point like configuration to another point like configuration one replaces the $e^{ - CM^2 } $
factor with $tr(e^{ - CM^2 } )$ where $\alpha 'M^2  = N$ and $N$
is the occupation number in the string oscillator description. Expanding the trace as 
\begin{equation}tr(e^{ - CM^2 } ) = \sum\limits_{N = 1}^\infty  {\rho (N)e^{ - CN/\alpha '} } \end{equation}
The Green's function becomes:
\begin{equation}\begin{array}{l}
 2\pi G(x = 0,t;0,0)\\
 = \displaystyle\frac{1}{\Gamma(s/2)}\int\limits_F {dC(\sum\limits_{n = 1}^\infty  {e^{ - C(\displaystyle\frac{{(n + A)^2 }}{{R^2 }}} )} \sum\limits_{N = 1}^\infty  {\rho (N)e^{ - CN/\alpha '} } C^{ - it/2a} }  \\ 
  = \sum\limits_{n = 1,N = 1}^\infty  {\rho (N)((a\displaystyle\frac{{n + A}}{R}} )^2  + \displaystyle\frac{{a^2 N}}{{\alpha '}})^{ - it/2a}  \\ 
 \end{array}\end{equation}

\subsubsection*{Two dimensional heterotic string}
The actual value of the degeneracy factor $\rho (N)$
depends on the string theory considered. 
For example the choice:
	\begin{equation}
tr(e^{ - CM^2 } ) = q\prod\limits_{n = 1}^\infty  {(1 + q^n )^{24} }  = 2^{ - 8} \displaystyle\frac{{\theta _2^{12} }}{{\theta _1^{'4} }} = \sum\limits_{N = 1}^\infty  {\rho _{12} (N)q^N  = } \sum\limits_{N = 1}^\infty  {\tau '(N)q^N } 
\end{equation}
where $q = e^{ - C/\alpha '} $
describes a sector in the fermionic formulation of $E_8  \times SO(8)$
or $SO(24)$
heterotic two dimensional string theory \cite{McGuigan:1991qp} and the subscript denotes the dimension of the root lattice of the group. Taking the leading term in the series over $n$
by sending $R \to 0$
and setting $a^2  = \alpha '$
we have:
	\begin{equation}
G(x = 0,t;0,0)|_{R = 0}  = \sum\limits_{N = 1}^\infty  {\rho _{12} (N)} N^{ - it/2a}  = \sum\limits_{N = 1}^\infty  {\tau '(N)} N^{ - it/2a} 
\end{equation}

\subsubsection*{Ramanujan Zeta function}

Another choice is 
\begin{equation}
tr(e^{ - CM^2 } ) = q\prod\limits_{n = 1}^\infty  {(1 - q^n )^{24} }  = 2^{-8}\theta _1^{'8}  = \sum\limits_{N = 1}^\infty  {\rho (N)q^N  = } \sum\limits_{N = 1}^\infty  {\tau (N)q^N } 
\end{equation}
This form is related to the Ramanujan zeta function \cite{Ivic}. 
	\begin{equation}
F(s) = \sum\limits_{n = 1}^\infty  {\tau (n)n^{ - s} } 
\end{equation}
The Ramanujan zeta function has similar properties to the Riemann zeta function with functional relation:
	\begin{equation}
(2\pi )^{ - s} \Gamma (s)F(s) = (2\pi )^{ - (12 - s)} \Gamma (12 - s)F(12 - s)
\end{equation}
The analogue of the Riemann hypothesis for the Ramanujan zeta function is the absence of zeros in the region $\displaystyle\frac{{12}}{2} < {\mathop{\rm Re}\nolimits} s < \displaystyle\frac{{13}}{2}$
. We can use the formula
 \begin{equation}
Tr(e^{ - CM^2 } ) = q\prod\limits_{m = 1}^\infty  {(1 - q^n )^{24} }  = 2^{-8}\theta _1 ^{'8} 
\end{equation}
where $q = e^{2i\pi \tau } $ to write the Green's function in theta form :
	\begin{equation}
\begin{array}{l}
g_5 (x,s) =2\pi G_5 (x,t;0,i\sigma )\\
 = \displaystyle\frac{1}{\Gamma(s/2)}\int\limits_F {d\tau \sum\limits_{n =  - \infty }^\infty  {e^{ - \pi \tau \displaystyle\frac{{(n + A)^2 }}{{R^2 }}} e^{i\displaystyle\frac{{(n + A)}}{R}x} 2^{-8}\theta _1 ^{'8} } } \tau ^{ i(t - i\sigma )/2a - 1}
 \end{array} 
\end{equation}
We can also introduce the string generalization of the states
	\begin{equation}
\begin{array}{l}
 |\left. {\nu '} \right\rangle  = \sum\limits_{n = 0,N}^\infty  {e^{ - \sigma E_{n,N} /2} ( - 1)^n |\left. {n,N} \right\rangle }  \\ 
 |\left. {\nu ,t} \right\rangle  = \sum\limits_{n = 0,N}^\infty  {e^{ - \sigma E_{n,N} /2} e^{ - itE_{n,N} } |\left. {n,N} \right\rangle }  \\ 
 \end{array}
\end{equation}
Then the amplitude for transition from the state $\nu $
to $\nu '$ is given by:
	\begin{equation}
A(\nu  \to \nu ') = \left\langle {\nu |\nu '} \right.,\left. t \right\rangle  = \sum\limits_{n,N = 1}^\infty  {e^{ - \sigma E_{n,N} } e^{ - itE_{n,N} } \tau (N)} 
\end{equation}
Using the dispersion relation (8.40) we have:

	\begin{equation}
E_{n,N}  = \displaystyle\frac{1}{a}\log (a\sqrt {\displaystyle\frac{{(n + A)^2 }}{{R^2 }} + \displaystyle\frac{N}{{\alpha '}}} )
\end{equation}
The transition amplitude is:
\begin{equation}
A(\nu  \to \nu ') = \left\langle {\nu |\nu '} \right.,\left. t \right\rangle  = \sum\limits_{n,N = 1}^\infty  {(a\sqrt {\displaystyle\frac{{(n + A)^2 }}{{R^2 }} + \displaystyle\frac{N}{{\alpha '}}} )^{( - it - \sigma )/a} \tau (N)} 
\end{equation}
To compare with the Kaluza-Klein neutrino oscillation we note that
	\begin{equation}
|\left. \nu  \right\rangle  = \sum\limits_{n = 0,N}^\infty  {\displaystyle\frac{{(\displaystyle\frac{1}{a})^{\sigma /a} }}{{\sqrt {(\displaystyle\frac{{(n + A)^2 }}{{R^2 }} + M^2 )} ^{\sigma /a} }}} |n,\left. N \right\rangle 
\end{equation}
Whereas the Kaluza Klein expansion is given by \cite{Lukas:2000rg}:
\begin{equation}
|\left. \nu  \right\rangle _{KK}  = \sum\limits_{n = 0,N}^\infty  {\displaystyle\frac{m}{{\sqrt {(\displaystyle\frac{{(n + R\mu _V )^2 }}{{R^2 }} + \mu _S ^2 )} }}} |n,\left. N \right\rangle 
\end{equation}
So that the form is the same when $m = \displaystyle\frac{1}{a},$ $\mu _V  = A/R,$ $\mu _V  = M$
and $\sigma  = a$. Again the time variation of this state is very different in our case because of the nontrivial logarithmic dispersion relation.

Taking the limit of $R \to 0$
only the first term contributes as the other terms are infinitely massive and we have the transition probability:
	\begin{equation}
P(\nu  \to \nu ',t) = |\left\langle {\nu |\nu '} \right.,\left. t \right\rangle |^2  = |\sum\limits_{n,N = 1}^\infty  {(\displaystyle\frac{{a^2 }}{{\alpha '}}N)^{( - it - \sigma )/2a} \tau (N)} |^2 
\end{equation}

\subsubsection*{Ten dimensional open fermionic string}
Finally the open ten dimensional fermionic string corresponds to the choice \cite{Fainberg:1988zg}:
	\begin{equation}
Tr(e^{ - CM^2 } ) = \displaystyle\frac{{\prod\limits_{n = 1}^\infty  {(1 + q^{n/2} )^8 } }}{{\prod\limits_{n = 1}^\infty  {(1 - q^n )^8 } }} = \displaystyle\frac{{\prod\limits_{n = 1}^\infty  {(1 + q^{n/2} )^4 } }}{{\prod\limits_{n = 1}^\infty  {(1 - q^{n/2} )^4 } }} = \displaystyle\frac{{\theta _2^2 }}{{\theta _1^{'2} }} = \sum\limits_{N = 0}^\infty  {\rho _o (N)e^{ - CN/\alpha '} } 
\end{equation}
In this case the dependent Green's function is given in the $(E,p)$
representation by:
	\begin{equation}
\begin{array}{l}
 G(x,t;0,0) =  \\ 
 \int {dEd^9 pdC} (\gamma ^0 \displaystyle\frac{1}{a}(e^{aE}  - 1) + p_i \gamma ^i )e^{ - iEt + ipx} e^{ - C( - \displaystyle\frac{1}{a}(e^{aE}  - 1))^2  + p^2 )}Tr(e^{ - CM^2 } )  \\ 
 \end{array}
\end{equation}
As we are interested in short distance behavior of the Green's function we examine the region of the integrand where $e^{aE}  \gg 1$
. Also we assume $x = 0$
so that:
\begin{equation}
\begin{array}{l}
 G(0,t;0,0)   \\ 
\sim \int {dEd^9 p} \gamma ^0 \displaystyle\frac{1}{a}e^{aE} e^{ - iEt} \int {dCe^{ - C( - \displaystyle\frac{1}{a}(e^{aE} )^2  + p^2  + i\varepsilon )} } \sum\limits_{N = 0}^\infty  {\rho _o (N)e^{ - CN/\alpha '} }  \\ 
  \sim \int {dE} \gamma ^0 \displaystyle\frac{1}{a}e^{aE} e^{ - iEt} \int {dCe^{ - C( - \displaystyle\frac{1}{a}(e^{aE} )^2  + i\varepsilon )} } \sum\limits_{N = 0}^\infty  {\rho _o (N)e^{ - CN/\alpha '} C^{ - 9/2} }  \\ 
 \end{array}
\end{equation}
To extract the $N$
dependence redefine $2a\bar E = 2aE - \log (N)$
and $\bar C = CN$
\begin{equation}
\begin{array}{l}
 G(0,t;0,0)  \\ 
  \sim \int {d\bar E} \gamma ^0 \displaystyle\frac{1}{a}e^{a\bar E} e^{ - i\bar Et} \int {d\bar Ce^{ - \bar C( - \displaystyle\frac{1}{a}(e^{a\bar E} )^2  + i\varepsilon )} e^{ - \bar C/\alpha '} } \bar C^{ - 9/2} \sum\limits_{N = 0}^\infty  {\rho _o (N)N^{ - \displaystyle\frac{{it}}{{2a}}+4 } }  \\ 
 \end{array}
\end{equation}
So that the short distance Green's function is again described by a Dirchlet expansion. The quantization in this case is coming from the string mass condition $M_N^2  = N/\alpha '$.

\section{Conclusions}

In this paper we represented the Lerch zeta function as a Green's function associated with a two dimensional Dirac equation and derived the Dirichlet functions including the Riemann zeta function as special cases of the relation. We then formulated the Riemann hypothesis in physical terms associated with the transition amplitude and dispersion relation $E = \displaystyle\frac{1}{a}\log (1 + ap)$
. The momentum $p$
was quantized $\displaystyle\frac{{2\pi n}}{L}$
because the spatial direction was a chosen to be a circle of circumference $L$
. The circle in turn obeyed the special relation $L = 2\pi a$
.Within this framework we obtained a particular representation of the Riemann hypothesis as a nonzero condition on the transition probability between two states that is analogous to the phenomena of neutrino mixing.

We discussed the parameter $\sigma $
from several viewpoints including mixing parameters, duality, susy breaking, zeros in path integrals and as an extra dimension.
We also studied the thermal partition function and thermal Green's function associated with the dispersion relation and derived the thermal divergence at temperature $\displaystyle\frac{1}{a}$
. We explained the relation of the thermal partition function of this theory to the thermal partition function of Julia, Spector and Bowick \cite{Spector:1988nn}, \cite{Julia:1993pz}, \cite{Bowick:1990mx}. We studied the fermionic string theory associated with a logarithmic oscillator, its matrix model and the relation of the periodization of the inverted oscillator to the Riemann zeros. We discussed the relationship between the ground rings of $(c = 1)$
string theory, elliptic curves, modular forms and Dirichlet series. Finally we derived several representations of the fermionic Green's function including the $(E,p)$
representation, point particle path integral representation, and string modifications. For the string modified Green's functions we derived Dirichlet series for the 2D heterotic string, the Ramanujan $\tau $
function and the open ten dimensional fermionic string. Other physical pictures and methods can be obtained by permuting the physical variables $(x,t,p,E)$
 or searching for Yang-Lee zeros of a one particle thermal partition function. One can also consider the Green's function of a nonstandard scalar particle. We discuss some of these in Appendix A , B and C. 

\appendix
\section{
 Permutations of $x,t,p,E$ and different physical realizations of the Green's function}

One can obtain different physical realizations of the two point function simply by permuting the dynamical variables $(x,t,p,E)$
. 

In the interpretation we have been using in this paper the point particle path integral representation of the Green's function is:
	\begin{equation}
G(x,t;0,0) = \int\limits_{x(0) = 0,t(0) = 0}^{x(1) = x,t(1) = t} {DxDtDEDpDe_0 D\chi _0 D\xi D\bar \xi } e^{ -i \int {Ld\tau } } 
\end{equation}
where
\begin{equation}
L = p\dot x - E\dot t + \displaystyle\frac{1}{2}e_0 (p^2  - \displaystyle\frac{1}{{a^2 }}(e^{aE}  - 1)^2 ) + i\chi _0 p\xi  + i\chi _0 \displaystyle\frac{1}{a}(e^{aE}  - 1)\bar \xi  - i\xi \dot \xi  - i\bar \xi  \dot{ \bar \xi} 
\end{equation}

The classical equations associated with this choice are given by:
\begin{equation}
\begin{array}{l}
 {\rm{Energy:}}\;\;\;\;       E = \displaystyle\frac{1}{a}\log (ap + 1) \\ 
 {\rm{momentum:}}\;\;\;\;p = \displaystyle\frac{1}{a}(e^{aE}  - 1) \\ 
 {\rm{velocity: }}\;\;\;\;     v = \displaystyle\frac{{\partial E}}{{\partial p}} = \displaystyle\frac{c}{{ap + 1}} = ce^{ - aE}  \\ 
 {\rm{position: }}\;\;\;\;   x = \displaystyle\frac{c}{{ap + 1}}t = e^{ - aE} ct \\ 
 {\rm{time: }} \;\;\;\;        t = \displaystyle\frac{{ap + 1}}{c}x = e^{aE} \displaystyle\frac{x}{c} \\ 
 \end{array}
\end{equation}
Finally we assumed periodic boundary conditions in the $x$
direction so that momentum was quantized $p = \displaystyle\frac{n}{R}$.

\subsection*{Sending $x \leftrightarrow t;p \leftrightarrow E$
}

This is a simple permutation exchanging space with time and momentum with energy.
The Green's function in this case becomes 
$G(t,x;0,0) = (\displaystyle\frac{R}{a})^{ix/a} \phi (\displaystyle\frac{t}{{2\pi R}},\displaystyle\frac{x}{a},\displaystyle\frac{R}{a})$
The bosonic piece of the point particle Lagrangian is:
\begin{equation}
L = E\dot t - p\dot x - \displaystyle\frac{{e_0 }}{2}(\displaystyle\frac{1}{{a^2 }}(e^{ap}  - 1)^2  - E^2 )
\end{equation}
The classical equations associated with the Lagrangian are:
\begin{equation}
\begin{array}{l}
 p = \displaystyle\frac{1}{a}\log (aE + 1) \\ 
 E = \displaystyle\frac{1}{a}(e^{ap}  - 1) \\ 
 v = \displaystyle\frac{{\partial E}}{{\partial p}} = \displaystyle\frac{c}{{aE + 1}} = ce^{ap}  \\ 
 t = \displaystyle\frac{c}{{aE + 1}}x = e^{ - ap} cx \\ 
 x = \displaystyle\frac{{aE + 1}}{c}t = e^{ap} \displaystyle\frac{t}{c} \\ 
 \end{array}
\end{equation}
In this case one takes time periodic so that $E = \displaystyle\frac{n}{R}$.

\subsection*{ Sending $x \leftrightarrow p;t \leftrightarrow E$
}
 The Green's function in this case is $G(p,E;0,0) = (\displaystyle\frac{R}{a})^{ip/a} \phi (\displaystyle\frac{E}{{2\pi R}},\displaystyle\frac{p}{a},\displaystyle\frac{R}{a}).$\\
The bosonic piece of the point particle Lagrangian is:
\begin{equation}
L =  - p\dot x + E\dot t - \displaystyle\frac{{e_0 }}{2}(\displaystyle\frac{1}{{a^2 }}(e^{at}  - 1)^2  - x^2 )
\end{equation}
The classical equations become
\begin{equation}
\begin{array}{l}
 t = \displaystyle\frac{1}{a}\log (ax + 1) \\ 
 x = \displaystyle\frac{1}{a}(e^{at}  - 1) \\ 
 v = \displaystyle\frac{{\partial E}}{{\partial p}} = \displaystyle\frac{{ax + 1}}{c} = ce^{at}  \\ 
 p = \displaystyle\frac{c}{{ax + 1}}E = e^{ - at} cE \\ 
 E = \displaystyle\frac{{ax + 1}}{c}p = e^{at} \displaystyle\frac{p}{c} \\ 
 \end{array}
\end{equation}
The trajectory \begin{equation}
x = \displaystyle\frac{1}{a}(e^{at}  - 1)
\end{equation}
comes from the Hamiltonian$E = axp + p$
which reduces to the Berry-Keating operator $E = apx$
 at large $x$ \cite{Berry}.
 The trajectory is also the geodesic associated with the metric $ds^2  =  - dt^2  + e^{ - 2at} dx^2 $
. In this case space is quantized as $x = \displaystyle\frac{n}{R}$
 and $a$
and $R$
have units of energy.

\subsection* {Sending $x \leftrightarrow E;t \leftrightarrow p$
}
Finally sending $x \leftrightarrow E;t \leftrightarrow p$
 the Green's function becomes\\
 $G(E,p;0,0) = (\displaystyle\frac{R}{a})^{iE/a} \phi (\displaystyle\frac{E}{{2\pi R}},\displaystyle\frac{p}{a},\displaystyle\frac{R}{a}).$ The bosonic piece of the point particle Lagrangian is:
\begin{equation}
L =  - E\dot t + p\dot x - \displaystyle\frac{{e_0 }}{2}(\displaystyle\frac{1}{{a^2 }}(e^{ax}  - 1)^2  - t^2 )
\end{equation}
The classical equations become:
\begin{equation}
\begin{array}{l}
 x = \displaystyle\frac{1}{a}\log (at + 1) \\ 
 t = \displaystyle\frac{1}{a}(e^{ax}  - 1) \\ 
 v = \displaystyle\frac{{\partial E}}{{\partial p}} = \displaystyle\frac{c}{{at + 1}} = ce^{ - ax}  \\ 
 E = \displaystyle\frac{c}{{at + 1}}p = e^{ - ax} cp \\ 
 p = \displaystyle\frac{{at + 1}}{c}E = e^{ax} \displaystyle\frac{E}{c} \\ 
 \end{array}
\end{equation}
In this case the trajectory is the null geodesic associated with the metric $ds^2  =  - e^{ - 2ax} dt^2  + dx^2 $
. Here time is quantized as $t = \displaystyle\frac{n}{R}$
and $a$
and $R$
have units of energy.

\section{One particle partition function}

The one particle partition function associated with the dispersion relation $E = \displaystyle\frac{1}{a}\log (a\displaystyle\frac{{2\pi n}}{L} + 1)$
 is given by:
\begin{equation}z(\beta ) = \sum\limits_{n = 0}^\infty  {e^{ - \beta E_n } }  = \sum\limits_{n = 0}^\infty  {(a\displaystyle\frac{{2\pi n}}{L}}  + 1)^{ - \beta /a}  = \left( {\displaystyle\frac{{a2\pi }}{L}} \right)^{ - \beta /a} \phi (0,\displaystyle\frac{\beta }{a},\displaystyle\frac{L}{{2\pi a}})\end{equation}
This reduces to the Riemann zeta function for the special 
case $\displaystyle\frac{L}{{2\pi a}} = 1.$

Because of the form of the dispersion relation the one particle partition function is equivalent to the partition function of a nonstandard quantum oscillator with energy frequency relation $E = \displaystyle\frac{1}{a}\log (a(n + \displaystyle\frac{1}{2})\omega  + 1).$ Note this relation reduces to the usual relation $E = (n + \displaystyle\frac{1}{2})\omega $
in the limit as $a$
goes to zero. The partition function of the nonstandard oscillator is:
	\begin{equation}
\begin{array}{l}
 z(\beta ,\mu ) = \sum\limits_{n = 0}^\infty  {e^{ - \beta E_n  - \beta \mu n} }  \\ 
  = \sum\limits_{n = 0}^\infty  {(a(n + \displaystyle\frac{1}{2})\omega  + 1)^{ - \beta /a} e^{ - \beta \mu n} }  = (a\omega )^{ - \beta /a} \phi (i\beta \mu ,\displaystyle\frac{\beta }{a},\displaystyle\frac{1}{2} + \displaystyle\frac{1}{{a\omega }}) \\ 
 \end{array}
\end{equation}
where we have introduced a chemical potential associated with the oscillator number operator. Continuing this function to the complex plane we see that when the special relation $a\omega  = 2$
is obeyed as well as $\beta \mu  = \pi i$
the Yang-Lee zeros of the partition function coincide with the zeta function zeros.
The Green's function nonstandard oscillator can be defined by the path integral:
	\begin{equation}
G(x,t;0,0) = \int\limits_{x(0) = 0}^{x(t) = x} {DxDpe^{i\int {d\tau L} } } 
\end{equation}
with $L = p\dot x - \displaystyle\frac{1}{a}\log (a\displaystyle\frac{1}{2}(\displaystyle\frac{{p^2 }}{m} + m\omega ^2 x^2 ) + 1).$
The classical equations associated with the nonstandard logarithmic oscillator are:
\begin{equation}
\begin{array}{l}
 E = \displaystyle\frac{1}{a}\log (\displaystyle\frac{a}{2}(\displaystyle\frac{{p^2 }}{m} + m\omega ^2 x^2 ) + 1) \\ 
 \dot x = \displaystyle\frac{{p/m}}{{\displaystyle\frac{a}{2}(\displaystyle\frac{{p^2 }}{m} + m\omega ^2 x^2 ) + 1}} = \displaystyle\frac{p}{{me^{aE} }} \\ 
 \dot p =  - \displaystyle\frac{x}{{\displaystyle\frac{a}{2}(\displaystyle\frac{{p^2 }}{m} + m\omega ^2 x^2 ) + 1}} =  - qme^{aE} \omega ^2 e^{ - 2aE}  =  - qm\omega ^2 e^{ - aE}  \\ 
 \displaystyle\frac{{p^2 }}{m} + m\omega ^2 q^2  = \displaystyle\frac{2}{a}(e^{aE}  - 1) \\ 
 x = x_0 \cos (\omega e^{ - aE} t) + \displaystyle\frac{{p_0 }}{{m\omega }}\sin (\omega e^{ - aE} t) \\ 
 p = p_0 \cos (\omega e^{ - aE} t) - x_0 \omega m\sin (\omega e^{ - aE} t) \\ 
 \end{array}
\end{equation}
Again in the nonstandard case the period depends on the energy and hence the amplitude through 
$T = \displaystyle\frac{{2\pi }}{{\omega e^{ - aE} }} = \displaystyle\frac{{2\pi }}{\omega }e^{aE}  = \displaystyle\frac{{2\pi }}{\omega }(1 + \displaystyle\frac{1}{2}m\omega ^2 x_0^2 a)$
for the case of $p_0  = 0$. The quantization yields the same eigenfunctions as the usual oscillator because of the expansion:
	\begin{equation}
H\psi _n  = E_n \psi _n  = \sum\limits_{k = 1}^\infty  {( - a)^{k - 1} (\omega (\hat n}  + \displaystyle\frac{1}{2})^k )\psi _n 
\end{equation}
Finally although the partition function of the nonstandard oscillator is the same as the one particle Weyl fermion partition function with dispersion relation $E = \displaystyle\frac{1}{a}\log (a\displaystyle\frac{{2\pi n}}{L} + 1)$
the underlying theories are different. To see this note that the Green's function for the nonstandard oscillator is:
\begin{equation}
G(x,t;0,0) = \sum\limits_{n = 0}^\infty  {e^{ - iE_n t} } H_n (x)H_n (0) = \sum\limits_{n = 0}^\infty  {(a(n + \displaystyle\frac{1}{2})\omega  + 1)^{ - it/a} } H_n (x)H_n (0)
\end{equation}
Whereas that for the nonstandard Weyl fermion the Green's function is:
\begin{equation}
G(x,t;0,0) = \sum\limits_{n = 0}^\infty  {e^{ - iE_n t} e^{2\pi in/L} }  = \sum\limits_{n = 0}^\infty  {(a\displaystyle\frac{{2\pi n}}{L}}  + 1)^{ - it/a} e^{2\pi in/L} 
\end{equation}
These are not the same although the partition function defined from the Green's function $z(\beta ) = \int {dxG(x,i\beta ;x,0)} $
are identical if one maps $\omega $
to $\displaystyle\frac{{2\pi }}{L}.$ Note the condition to obtain the Riemann zeta function $a\omega  = 2$ is far from the limit to obtain the standard oscillator $a\omega  \ll 1.$

\section{Scalar field} 

\subsection*{Green's function}

Most of the discussion has been with a fermion field obeying the dispersion relation
$p = \displaystyle\frac{1}{a}(e^{aE}  - 1)$. Many of the considerations apply to a scalar field $\Phi (x,t)$
as well. Also the theory of a single Weyl fermion has a gravitational anomaly. Adding left moving scalar matter to the theory can cancel this anomaly as happens in the Type 0B string model.
Taking the square of the Dirac operator we obtain the nonstandard Klein Gordon equation

\begin{equation}
(\partial _x^2  + \displaystyle\frac{1}{{a^2 }}(e^{ai\partial _t }  - 1)^2 )\Phi (x,t) = 0
\end{equation}
On a circle momentum is quantized as $p = \displaystyle\frac{n}{R}.$ Forming a series expansion for the scalar field:
\begin{equation}
\Phi (x,t) = \sum\limits_n {\displaystyle\frac{1}{{\sqrt {2E_n } }}} a_n \psi _n (x,t)
\end{equation}
Where $a_{ - n} $
and $a_n $
are creation and annihilation operators and $\psi _n $
are the mode solutions $\psi _n (x,t) = e^{ - iE_n t} e^{i\displaystyle\frac{n}{R}x} $.
The one particle energies are:
\begin{equation}
E_n  = \displaystyle\frac{1}{a}\log (a\displaystyle\frac{n}{R} + 1)
\end{equation}
The scalar field expansion is given by:
	\begin{equation}
\Phi (x,t) = \sum\limits_{n = 0}^\infty  {\displaystyle\frac{{a_n }}{{\sqrt {2\displaystyle\frac{1}{a}\log (a\displaystyle\frac{n}{R} + 1)} }}} (a\displaystyle\frac{n}{R} + 1)^{ - it/a} e^{in\displaystyle\frac{x}{R}}  + c.c
\end{equation}
The bosonic or scalar Green's function is denoted by $K$
and is expressed as:
	\begin{equation}
K(x,t;0,0) = \left\langle {0|\Phi (x,t)\Phi (0,0)|\left. 0 \right\rangle } \right. = \sum\limits_{n = 0}^\infty  {\displaystyle\frac{{(a\displaystyle\frac{n}{R} + 1)^{ - it/a} e^{i\displaystyle\frac{n}{R}x} }}{{2\displaystyle\frac{1}{a}\log (a\displaystyle\frac{n}{R} + 1)}}} 
\end{equation}
The two point function between the scalar field and its time derivative  is somewhat simpler and can be expressed in terms of the Lerch zeta function:
	\begin{equation}
\left\langle {0|i\partial _t } \right.\Phi (x,t)\Phi (0,0)|\left. 0 \right\rangle  = \displaystyle\frac{1}{2}\sum\limits_{n = 0}^\infty  {(a\displaystyle\frac{n}{R} + 1)^{ - it/a} e^{i\displaystyle\frac{n}{R}x} }  = \displaystyle\frac{1}{2}(\displaystyle\frac{R}{a})^{it/a} \phi (\displaystyle\frac{x}{{2\pi R}},\displaystyle\frac{t}{a},\displaystyle\frac{R}{a})
\end{equation}

\subsection*{Statistical Mechanics}

The statistical mechanics of the scalar field is determined by the Bose-Einstein distribution with the multiparticle Hamiltonian:
	\begin{equation}
H = \sum\limits_{n = 0}^\infty  {\varepsilon _n a_{ - n} a_n } 
\end{equation}
and the one particle energies $a\epsilon_n=\log(1+an/R)$. The partition function is given by:
	\begin{equation}
Z = tr(e^{ - \beta H} ) = \prod\limits_{n = 0}^\infty  {(1 - e^{ - \beta \varepsilon _n } )^{ - 1}  = } \prod\limits_{n = 0}^\infty  {(1 - (a\displaystyle\frac{n}{R} + 1)^{ - \beta /a} )^{ - 1} } 
\end{equation}
As in the Fermi-Dirac case one can develop a series representation for the free energy. Taking the logarithm of (C.8) we have:
	\begin{equation}
\log Z =  - \sum\limits_{n = 0}^\infty  {\log (1 - (a\displaystyle\frac{n}{R} + 1)^{ - \beta /a} )} 
\end{equation}
Now we can expand the logarithm appearing in each term of the sum using the formula $ - \log (1 - y) = \sum\limits_{m = 1}^\infty  {\displaystyle\frac{{y^m }}{m}} $
 to obtain the series representation:
	\begin{equation}
\log Z = \sum\limits_{n = 0}^\infty  {\sum\limits_{m = 1}^\infty  {\displaystyle\frac{1}{m}} } (a\displaystyle\frac{n}{R} + 1)^{ - \beta m/a}  = \sum\limits_{m = 1}^\infty  {\displaystyle\frac{1}{m}} (\displaystyle\frac{R}{a})^{\beta m/a} \zeta _H (\displaystyle\frac{{\beta m}}{a},\displaystyle\frac{R}{a})
\end{equation}
For the special case $R = a$
relevant to the Riemann zeta function we have:
	\begin{equation}
\log Z(R = a) = \sum\limits_{m = 1}^\infty  {\displaystyle\frac{1}{m}} \zeta (\displaystyle\frac{{\beta m}}{a})
\end{equation}
The thermal divergence occurs at the critical inverse temperatures $\beta _c  = \displaystyle\frac{a}{m}$
with $m$
an integer.

The multiparticle density of states can be obtained from the inverse Laplace transform of the partition function as
	\begin{equation}
\sigma (E) = \int\limits_{ - i\infty  + c}^{i\infty  + c} {d\beta Z(\beta )e^{\beta E} } 
\end{equation}
For the special case $R = a$
the total energy of the multiparticle bosonic system is given by:
	\begin{equation}
E_{tot}  = \displaystyle\frac{1}{a}\sum\limits_{n = 2}^\infty  {N_n \log n} 
\end{equation}
Where $N_n $
is the number of particles of momentum $n$
. As in the Fermi-Dirac case the quantity $\sigma (E)$
has interpretation in terms of multiplicative number theory. $\sigma (E)$
is the number of ways of writing $e^{aE} $
as a product irrespective of order and allowing repeat factors.

For example we can enumerate the states that yield total energy $E = \displaystyle\frac{1}{a}\log (120)$
\begin{equation}
\begin{array}{l}
 120 = 5! =  \\ 
 5 \cdot 3 \cdot 2 \cdot 2 \cdot 2;     a_{ - 5} a_{ - 3} a_{ - 2}^3 |\left. 0 \right\rangle  \\ 
 15 \cdot 2 \cdot 2 \cdot 2;       a_{ - 15} a_{ - 2}^3 |\left. 0 \right\rangle  \\ 
 10 \cdot 3 \cdot 2 \cdot 2;       a_{ - 10} a_{ - 3} a_{ - 2}^2 |\left. 0 \right\rangle  \\ 
 6 \cdot 5 \cdot 2 \cdot 2;         a_{ - 6} a_{ - 5} a_{ - 2}^2 |\left. 0 \right\rangle  \\ 
 5 \cdot 4 \cdot 3 \cdot 2;         a_{ - 5} a_{ - 4} a_{ - 3} a_{ - 2} |\left. 0 \right\rangle  \\ 
 30 \cdot 2 \cdot 2;          a_{ - 30} a_{ - 2}^2 |\left. 0 \right\rangle  \\ 
 20 \cdot 3 \cdot 2;          a_{ - 20} a_{ - 3} a_{ - 2} |\left. 0 \right\rangle  \\ 
 15 \cdot 4 \cdot 2;          a_{ - 15} a_{ - 4} a_{ - 2} |\left. 0 \right\rangle  \\ 
 12 \cdot 5 \cdot 2;          a_{ - 12} a_{ - 5} a_{ - 2} |\left. 0 \right\rangle  \\ 
 10 \cdot 4 \cdot 3;          a_{ - 10} a_{ - 4} a_{ - 3} |\left. 0 \right\rangle  \\ 
 8 \cdot 5 \cdot 3;            a_{ - 8} a_{ - 5} a_{ - 3} |\left. 0 \right\rangle  \\ 
 6 \cdot 5 \cdot 4;            a_{ - 6} a_{ - 5} a_{ - 4} |\left. 0 \right\rangle  \\ 
 60 \cdot 2;             a_{ - 60} a_{ - 2} |\left. 0 \right\rangle  \\ 
 40 \cdot 3;             a_{ - 40} a_{ - 3} |\left. 0 \right\rangle  \\ 
 30 \cdot 4;            a_{ - 30} a_{ - 4} |\left. 0 \right\rangle  \\ 
 24 \cdot 5;            a_{ - 24} a_{ - 5} |\left. 0 \right\rangle  \\ 
 20 \cdot 6;            a_{ - 20} a_{ - 6} |\left. 0 \right\rangle  \\ 
 15 \cdot 8;            a_{ - 15} a_{ - 8} |\left. 0 \right\rangle  \\
12 \cdot 10;            a_{ - 12} a_{ - 10} |\left. 0 \right\rangle  \\
120;              a_{ - 120} |\left. 0 \right\rangle  \\ 
 \end{array}
\end{equation}
Thus $\sigma (\displaystyle\frac{{\log (120}}{a}) = 20$
separate states. The bosonic scalar field has more states than the fermionic case as it allows repeats.


\begin{thebibliography}{100}


 
\bibitem{Khuri:2001yd}
  N.~N.~Khuri,
  Math.\ Phys.\ Anal.\ Geom.\  {\bf 5}, 1 (2002)
  [arXiv:hep-th/0111067].
 
\bibitem{Berry}
Berry, M V and Keating, J P, 1999 SIAM Review 41 236-266, ``The Riemann zeros and eigenvalue asymptotics''.

\bibitem{Connes:2004es}
  A.~Connes and M.~Marcolli,
  arXiv:math.nt/0404128.
 
\bibitem{Elizalde:2001ee}
  E.~Elizalde, V.~Moretti and S.~Zerbini,
  Int.\ J.\ Mod.\ Phys.\ A {\bf 18}, 2189 (2003)
  [arXiv:math-ph/0109006].
 

  \bibitem{Titchmarsh}
E.~C.~Titchmarsh, The Theory of the Riemann Zeta-Function, Oxford (1951).

\bibitem{Edwards}
H.~M.~Edwards, Riemann's Zeta Function, Dover, NY (1974).

\bibitem{Ivic}
A.~Ivic, The RiemannZeta-Function, Theory and Applications, Dover, NY (1985).

\bibitem{Laurincikas}
A.~Laurincikas and R.~Garunkstis, The Lerch Zeta Function, Kluwer (2002).

\bibitem{bateman}
A.~Erdélyi, W.~ Magnus, F.~Oberhettinger and F.~.G.~Tricomi, Higher Transcendental Functions, Malabar, FL: Krieger,(1981)

 
\bibitem{Witten:2003nn}
  E.~Witten,
  Commun.\ Math.\ Phys.\  {\bf 252}, 189 (2004)
  [arXiv:hep-th/0312171].
\bibitem{Lukas:2000wn}
  A.~Lukas, P.~Ramond, A.~Romanino and G.~G.~Ross,
  Phys.\ Lett.\ B {\bf 495}, 136 (2000)
  [arXiv:hep-ph/0008049].
 
\bibitem{Lukas:2000rg}
  A.~Lukas, P.~Ramond, A.~Romanino and G.~G.~Ross,
  JHEP {\bf 0104}, 010 (2001)
  [arXiv:hep-ph/0011295].
 
\bibitem{Ng:2003rv}
  J.~N.~Ng,
  arXiv:hep-ph/0311352.
 
 
\bibitem{Dienes:2000ph}
  K.~R.~Dienes and I.~Sarcevic,
  Phys.\ Lett.\ B {\bf 500}, 133 (2001)
  [arXiv:hep-ph/0008144].
 
\bibitem{Witten:1981nf}
  E.~Witten,
  Nucl.\ Phys.\ B {\bf 188}, 513 (1981).

\bibitem{Greene:1989ya}
 B.~R.~Greene, A.~D.~Shapere, C.~Vafa and S.~T.~Yau,
 Nucl.\ Phys.\ B {\bf 337}, 1 (1990).

\bibitem{Moore:1987ue}
  G.~W.~Moore,
  Nucl.\ Phys.\ B {\bf 293}, 139 (1987)
  [Erratum-ibid.\ B {\bf 299}, 847 (1988)].
 
\bibitem{Nair:1986zn}
  V.~P.~Nair, A.~D.~Shapere, A.~Strominger and F.~Wilczek,
  Nucl.\ Phys.\ B {\bf 287}, 402 (1987).
 
\bibitem{Gozzi:1988uj}
  E.~Gozzi and M.~Reuter,
  Nucl.\ Phys.\ B {\bf 320}, 160 (1989).
 
\bibitem{Giannakis:1988yh}
  I.~Giannakis, C.~R.~Ordonez, M.~A.~Rubin and R.~Zucchini,
  Int.\ J.\ Theor.\ Phys.\  {\bf 28}, 3 (1989).
 

\bibitem{Witten:1982fp}
 E.~Witten,
 Phys.\ Lett.\ B {\bf 117}, 324 (1982).
 
\bibitem{Axenides:1993pn}
  M.~Axenides, A.~Johansen and H.~B.~Nielsen,
  Mod.\ Phys.\ Lett.\ A {\bf 9}, 623 (1994)
  [arXiv:hep-ph/9309278].
 
\bibitem{Kikukawa:1997qh}
  Y.~Kikukawa and H.~Neuberger,
  Nucl.\ Phys.\ B {\bf 513}, 735 (1998)
  [arXiv:hep-lat/9707016].
 
\bibitem{Callan:1984sa}
  C.~G.~.~Callan and J.~A.~Harvey,
  Nucl.\ Phys.\ B {\bf 250}, 427 (1985).
 

\bibitem{Kaplan:1992bt}
  D.~B.~Kaplan,
  Phys.\ Lett.\ B {\bf 288}, 342 (1992)
  [arXiv:hep-lat/9206013].
 
 
\bibitem{Neuberger:2003nu}
  H.~Neuberger,
  arXiv:hep-lat/0311040.
 
\bibitem{Spector:1988nn}
  D.~Spector,
  Commun.\ Math.\ Phys.\  {\bf 127}, 239 (1990).
 
\bibitem{Julia:1993pz}
  B.~L.~Julia,
LPTENS-93-05
B.L. Julia, Physica A 203, 425 (1994).
 
\bibitem{Bowick:1990mx}
  M.~J.~Bowick,
SU-4238-444
{\it Presented at 2nd Workshop on Thermal Field Theories and Their Applications, Taukuba, Japan, Jul 23-27, 1990}
 
\bibitem{Andrews}
G.~E.~Andrews, Number Theory, Dover, NY (1971).
 
\bibitem{Moore:1991sf}
  G.~W.~Moore,
  Nucl.\ Phys.\ B {\bf 368}, 557 (1992).

\bibitem{Berenstein:2004kk}
  D.~Berenstein,
  JHEP {\bf 0407}, 018 (2004)
  [arXiv:hep-th/0403110].

\bibitem{Barton:1984ey}
  G.~Barton,
  Annals Phys.\  {\bf 166}, 322 (1986).
 
\bibitem{Gross:1990st}
  D.~J.~Gross and I.~R.~Klebanov,
  Nucl.\ Phys.\ B {\bf 352}, 671 (1991).
 
\bibitem{Freund:1991md}
  P.~G.~O.~Freund,
  J.\ Math.\ Phys.\  {\bf 33}, 1148 (1992).

\bibitem{Klebanov:1991qa}
  I.~R.~Klebanov,
  arXiv:hep-th/9108019.

\bibitem{Takayanagi:2004jz}
  T.~Takayanagi,
  JHEP {\bf 0405}, 063 (2004)
  [arXiv:hep-th/0402196].

\bibitem{Creutz:1980gp}
  M.~Creutz and B.~Freedman,
  Annals Phys.\  {\bf 132}, 427 (1981).
  

\bibitem{Imbimbo:1995yv}
  C.~Imbimbo and S.~Mukhi,
  Nucl.\ Phys.\ B {\bf 449}, 553 (1995)
  [arXiv:hep-th/9505127].
 
\bibitem{Imbimbo:1995ns}
  C.~Imbimbo and S.~Mukhi,
  arXiv:hep-th/9511127.
 
\bibitem{Hyun:2005fq}
  S.~Hyun, K.~Oh, J.~D.~Park and S.~H.~Yi,
  arXiv:hep-th/0502075.
  
\bibitem{Ghoshal:1995wm}
  D.~Ghoshal and C.~Vafa,
  Nucl.\ Phys.\ B {\bf 453}, 121 (1995)
  [arXiv:hep-th/9506122].
 
\bibitem{Milne}
J.~Milne, Elliptic Curves, online course notes.\\
 http://www.jmilne.org/math/CourseNotes/math679.html (1996).
 
\bibitem{Ribet}
K.~Ribet, Ann.Fac.Sci.Toulouse.Math.11,116(1990).

\bibitem{Shimura}
G.~Shimura and Y.~Taniyama, Complex Multiplication of Abelian Varieties and Its Applications to Number Theory. Tokyo: Mathematical Society of Japan, (1961).

\bibitem{Douglas:2003up}
  M.~R.~Douglas, I.~R.~Klebanov, D.~Kutasov, J.~Maldacena, E.~Martinec and N.~Seiberg,
  arXiv:hep-th/0307195.

\bibitem{Ghoshal:1992kx}
  D.~Ghoshal, D.~P.~Jatkar and S.~Mukhi,
  Nucl.\ Phys.\ B {\bf 395}, 144 (1993)
  [arXiv:hep-th/9206080].

\bibitem{Naka:2000uy}
  M.~Naka and M.~Nozaki,
  Nucl.\ Phys.\ B {\bf 599}, 334 (2001)
  [arXiv:hep-th/0010002].

\bibitem{Brink:1976sz}
  L.~Brink, S.~Deser, B.~Zumino, P.~Di Vecchia and P.~S.~Howe,
  Phys.\ Lett.\ B {\bf 64}, 435 (1976).
 
\bibitem{Brink:1976uf}
  L.~Brink, P.~Di Vecchia and P.~S.~Howe,
  Nucl.\ Phys.\ B {\bf 118}, 76 (1977).
 
 
\bibitem{Fainberg:1987jr}
  V.~Y.~Fainberg and A.~V.~Marshakov,
  Nucl.\ Phys.\ B {\bf 306}, 659 (1988).
 
\bibitem{Fradkin:1991ci}
  E.~S.~Fradkin and D.~M.~Gitman,
  Phys.\ Rev.\ D {\bf 44}, 3230 (1991).
 
\bibitem{Ambjorn:1989ba}
  J.~Ambjorn, B.~Durhuus and T.~Jonsson,
  Nucl.\ Phys.\ B {\bf 330}, 509 (1990).
 
\bibitem{Mikovic:1990dz}
  A.~Mikovic,
  Phys.\ Lett.\ B {\bf 261}, 41 (1991).


\bibitem{Jacobson:1983xt}
  T.~Jacobson,
  J.\ Phys.\ A {\bf 17}, 2433 (1984).
 
\bibitem{Feynman:1965}
R.~P.~Feynman and A.~R.~Hibbs, Quantrum Mechanics and Path Integrals, MacGraw-Hill, NY (1965).
 
\bibitem{Gersch:1981fv}
  H.~A.~Gersch,
  Int.\ J.\ Theor.\ Phys.\  {\bf 20}, 491 (1981).
 
\bibitem{Jacobson:1983qi}
  T.~Jacobson and L.~S.~Schulman,
  J.\ Phys.\ A {\bf 17}, 375 (1984).
  
\bibitem{Polyakov:1987ez}
  A.~M.~Polyakov,
  ``Gauge Fields And Strings,''(1987).


\bibitem{Henkel:1998zj}
  M.~Henkel and D.~Karevski,
  J.\ Phys.\ A {\bf 31}, 2503 (1998)
  [arXiv:cond-mat/9711265].
 
\bibitem{McGuigan:1991qp}
  M.~D.~McGuigan, C.~R.~Nappi and S.~A.~Yost,
  Nucl.\ Phys.\ B {\bf 375}, 421 (1992)
  [arXiv:hep-th/9111038].


\bibitem{Fainberg:1988zg}
  V.~Y.~Fainberg and A.~V.~Marshakov,
  Phys.\ Lett.\ B {\bf 211}, 81 (1988).

 \end{thebibliography}
\end{document}